\documentclass[longauth]{aa}

\usepackage{lscape}
\usepackage{natbib}
\usepackage{txfonts}
\usepackage{gensymb}
\usepackage[colorlinks=true,citecolor=blue]{hyperref}
\usepackage{graphicx}
\usepackage{tabularx}
\bibpunct{(}{)}{;}{a}{}{,} 

\def\m2s2{\,m$^{2}$\,s$^{-2}$} 

\begin{document}

   \title{New binaries from the SHINE survey\thanks{Full Tables 1–11 are only available at the CDS via anonymous ftp to \url{cdsarc.u-strasbg.fr (130.79.128.5) or via http://cdsarc.u-strasbg.fr/viz-bin/cat/J/A+A/663/A144}} \thanks{Based on data collected at the European Southern Observatory, Chile. \textit{SHINE datasets}: ESO Programmes 095.C-0298, 096.C-0241, 097.C-0865, 098.C-0865, 099.C-0209, 1100.C-0481, 104.C-0416. 
   \textit{Additional datasets:} ESO Programmes 074.C-0037, 076.C-0010, 077.C-0012, 079.C-0046, 083.A-9003, 090.A-9010, 095.C-0.389, 098.C-0739, 1101.C-0557, 103.C-0.628}}
    
\titlerunning{New binaries from SHINE}

\author{
    M.~Bonavita \inst{\ref{padova},\ref{supa_edin},\ref{ifa_uoe}} \and
    R.~Gratton\inst{\ref{padova}} \and 
    S.~Desidera\inst{\ref{padova}} \and
    V. Squicciarini\inst{\ref{padova}, \ref{unipd}} \and
    V.~D'Orazi\inst{\ref{padova}} \and
    A.~Zurlo\inst{\ref{lam},\ref{diegoportales1},\ref{diegoportales2}} \and
    B.~Biller\inst{\ref{ifa_uoe},\ref{supa_edin},\ref{mpia}} \and
    G.~Chauvin\inst{\ref{ipag},\ref{umi}} \and
    C.~Fontanive\inst{\ref{bern},\ref{padova}} \and
    M.~Janson\inst{\ref{mpia},\ref{stock}} \and
    S.~Messina\inst{\ref{catania}} \and
    F.~Menard\inst{\ref{ipag}} \and
    M.~Meyer\inst{\ref{umich},\ref{eth}} \and
    A.~Vigan\inst{\ref{lam}} \and
    H.~Avenhaus\inst{\ref{eth}} \and
    R.~Asensio~Torres\inst{\ref{mpia}} \and
    J.-L.~Beuzit\inst{\ref{lam},\ref{ipag}} \and
    A.~Boccaletti\inst{\ref{lesia}} \and
    M.~Bonnefoy\inst{\ref{ipag}} \and
    W.~Brandner\inst{\ref{mpia}} \and
    F.~Cantalloube\inst{\ref{mpia}} \and
    A.~Cheetham\inst{\ref{geneva}} \and
    M.~Cudel\inst{\ref{ipag}} \and
    S.~Daemgen\inst{\ref{eth}} \and
    P.~Delorme\inst{\ref{ipag}} \and
    C.~Desgrange\inst{\ref{diegoportales1},\ref{cral}} \and
    C.~Dominik\inst{\ref{ams}} \and
    N.~Engler\inst{\ref{eth}}
    P.~Feautrier\inst{\ref{ipag}} \and 
    M.~Feldt\inst{\ref{mpia}} \and
    R.~Galicher\inst{\ref{lesia}} \and
    A.~Garufi\inst{\ref{eth},\ref{arcetri}} \and
    D.~Gasparri\inst{27} \and
    C.~Ginski\inst{\ref{leiden}} \and
    J.~Girard\inst{\ref{eso_chile}} \and
    A.~Grandjean\inst{\ref{ipag}} \and
    J.~Hagelberg\inst{\ref{geneva}} \and
    Th.~Henning\inst{\ref{mpia}} \and
    S.~Hunziker\inst{\ref{eth}} \and
    M.~Kasper\inst{\ref{eso_garching}} \and
    M.~Keppler\inst{7} \and
    E.~Lagadec\inst{\ref{oca}} \and
    A.-M.~Lagrange\inst{\ref{ipag}} \and
    M.~Langlois\inst{\ref{cral},\ref{lam}} \and
    J.~Lannier\inst{\ref{ipag}} \and
    C.~Lazzoni\inst{\ref{padova}, \ref{unipd}} \and
    H.~Le Coroller\inst{\ref{lam}} \and
    R.~Ligi\inst{\ref{brera}} \and
    M.~Lombart\inst{\ref{umi},\ref{cral}} \and
    A.-L.~Maire\inst{\ref{star},\ref{mpia}} \and
    S.~Mazevet\inst{\ref{meudon}}
    D.~Mesa\inst{\ref{padova}} \and
    D.~Mouillet\inst{\ref{ipag}}
    C.~Moutou\inst{\ref{irap},\ref{lam}} \and
    A.~M\"uller\inst{\ref{mpia}} \and
    S.~Peretti\inst{\ref{geneva}}
    C.~Perrot\inst{\ref{lesia},\ref{valpo1},\ref{valpo2}} \and
    S.~Petrus\inst{\ref{ipag}} \and
    A.~Potier\inst{\ref{lesia}} \and
    J.~Ramos\inst{\ref{mpia}} \and
    E.~Rickman\inst{\ref{esa}} \and
    D.~Rouan\inst{\ref{ipag}} \and
    G.~Salter\inst{\ref{lam}} \and
    M.~Samland\inst{\ref{mpia},\ref{stock}} \and
    T.~Schmidt\inst{\ref{lesia}} \and
    E.~Sissa\inst{\ref{padova}} \and
    T.~Stolker\inst{\ref{ams}} \and
    J.~Szul\'agyi\inst{\ref{eth}} \and
    M.~Turatto\inst{\ref{padova}} \and
    S.~Udry\inst{\ref{geneva}} \and
    F.~Wildi\inst{\ref{geneva}}
}

\authorrunning{M. Bonavita et al.}

\institute{
    INAF - Osservatorio Astronomico di Padova, Vicolo della Osservatorio 5, 35122, Padova, Italy \label{padova} \email{\href{mailto:mbonav@roe.ac.uk}{mbonav@roe.ac.uk}}
    \and
    Scottish Universities Physics Alliance (SUPA), Institute for Astronomy, University of Edinburgh, Blackford Hill, Edinburgh EH9 3HJ, UK \label{supa_edin}
    \and
    Institute for Astronomy, University of Edinburgh, EH9 3HJ, Edinburgh, UK \label{ifa_uoe}
    \and 
    Universit\'a degli studi di Padova, Dipartimento di Fisica ed Astronomia "Galileo Galilei", Via G. Marzolo, 8 - 35131 Padova, Italy
    \label{unipd}
    \and
    N\'ucleo de Astronom\'ia, Facultad de Ingenier\'ia y Ciencias, Universidad Diego Portales, Av. Ejercito 441, Santiago, Chile \label{diegoportales1}
    \and
    Escuela de Ingenier\'ia Industrial, Facultad de Ingenier\'ia y Ciencias, Universidad Diego Portales, Av. Ejercito 441, Santiago, Chile \label{diegoportales2}
    \and
    Aix Marseille Univ, CNRS, CNES, LAM, Marseille, France \label{lam} 
    \and
    Max Planck Institute for Astronomy, K\"onigstuhl 17, D-69117 Heidelberg, Germany \label{mpia}
    \and 
    Department of Astronomy, Stockholm University, SE-10691 Stockholm, Sweden \label{stock}
    \and
    Univ. Grenoble Alpes, CNRS, IPAG, F-38000 Grenoble, France \label{ipag}
    \and
    Unidad Mixta Internacional Franco-Chilena de Astronom\'{i}a, CNRS/INSU UMI 3386 and Departamento de Astronom\'{i}a, Universidad de Chile, Casilla 36-D, Santiago, Chile \label{umi}
    \and
    Geneva Observatory, University of Geneva, Chemin Pegasi 51, 1290 Versoix, Switzerland \label{geneva}
    \and
    CRAL, CNRS, Université Lyon 1,Université de Lyon, ENS, 9 avenue Charles Andre, 69561 Saint Genis Laval, France \label{cral}
    \and
    LESIA, Observatoire de Paris, Université PSL, Universit\'e de Paris, CNRS, Sorbonne Université, 5 place Jules Janssen, 92195 Meudon, France \label{lesia}
    \and
    Anton Pannekoek Institute for Astronomy, Science Park 9, NL-1098 XH Amsterdam, The Netherlands \label{ams}
    \and
    Center for Space and Habitability, University of Bern, 3012 Bern, Switzerland \label{bern}
    \and
    Department of Astronomy, University of Michigan, Ann Arbor, MI 48109, USA \label{umich}
    \and
    Institute for Particle Physics and Astrophysics, ETH Zurich, Wolfgang-Pauli-Strasse 27, 8093 Zurich, Switzerland \label{eth}
    \and
    STAR Institute, University of Li\`ege, All\'ee du Six Ao\^ut 19c, B-4000 Li\`ege, Belgium \label{star}
    \and
    INAF - Catania Astrophysical Observatory, via S. Sofia 78, I-95123 Catania, Italy \label{catania}
    \and
    Univ. de Toulouse, CNRS, IRAP, 14 avenue Belin, F-31400 Toulouse, France \label{irap}
    \and
    European Southern Observatory (ESO), Karl-Schwarzschild-Str. 2,85748 Garching, German \label{eso_garching}
    \and
    Universit\'e C\^ote d’Azur, OCA, CNRS, Lagrange, France \label{oca}
    \and
    Leiden Observatory, Leiden University, PO Box 9513, 2300 RA Leiden, The Netherlands \label{leiden}
    \and 
    INAF - Osservatorio Astronomico di Brera, Via E. Bianchi 46, 23807 Merate, Italy \label{brera}
    \and
    European Southern Observatory, Alonso de C\`ordova 3107, Vitacura, Casilla 19001, Santiago, Chile \label{eso_chile}
    \and 
    Space Telescope Science Institute, 3700 San Martin Drive, Baltimore, MD, 21218, USA \label{stsci}
    Instituto de F\'isica y Astronom\'ia, Facultad de Ciencias, Universidad de Valpara\'iso, Av. Gran Breta\~na 1111, Valpara\'iso, Chile \label{valpo1}
    \and
    NOVA/UVA \label{nova}
    \and
    European Space Agency (ESA), ESA Office, Space Telescope Science Institute, 3700 San Martin Drive, Baltimore, MD21218, USA \label{esa}
    \and 
    INAF - Osservatorio Astrofisico di Arcetri \label{arcetri}
    \and 
    Laboratoire Univers et Théories, Université Paris Diderot, Observatoire de Paris, PSL University, 5 Place Jules Janssen, 92195 Meudon, France \label{meudon}}

\date{Received 08/02/2021 / Accepted 25/03/2021}

\abstract{
We present the multiple stellar systems observed within the SpHere INfrared survey for Exoplanet (SHINE). SHINE searched for sub-stellar companions to young stars using high contrast imaging. Although stars with known stellar companions within the SPHERE field of view ($<5.5$ arcsec) were removed from the original target list, we detected additional stellar companions to 78 of the 463 SHINE targets observed so far. Twenty-seven per cent of the systems have three or more components. Given the heterogeneity of the sample in terms of observing conditions and strategy, tailored routines were used for data reduction and analysis, some of which were specifically designed for these datasets.
We then combined SPHERE data with literature and archival data, TESS light curves, and Gaia parallaxes and proper motions for an accurate characterisation of the systems. Combining all data, we were able to constrain the orbits of 25 systems.
We carefully assessed the completeness of our sample for separations between 50-500 mas (corresponding to periods of a few years to a few decades), taking into account the initial selection biases and recovering part of the systems excluded from the original list due to their multiplicity. This allowed us to compare the binary frequency for our sample with previous studies and highlight interesting trends in the mass ratio and period distribution. 
We also found that, when such an estimate was possible, the values of the masses derived from dynamical arguments were in good agreement with the model predictions.
Stellar and orbital spins appear fairly well aligned for the 12 stars that have enough data, which favours a disk fragmentation origin.
Our results highlight the importance of combining different techniques when tackling complex problems such as the formation of binaries and show how large samples can be useful for more than one purpose.}

   \keywords{Planetary systems - (Stars:) binaries: visual}

   \maketitle
   
\section{Introduction}
\label{sec:intro}
Multiple stellar systems are common in our solar neighbourhood \citep{Duquennoy1991, Raghavan2010, Duchene2013} and in the Galaxy, regardless of the environment. We observe binaries in sparse young star-forming regions \citep[SFRs;][]{Ghez1997, Nguyen2012} as well as in older, much denser populations, such as globular clusters \citep{Sollima2007}. More than 70\% of massive early-type stars \citep{Kouwenhoven2007, Peter2012} and 50\%--60\% of solar-type stars \citep{Duquennoy1991, Raghavan2010, Duchene2013, Moe2017} are observed in binary or higher-order multiple systems, with the fraction decreasing to 30\%--40\% for M stars \citep{Fischer1992, Delfosse2004, Janson2012}. An even higher fraction of binaries have been observed in low-density SFRs \citep{Duchene1999, Kraus2008, Kraus2011}, but it is as yet unclear if this excess extends over all masses and separations or is instead limited to the smallest masses or the widest separation range. 

There is still considerable debate on the main mechanism(s) leading to binary formation \citep[see e.g.][]{Tohline2002, Kratter2011, Duchene2013}. The favoured scenarios are (turbulent) core fragmentation of clouds for separations higher than 500 au \citep{Offner2010, Offner2016} and disk fragmentation for separations lower than 500 au \citep{Kratter2010}, with the two mechanisms not to be thought of as mutually exclusive. A nice example of multiple star formation caught in the act with both mechanisms likely working simultaneously on different scales is L1448 IRS3 \citep{Reynolds2020}. The values of the separation mentioned above only apply to solar-type stars, while higher- or lower-mass stars could behave differently (see e.g. \citealt{Andrews2009, White2001}). Disk fragmentation is expected to be more efficient around massive stars because of the larger value of the accretion rate from the natal cloud and hence the larger expected disk-to-star mass ratio during early phases of formation, when binaries are likely to form (see e.g. \citealt{Andrews2009, Lodato2008, Schib2020}). In disk fragmentation, mass accretion onto the secondary may be favoured with respect to accretion onto the primary \citep{Bate2002}; if the disk survives long enough, this would lead to a preference for equal mass binaries \citep{Kratter2010} that are observed to be over-represented over a wide range of periods (see e.g. \citealt{Lucy1979, Raghavan2010}). On the other hand, the disk may disperse before this condition is met; hence, the final mass ratio is not firmly established and may well be variable from case to case. 
It is difficult to accurately predict the outcome of binary formation from disk fragmentation due to the huge range of parameters involved and the complexity of the basic mechanisms that are often poorly understood (see \citealt{Bate2018, Schib2020} and the discussion in \citealt{TokovininMoe2020}). Many uncertainties remain regarding the range of disk-to-star mass ratios, the accretion of mass onto the disk from the parental cloud, the threshold for the onset of disk instabilities, the migration of secondaries within the disk, the accretion rates on the stars, the loss of angular momentum related to magnetohydrodynamic winds, and the role of ternary or higher-multiplicity systems. Exploration of the wide range of parameters with detailed hydrodynamical models is currently extremely expensive in terms of computational time. If different mechanisms truly have different effects on the final distribution of the system parameters, for example, on the distribution of mass ratios as a function of separation, an accurate characterisation of the binary population is a key requirement for constraining binary formation models. 

Moreover, given that the typical size of protoplanetary disks is close to the peak of the log-normal distribution of binary separation \citep{Duquennoy1991, Raghavan2010, Moe2017, Najita2018, Ansdell2018, Cieza2019}, the majority of young stellar objects are part of multiple systems. For this reason, understanding binary star formation and the role played by stellar companions on protoplanetary disks is a key aspect of a complete understanding of planet formation and evolution \citep{Bonavita2020, Hirsch2020, Fontanive2019}. Recent discussions of the interplay between multiplicity and protoplanetary disks combining ALMA (Atacama Large Millimeter/submillimeter Array) and high contrast imaging data can be found in \citet{Zurlo2020, Zurlo2020b}.

In order to contribute to this discussion, in this paper we present a sample of 78 multiple systems observed in the context of the SpHere INfrared survey for Exoplanet (SHINE; for details see \citealt{Chauvin2017, vigan2020, Desidera2021, Langlois2021}) with the Spectro-Polarimetric High-contrast Exoplanet REsearch instrument mounted at the Very Large Telescope \citep[SPHERE@VLT][]{sphere}. While the observations were not acquired for the specific purpose of observing binaries, we show in the discussion that the very high spatial resolution and contrast of our data results in a very complete sample of binaries at projected separations from a few to a few tens of au, a region corresponding to the expected size of the protostellar disks. This region, which is close to the peak of the log-normal distribution of periods \citep{Duquennoy1991, Raghavan2010, Moe2017}, is difficult to observe with radial velocities (RVs; because of the long periods), with seeing-limited data (because of the resolution), or with speckle interferometry data (because of the contrast). Our excellent completeness allows a discussion of the mass ratio distribution. Furthermore, we searched the literature looking for additional information; this information allowed the orbital parameters for about 30\% of the observed systems to be constrained, which was useful for an early statistical discussion of the distribution of orbital parameters that can be compared with different formation scenarios. By construction, our sample consists of young stars in sparsely populated environments. This implies that the systems we consider are not expected to be significantly influenced by their neighbours. The observed distributions should thus reflect the properties of these systems at their birth. Finally, an extensive comparison with data from the Transiting Exoplanet Survey Satellite \citep[TESS][]{tess2015} and Gaia data provides a better picture of these systems over a very wide range of periods.

The paper is organised as follows: The properties of the systems in our sample are summarised in Sect.~\ref{sec:sample}; Sect.~\ref{sec:data_red} describes the observations and the method used for the data reduction; our main results are presented in Sect.~\ref{sec:results} and discussed in Sect.~\ref{sec:discussion}; finally, Sect.~\ref{sec:conclusions} draws the final conclusions, raises outstanding questions, and discusses future work aimed at answering said questions. In Appendix~\ref{app:prot} we report on data for these binaries obtained with TESS, and in Appendix \ref{app:targets} we discuss in some detail each of the 78 stellar systems considered in this paper.

\begin{table*}\renewcommand{\arraystretch}{1.2}
\caption{Sample characteristics (full table available through CDS)\label{tab:master}}
\resizebox{\linewidth}{!}{
\begin{tabular}{rllcccllll}
\hline \hline
ID  & RA (2000) & Dec.(2000) & Parallax  & \multicolumn{2}{c}{Proper Motion} & $J$ mag & $H$ mag & $K$ mag & SpType     \\
    & (h m s)   & (d m s )  & (mas)     &  RA (mas/yr) & Dec. (mas/yr)        &       &       &       &            \\
\hline \hline
HIP 2729    & 00 34 51.2019 & -61 54 58.129 & 22.512 $\pm$ 0.021 & 88.69 $\pm$ 0.04 & -52.66 $\pm$ 0.04 & 7.34 & 6.53 & 6.72 & K4Ve \\
AF Hor      & 02 41 47.3054 & -52 59 30.645 & 23.075 $\pm$ 0.044 & 93.52 $\pm$ 0.14 & -11.58 $\pm$ 0.17 & 8.48 & 7.64 & 7.85 & M2Ve \\
TYC 8491-0656-1 & 02 41 46.8356 & -52 59 52.395 & 22.937 $\pm$ 0.046 & 96.77 $\pm$ 0.06 & -14.16 $\pm$ 0.07 & 7.58 & 6.76 & 6.93 & K6Ve \\
TYC 8497-0995-1 & 02 42 33.0255 & -57 39 36.830 & 20.109 $\pm$ 0.010 & 84.95 $\pm$ 0.77 & -9.24 $\pm$ 0.82 & 8.56 & 7.78 & 7.97 & K5Ve \\
HIP 17157   & 03 40 29.3861 & -47 55 30.550 & 39.127 $\pm$ 0.063 & 91.54 $\pm$ 0.06 & 102.45 $\pm$ 0.11 & 7.13 & 6.33 & 6.52 & K7V \\
HIP 17797   & 03 48 35.8772 & -37 37 12.541 & 18.809 $\pm$ 0.222 & 74.44 $\pm$ 0.71 & -9.09 $\pm$ 0.87 & 3.9 & 4.626 & 4.824 & B9.5 \\
\hline
\end{tabular}}
\end{table*}

\section{Sample properties}
\label{sec:sample}

SHINE is a large direct imaging planet-search survey started at the VLT in 2015 in the framework of the SPHERE Guaranteed Time Observations (GTO) carried on by the SPHERE Consortium. The survey concept and the selection of the sample are described in detail in \cite{Desidera2021} and the observations and data reduction procedures in \cite{Langlois2021}, while the statistical analysis and inference on planet population from the first 150 stars (F150 sample) is presented in \citet{vigan2020}.

The targets for SHINE were chosen from an extended list of $\sim$ 800 young, nearby stars, optimised for detectability of planets with SPHERE \citep[see][for details]{Desidera2021}. A total of 463 of these targets were actually observed. Any known sub-arcsecond or spectroscopic binaries were excluded from the input list once it was frozen (mid 2014) both to avoid complications related to the impact on adaptive optics (AO) of a bright stellar companion and due to the focus on the main survey on single stars or members of wide binaries \footnote{A few known binaries were observed as special objects during the SHINE-GTO survey. They are not part of the statistical analysis, i.e. the sample of 463 stars that we consider here. Examples of binaries among special objects include HD142527 \citep{claudi2019}, V4046 Sgr \citep{dorazi2019},  HD 100453 \citep{benisty2017}, and HD1160 \citep{maire2016pztel}. A small filler-like programme on known binaries in young moving groups for dynamical mass determination was also performed \citep{rodet2018}. These binaries are also not considered in this paper}. 
Nevertheless, thanks to the unique sensitivity of SPHERE down to very close separations we identified 78 multiple systems among the SHINE targets. Of these 56 are newly discovered pairs, and the remaining are systems that either escaped the first selection or were discovered after the sample had already been frozen.\footnote{To be consistent with the initial selection bias, all the targets in the binary sample presented here have been excluded from the SHINE statistical sample}

Given the original selection bias against known binaries (both visual and spectroscopic) we expect the sample of binaries identified in this paper is highly skewed towards low separation, faint companions. This will be further discussed in Sect. 5. In the following subsection, we present the determination of the stellar properties for the objects in our sample, summarised in Table~\ref{tab:age_per}. 

\subsection{Distance and proper motion}
\label{sec:dist}
In nearly all the cases, distance and proper motion values were originally taken from the second Gaia Data Release \citep[Gaia DR2][]{gaia2018} and then updated once the Early Third Data Release \citep[EDR3][]{edr3} became available. Gaia parameters are missing for one object, HIP 107948, for which we adopted the Hipparcos values. For some other objects, the nominal Gaia errors on both parallax and proper motion are largely underestimated because of the effect of the companion, which is not taken into account in the astrometric solution. In particular, for TYC 7133-2511-1, we adopted the mean distance of the Cometary Globule CG 30 group \citep[CG30][]{yep2020} rather than Gaia ones (see Appendix \ref{app:targets} for a detailed discussion of this issue). Other interesting individual cases are discussed in Appendix \ref{app:targets}. This issue should be overcome in the final Gaia data release, which will include the presence of companions in the astrometric solution.

\subsection{Radial velocities}
\label{sec:rv}
Available RV time series  were also used when available to us to derive the orbital solutions for 24 of our targets, as detailed in Sect. \ref{sec:orbital_fit}. 
An in depth analysis, including a full orbital solution, for HIP~36985 and HIP~113201 is presented in \cite{Biller:2022}.

\subsection{Stellar ages}
\label{sec:ages}
The stellar ages, reported in Table~\ref{tab:age_per}, were derived using the methods described in \citet{desidera2015} and \cite{Desidera2021}. As a result, the stellar ages provided here are in the same scale as those in \citet{bonavita2016}, \citet{Vigan2017}, and \cite{Desidera2021}.
Membership to groups, as derived using the BANYAN $\Sigma$ online tool \footnote{\url{http://www.exoplanetes.umontreal.ca/banyan/banyansigma.php}} \citep{gagne2018} is the prime age method for our targets. The adopted ages for young moving groups are mostly based on \citet{bell2015} and are discussed in \citet{bonavita2016} and \cite{Desidera2021}. The binarity of all our targets adds significant uncertainties in several cases, leading to ambiguous results (e.g. highly significant membership in a given group or not depending on the adopted proper motion, RV, or parallax). These cases are discussed individually in Appendix~\ref{app:targets}. 
For field objects, or stars with ambiguous membership, the age is derived from indirect age indicators such as the equivalent width of 6708\AA\ Lithium doublet, rotation period, X-ray emission, chromospheric activity, and on isochrone fitting. When possible, we took advantage of the multiplicity of the objects considering the indicators of the components of the systems -- including wider companion outside the SPHERE field of view (FoV) -- to improve the reliability of the derived ages.

For several targets we obtained new measurements of spectroscopic parameters on the basis of spectra acquired for this purpose or as part of other programmes, using the FEROS \citep[The Fiber-fed Extended Range Optical Spectrograph][]{FEROS} and HARPS \citep[High Accuracy Radial velocity Planet Searcher][]{HARPS} spectrographs at La Silla Observatory. These measurements, performed on spectra reduced with the instrument pipelines, are used for the stellar characterisation, are presented in the notes on individual targets (Appendix~\ref{app:targets}).

When available, we also made use of data from TESS \citep{tess2015} to determine rotational periods for the stars and obtain an independent estimate of the age using gyrochronology. The details of the TESS data analysis is presented in Appendix~\ref{app:prot}, while the results obtained for the single targets are included in the notes in Appendix~\ref{app:targets}.

\subsection{Stellar masses}
\label{sec:mstar}
The masses for all the components of our systems were determined following the approach used for the targets of the BEAST \citep[The B-star Exoplanet Abundance Study][]{beast} survey. 
For objects with masses smaller than $1.4 M_\odot$ we used the BT-Settl pre-main-sequence isochrones \citep{2014IAUS..299..271A}, while for higher-mass objects we used the empirical tables by \cite{2013ApJS..208....9P} instead. For all our targets we used the distances from Sect. \ref{sec:dist} for the conversion to absolute magnitude. 
When more than one photometric measurement was available, we retrieved the mass using each one separately, obtaining values always compatible within the errors. 
For these objects the mass value adopted is the average of the single measurements. 
The method was applied using the adopted value of the age as well as with the minimum and maximum age, which allowed limits  to be put on the mass estimates.

\begin{table*}\renewcommand{\arraystretch}{1.5}
\caption{Values of the adopted age (Age$^{max}_{min}$) and rotation period (P$_{Rot}$), derived as described in Sect.~\ref{sec:ages}, for all the stars in our sample (full version avaliable through CDS). \label{tab:age_per}}
\resizebox{0.6\linewidth}{!}{    
\begin{tabular}{rllllll}
\hline \hline 
ID  & Age$^{max}_{min}$  & MG  & Age$_{rot}$    & $P_{rot}^{TESS}$  & $P_{rot}^{lit}$   & Notes \\
    &  (Myr)                     &     & (Myr)          & (days)            &  (days)           &  \\
\hline
HIP 2729 & 45$_{35}^{50}$ & TUC & -- & 0.3767 - 0.3263 & 0.377$^1$ & F, AB \\
AF Hor & 45$_{35}^{50}$ & TUC & 28 & 3.39 & -- & 0.0 \\
TYC 8491-0656-1 & 45$_{35}^{50}$ & TUC & -- & 0.5196 - 0.5603 & 0.560, 0519$^{1,2}$ & F, AB \\
TYC 8497-0995-1 & 45$_{35}^{50}$ & TUC & 135 & 7.408 & 7.38, 7.412$^{1,3}$ & 0.0 \\
HIP 17157 & 150$_{100}^{200}$ & --- & 170 & 7.694 & -- & 0.0 \\
HIP 17797 & 45$_{35}^{50}$ & TUC & 63 & 6.897, 11.766 & -- & B \\
\hline \hline
    \end{tabular}}\\
    
{\footnotesize \textbf{Notes:} The ages of the young moving groups (MG; Lower Centaurus Crux: LCC; Upper Centaurus-Lupus: UCL; Tucana-Horologium: TUC;  AB Doradus: ABDO; Columba: COL; Argus: ARG; Carina-Near: CANE; Beta Pictoris: BPIC; Upper Scorpius: US) are adopted from \citet{bonavita2016} and further discussed in Desidera et al. 2021. For Octans-Near (OCNE), the age is derived in this work from comparison of the age indicators with respect to those of the groups above and reference open clusters (Pleiades, Hyades). For CG30 group the age is taken from \citet{yep2020}. The ages in this work are therefore homogeneous with those adopted in \citet{Vigan2017,vigan2020}. \\
\noindent The ages of the objects with no clear moving group membership indications have been estimated following the approach described in Desidera et al. (2020, submitted, see Sect.\ref{sec:ages} for details). \\
\noindent For the objects with available TESS data we also include the derived value of (P$_{Rot}$) and the corresponding value of the gyrocronologic age (Age$_{rot}$). The values of P$_{Rot}$ retrieved from the literature are also listed with the corresponding reference ($^1$\citet{kiraga2012}; $^2$\citet{oelkers2018}, $^3$\citet{messina2010}, $^4$\citet{messina2011}, $^5$\citet{messina2017}, $^6$\citet{wright2011}, $^7$\citet{Desidera2021}.). \\
\noindent The note in the last column refers to the nature of the estimated period ({\it F:} fast rotator; {\it P:} retrieved period is likely to be from pulsations; {\it B:} retrieved period is likely the one of the listed companion; {\it AB:} period retrieved for both components). }
\end{table*}

\begin{table*}
\caption{Summary of the SPHERE setup used for our observations (full table available through CDS).}\label{t:sphere_obs}
    \begin{tabular}{rlllllllrr}
\hline \hline
ID  & Obs Date      & Mode  & \multicolumn{2}{c}{DITxNDIT (sec)}  & ND Filt &  FoV Rot.  & Seeing    & $\tau_0$ & Strehl  \\
    & (JD - 245000) &       &   $IFS_{Science}$ & $IFS_{PSF}$   &  & (deg)    & (arcsec)  & (ms)      & \\
\hline 
HIP 2729 & 57357.01 & IRDIFS & 96 x 48 & 32 x 5 & ND 2.0 & 28.8 & 0.50 & N/A &   \\
HIP 2729 & 58378.19 & IRDIFS & 64 x 64 & 4 x 31 & ND 1.0 & 26.4 & 0.93 & 3.5 & 0.73 \\
AF Hor & 57323.19 & IRDIFS & 64 x 47 & 16 x 8 & ND 1.0 & 25.3 & 0.95 & 1.6 &   \\
TYC 8491-0656-1 & 57322.18 & IRDIFS & 64 x 46 & 16 x 8 & ND 2.0 & 33.4 & 1.62 & 1.0 &   \\
TYC 8491-0656-1 & 58088.09 & IRDIFS & 96 x 32 & 4 x 31 & OPEN & 24.4 & 0.54 & 6.1 &   \\
TYC 8497-0995-1 & 57356.08 & IRDIFS & 64 x 55 & 8  0 & ND 1.0 & 27.5 & 1.62 & 11.5 &   \\
TYC 8497-0995-1 & 58087.07 & IRDIFS & 96 x 32 & 16 x 12 & ND 1.0 & 20.8 & 0.69 & 5.3 &   \\
HIP 17157 & 57675.26 & IRDIFS & 64 x 53 & 2 x 47 & ND 1.0 & 34.3 & 0.42 & 2.8 &   \\
HIP 17797 & 57709.28 & IRDIFS & 32 x 47 & 4 x 21 & ND 2.0 & 7.4 & 3.64 & 0.7 &   \\
HIP 17797 & 58089.12 & IRDIFS & 32 x 160 & 16 x 10 & OPEN & 78.7 & 2.68 & 2.6 &   \\
\hline 
\end{tabular}
\end{table*}

\section{Observations and data reduction} 
\label{sec:data_red}

\begin{figure*}
   \centering
   \includegraphics[width=0.48\textwidth]{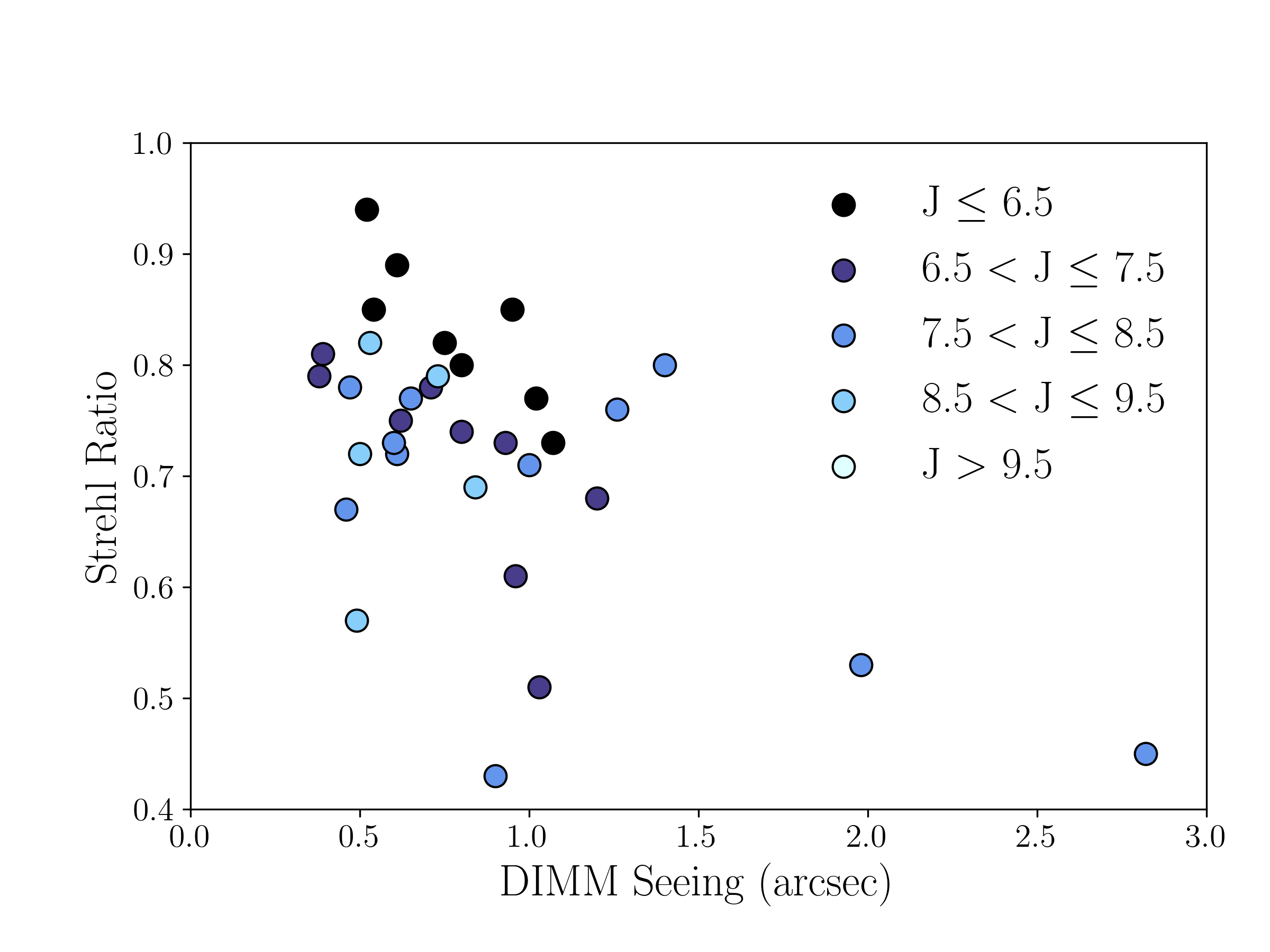}
   \includegraphics[width=0.48\textwidth]{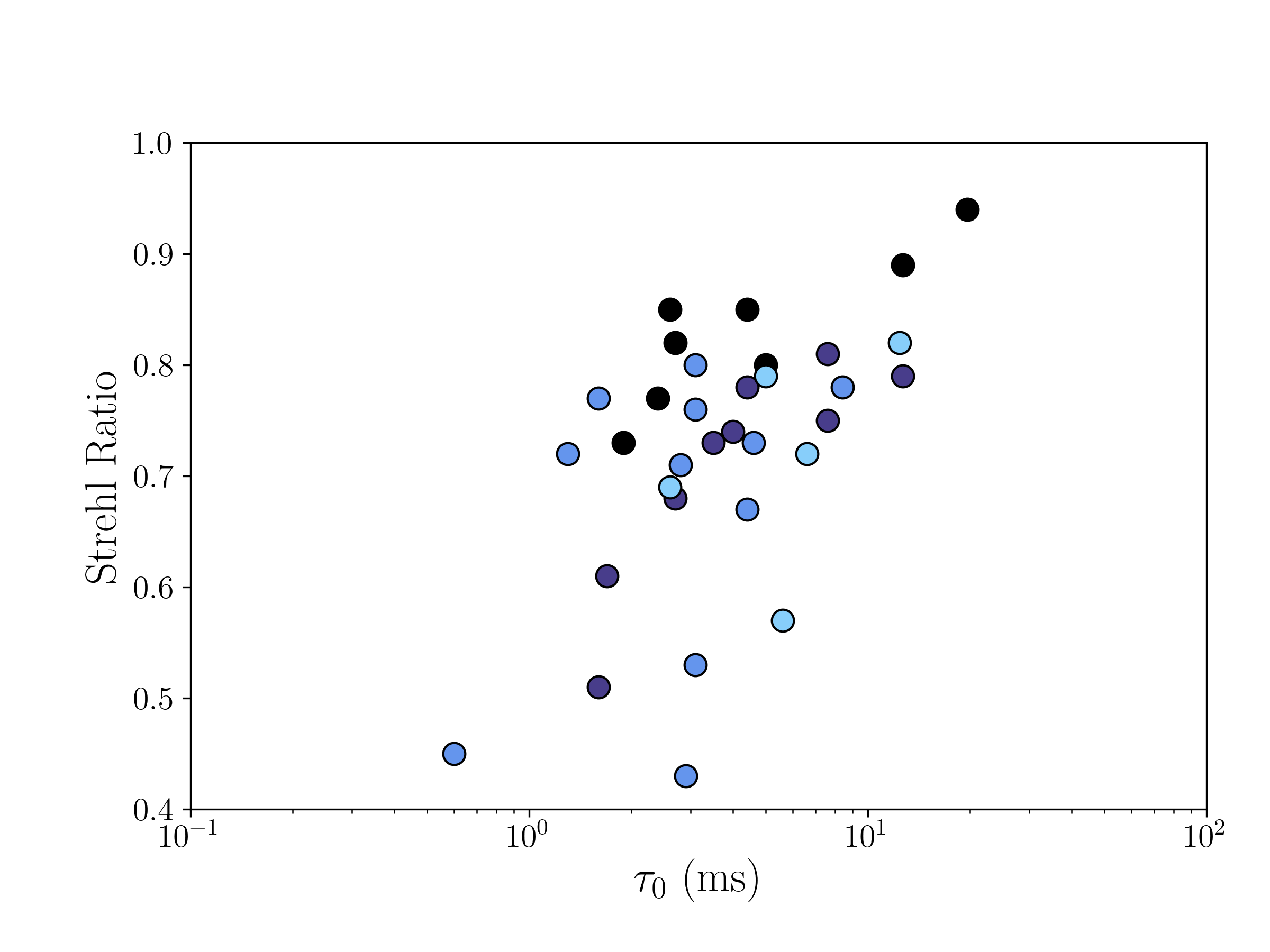}
   \caption{Strehl ratio (SR) as a function of seeing FWHM as measured by DIMM at Paranal (left panel) and atmospheric coherence time $\tau_0$ (right panel); the points show the average values during each individual observation of a target and are plotted with different colours depending on the stars' $J $ magnitude.}
\label{fig:strehl}%
    \end{figure*}

All observations were performed with VLT/SPHERE \citep{sphere} with the two near-infrared (NIR) subsystems, IFS (Integral Field Spectrograph \citealt{ifs}) and IRDIS (InfraRed Dual-band Imager and Spectrograph \citealt{irdis}), observing in parallel (IRDIFS Mode), with IRDIS in dual-band imaging (DBI) mode \citep{Vigan2010}. In a few cases the IRDIFS-EXT mode was used, which enables covering the Y-, J-, H-, and K-band in a single observation, providing a high level of spectral content for subsequent analyses.
A summary of the observing parameters and conditions is given in Table~\ref{t:sphere_obs}. The median full width at half maximum (FWHM) of the seeing as measured by the Paranal Differential Image Motion Monitor (DIMM) over the whole set of observations was 0.80~arcsec. The median value for the atmospheric coherence time $\tau_0$ was 3.5 ms. The median value of the Strehl ratio (SR) delivered by the SPHERE extreme adaptive optics system \citep[SAXO][]{Fusco16, sphere}, available only for about a third of the whole set, is 0.76. Figure~\ref{fig:strehl} shows the run of SR as a function of the seeing FWHM and of $\tau_0$; we used different symbols for stars in different ranges of magnitude. As expected, there is a clear correlation between atmospheric conditions, stellar magnitude, and SR; the best results are obtained considering $\tau_0$. The observed correlations reproduce well what is obtained for single stars \citep[see][]{Langlois2021}, that is, there was no significant degradation of the SAXO performances when observing binaries rather than single stars.\\
\indent The observing strategy for our targets was the same as the one used for the SHINE survey \citep[see e.g.][]{Chauvin2017}, so each observation was set to include
(i) a \emph{point spread function} (PSF) sub-sequence of off-axis unsaturated images obtained using a neutral density filter to avoid saturation (reported in column 5 of Table~\ref{t:sphere_obs}); (ii) a \emph{star centre} coronagraphic observation with four symmetric satellite spots used to achieve an accurate determination of the star position behind the coronagraphic mask for the following deep coronagraphic sequence;  (iii) the \emph{deep coronagraphic sub-sequence}, acquired with the apodised-pupil Lyot coronagraph implemented in SPHERE \citep{carbillet2011,Guerri2011}; and (iv) a new star centre sequence, a new PSF registration, and a short sky observing sequence for the fine correction of the hot pixel variation during the night. 

Nevertheless, given the focus of SHINE on single stars, a full dataset was only available for a subset of our targets for which the stellar companion was not obviously detected in the first PSF sequence. In a large fraction of cases only the PSF, and sometimes the centring, sequence was available. For this reason it was not always possible to simply reduce the data using the SPHERE Data Reduction and Handling (DRH) automated pipeline \citep{Pavlov2008} at the SPHERE Data Center (SPHERE-DC, see \citealt{delorme2017}), which assume that the whole sequence is available. In addition, many  stellar companions detected throughout this paper fall behind the field mask of the coronagraph, and are then not detectable on the deep coronagraphic sequence. To handle these cases we used a number of tailored routines, which are described in the following sections. 

\subsection{Standard SHINE reduction} \label{sec:specal}
Stellar companions are very bright in SHINE observations. The standard procedures devised to detect and measure faint companions based on differential imaging are adequate to detect and measure only those stellar companions with a rather high contrast. For these stellar companions, we could still adopt the usual reduction provided by the SPHERE Data Centre \citep{delorme2017, Galicher2018}. While various analysis techniques were run for these targets, the parameters considered in this paper for the stellar companions are those obtained using a procedure that simply rotates the images for the parallactic angle and then applies a median to produce the final image. This avoids concerns due to self-subtraction due to aggressive differential imaging procedures that are not required for these bright companions. 

\subsection{Manual detections on non-coronagraphic observations} \label{sec:non_coro_manual}
In several cases the companions are so bright that they were already detected in the quick look images at the telescope. Since SHINE focused on single stars or very wide binaries \citep[see][]{Desidera2021}, observations of these targets were interrupted (to save telescope time) and only the acquisition (non-coronagraphic) images were available. Such observations were not reduced using the standard procedure devised at the SPHERE Data Centre \citep{delorme2017}. In addition, in a few cases the full dataset (including the long sequence with the star behind the coronagraph) is available, but the bright secondary image actually saturated the detector (the rest of the image was still usable to detect additional close companion candidates).

In both cases, we used the non-coronagraphic (flux) calibrations -- where suitable neutral density filters (NDFs) were used to avoid saturation -- to extract the relative position and contrast of the companions relative to the primaries. This was done on the raw images using an IDL procedure reproducing the aperture magnitude algorithm of DAOPHOT \citep{Stetson1987}. 
In a few cases where the components are separated by less than the FWHM of the diffraction peak, we fitted the image assuming it is the sum of two typical PSFs and optimising the least square sum of residuals using as free parameters the intensities and positions with an Amoeba downhill approach. 
Astrometry was then obtained using the procedure recommended by \cite{Maire2016}. 

\subsection{Automatic detections on non-coronagraphic observations} \label{sec:non_coro_auto}
Very close companions (separation $<$0.1 arcsec) are behind the coronagraphic mask in the science exposures; this makes their detection difficult and derivation of astrometric and photometric properties biased. However, we may detect bright (usually stellar) close companions on the flux calibration, where the star is offset with respect to the coronagraphic mask. Typically two such images are acquired, one before and one after the science sequence. We can exploit this making a differential image that cancels static aberrations. We prepared a fast automatic procedure that allowed a contrast map to be derived from this differential image and close companions to be detected for all IFS SHINE observations. After some fine-tuning of the parameters, we retrieved 24 (stellar) close companions, nine of which (at separation in the range 30-60 mas) are new detections. All the new detections are around stars that have large discrepancies in the proper motion determinations from Hipparcos, GAIA DR1, and GAIA DR2.\\
\indent We devised an automatic procedure that uses the data cubes contained in the flux calibration files, output of the {\it convert} routine run at the SPHERE Data Center \citep{delorme2017} to create differential images in different bands. Whenever two or more exposures are available, the procedure uses the first and last ones; else, only one data cube was used. We also did not consider data cubes when the field rotation of the science sequence is smaller than 15 degree. For all data cubes that satisfy these criteria, we executed the following steps: (i) The initial and final 3D data cubes (x, y, wavelength: $dc_1$ and $dc_2$) are accurately re-centred using a Gaussian fitting of the images collapsed along the x and y coordinate for each wavelength. (ii) Two collapsed images ($img_1$ and $img_2$) are created for each of the $Y$ (1.0-1.1~$\mu$m), $J$ (1.2-1.3~$\mu$m), and $H$ (1.5-1.65~$\mu$m) bands from $dc_1$ and $dc_2$, respectively. The $H-$band images are only available for observations in the \texttt{IRDIFS-EXT} mode. (iii) For each band, $img_1$ and $img_2$ are normalised at the their peak value. (iv) A differential image is created: $imgd_1=img_1-img_2$. (v) This differential image is rotated by the first and last angles contained, creating two images. (vi) The final differential image is the mean of these two images. They are stored in files called {\it nocoro$_X$.fits}, where X=Y, J or H for the $Y$, $J$, and $H-$band, respectively.

The next step is the automatic detection of candidates. For this purpose, we used the $J-$band images. The procedure looks for the maximum within the ring from 5 to 18 pixels from centre (that is from 37 to 134 mas), taking care that this is above the limiting contrast at the appropriate separation (see next subsection). Then it accurately determines the position of the peak in the $Y$ and $J-$band. \\
\indent We used a number of criteria to eliminate false alarms automatically. First, we derived the ratio between the distance from the centre of the images of the peak in $J$ and $Y-$band. If the ratio is between 1.1 and 1.3, the candidate is discarded because the value is close to the ratio (=1.19) between the central wavelengths of the $Y-$ and $J-$bands, and it can then be attributed to diffraction effects or to a speckle. In addition, the candidate is kept only if the centre is not offset by more than 2 pixels from the peak value, if the sigma of the Gaussian fitting is between 0.7 and 4, and if the central peak is positive. Finally, the candidate should have a parameter called $qual>0.2$. $qual$ is equal to the ratio between the square of the value at the peak of the candidate companion, divided by the product of the values of the differential image in position with PA equal to $\pm$ the field rotation;  it should be noted that if the values in both these positions are positive, $qual$ is multiplied by -1.  The rationale behind the use of this parameter is that we expect that a real companion would appear in the final differential image as a positive peak, surrounded by two negative peaks at symmetric positions with PA values differing by the total field rotation between the two images.

These criteria are very effective in reducing the number of false alarms to very manageable values. We made a final selection after a visual inspection of the images. In practice, the automatic procedure detected 28 candidates over 660 sequences that satisfy the criteria for using this procedure. We eliminated six candidates by visual inspection; this means that the automatic procedure has an efficiency of 79\% in detecting good candidates. We missed the automatic detection of TYC~8400-567-1, because of the small field rotation, but the object is obvious in the differential image. We then added three more detections that were all slightly below threshold in the $qual$\ factor. They are around stars having a candidate in a better observation (HIP~37918, HIP~55334) or for which we expect a companion from strong variations in the proper motion (HIP~109285). This makes up our final list of 26 detections around 21 stars.  We display the corresponding differential images in Fig.~\ref{fig:gallery}. We notice that by construction, there is only one candidate per observation. We display in the bottom part of Fig.~\ref{fig:gallery} the six cases eliminated by visual inspection. 

Repeating the same selection, but using in addition the $H-$band, results in two more detections (HIP~63041 and a second epoch for HIP~78092). Lowering the threshold to 4.5 only adds three reasonable candidates: HIP~25434 and HIP~61087, which correspond to stars with large proper motion anomaly (PMA) in \citealt{Kervella2019}, and a second epoch for TYC~6872-1011-1. Hereafter, we consider the companions of HIP~63041, HIP~25434 and HIP~61087 as detected with this procedure, making a total of 31 detections around 24 stars. 

\begin{figure}[h]
    \centering
    \includegraphics[width=8.5cm]{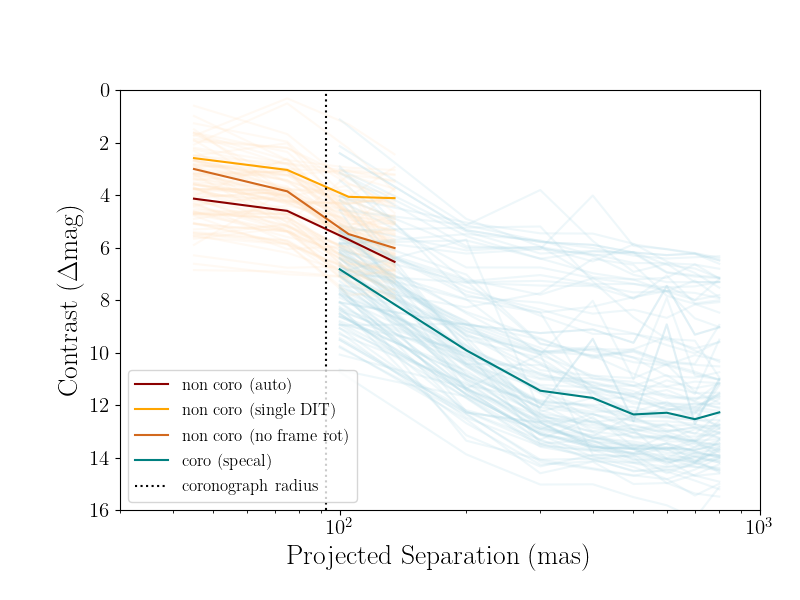}
    \caption{Limiting contrast (in $\Delta mag$) vs. projected separation achieved in the coronagraphic (blue lines) and non-coronagraphic (yellow lines) images. The solid lines show the median contrast obtained with the various methods described in Sect.~\ref{sec:detlim}. The dashed vertical line marks the coronagraphic radius. We note that objects with multiple epochs will appear more than once.}
    \label{fig:cnt_lim}
\end{figure}

\begin{table*}
\caption{IFS contrast limits (expressed as $\Delta mag$) for all the available datasets. \textit{Red.Nc} shows the reduction method used to obtain the limits from the non-coronagraphic images (see Sect.~\ref{sec:detlim} for details): \textit{NcA:} no-coro auto; \textit{sDIT:} single DIT; \textit{NfR:} no field rotation (full table available through CDS). \label{tab:ifs_ccurves}} 

\resizebox{\linewidth}{!}{    
    \begin{tabular}{l|ll|llll|llllllll}
\hline \hline
ID  & JD  &  Red.Nc & \multicolumn{4}{|c|}{Contrast (nocoro)}  & \multicolumn{8}{c}{Contrast (nocoro)}  \\
   &     &         & 45mas & 75mas & 105mas & 135mas          & 100mas & 200mas & 300mas & 400mas & 500mas & 600mas & 700mas & 800mas  \\
 \hline
HIP 2729 & 57357.01 & NcA & 2.84 & 3.22 & 4.18 & 4.3 & 6.28 & 9.89 & 11.51 & 12.16 & 12.47 & 12.94 & 13.07 & 13.25 \\
HIP 2729 & 58378.19 & NcA & 2.84 & 4.84 & 6.16 & 7.31 & 6.92 & 11.02 & 12.35 & 13.05 & 13.68 & 13.15 & 13.62 & 13.35 \\
AF Hor & 57323.19 & sDIT & 0.6 & 1.68 & 3.55 & 4.41 & 6.94 & 9.27 & 10.67 & 10.88 & 11.26 & 11.31 & 11.54 & 11.53 \\
TYC 8491 0656 1 & 57322.18 & sDIT & 1.95 & 2.2 & 3.59 & 4.02 & 8.07 & 11.44 & 13.13 & 13.97 & 14.36 & 14.82 & 15.45 & 15 \\
TYC 8491 0656 1 & 58088.09 & NcA & 1.5 & 3.95 & 5.12 & 6.42 & 7.62 & 11.35 & 12.61 & 13.01 & 13.2 & 13.18 & 13.37 & 13.55 \\
TYC 8497 0995 1 & 57356.08 & NcA & 3.44 & 3.1 & 5.22 & 5.54 & 6.24 & 8.85 & 9.82 & 10.99 & 11.12 & 11.66 & 11.73 & 11.63 \\
TYC 8497 0995 1 & 58087.07 & NcA & 5.31 & 5.18 & 6.4 & 6.97 & 6.61 & 10.78 & 12.09 & 12.36 & 12.61 & 12.62 & 12.72 & 13.06 \\
HIP 17157 & 57675.26 & NcA & 3.77 & 3.82 & 5.7 & 6.55 & 8.53 & 11.57 & 13.11 & 13.62 & 13.66 & 14.19 & 13.98 & 14.19 \\
HIP 17157 & 57675.26 & NcA & 3.77 & 3.82 & 5.7 & 6.55 & 8.53 & 11.57 & 13.11 & 13.62 & 13.66 & 14.19 & 13.98 & 14.19 \\
\hline 
 \end{tabular}}
\end{table*}

\subsection{Detection limits}
\label{sec:detlim}

In order to evaluate our sensitivity to stellar companions, we determined detection limits for point sources; we note that here we do not consider sub-stellar companions, so that we focus on rather bright objects. Whenever the standard SPHERE sequence was available, including flux and centre calibration and science exposure acquired in pupil stabilised mode, we used the normal procedure to derive detection limits outside the coronagraphic field masks that makes use of the SPECAL software as described in \citet{Galicher2018} and used in the F150 survey \citep{Langlois2021}. The detection limits considered here were obtained using the Template Locally Optimised Combination of Images \citep[TLOCI;][]{tloci} for IRDIS and the ASDI-PCA \citep[Angular Spectral Differential Imaging with Principal Component Analysis);][]{Galicher2018} for IFS. Since this procedure was devised to detect sub-stellar companions, these detection limits are usually much deeper than required to detect stellar companions, so we are confident that we detected all stellar companions at separation larger than 120 mas and within the IRDIS FoV (that is, within 5.5 arcsec).

The limiting contrast is usually derived by considering the standard deviation in a series of rings with increasing radii. 
The presence of a companion strongly modifies the standard deviation within each ring, especially at close separations.  
For this reason, a proper derivation of a limiting contrast on the non-coronagraphic images is a tricky issue. We therefore adopted the following simplified procedure. First, we transformed the image from Cartesian to polar coordinates. Second, we considered the separation from 5 to 18 pixels from centre (that is, from 37 to 134 mas). At each separation, we divided the image into eight sectors and estimated the standard deviation within each sector.
Third, we assumed that the limiting contrast is a threshold times the median of the standard deviations obtained for each sector.
Finally, we slightly smoothed the final detection curve; we tried various threshold values, finding that there are very few false alarms for threshold=5.0 and that essentially the same detections are obtained with a threshold value in the range from 5 to 6.

A similar procedure was adopted for those cases where only part of the required dataset was available, and we could not run procedures that exploit angular differential imaging (ADI) or the field rotation between the different acquisition of flux calibrations because a single DIT (Detector Integration Time) was available. In these cases the detection limits are much shallower than for those cases where the complete sequence was available, with typical values of about 7 magnitude at separation larger than 200 mas. Still, this limiting contrast is enough to detect almost all stellar companions.

\begin{figure}[t]
    \centering
    \includegraphics[width=8.5cm]{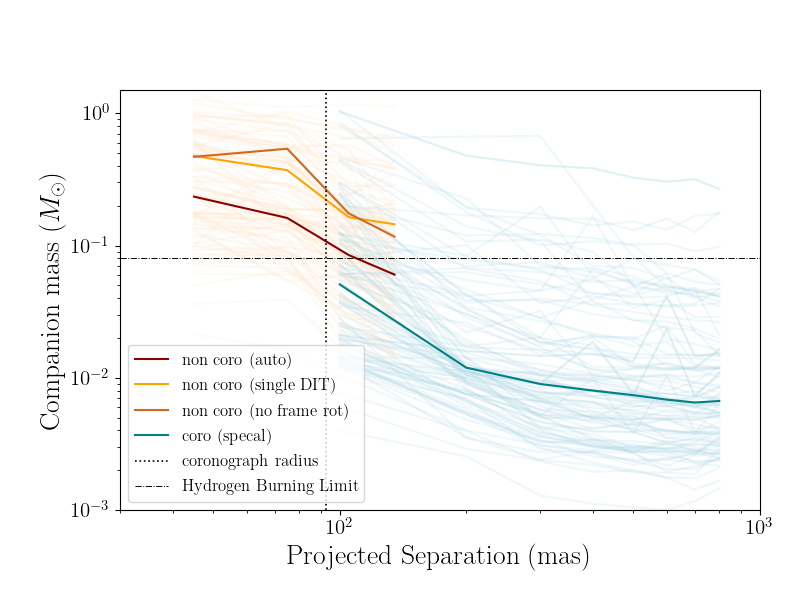}
    \includegraphics[width=8.5cm]{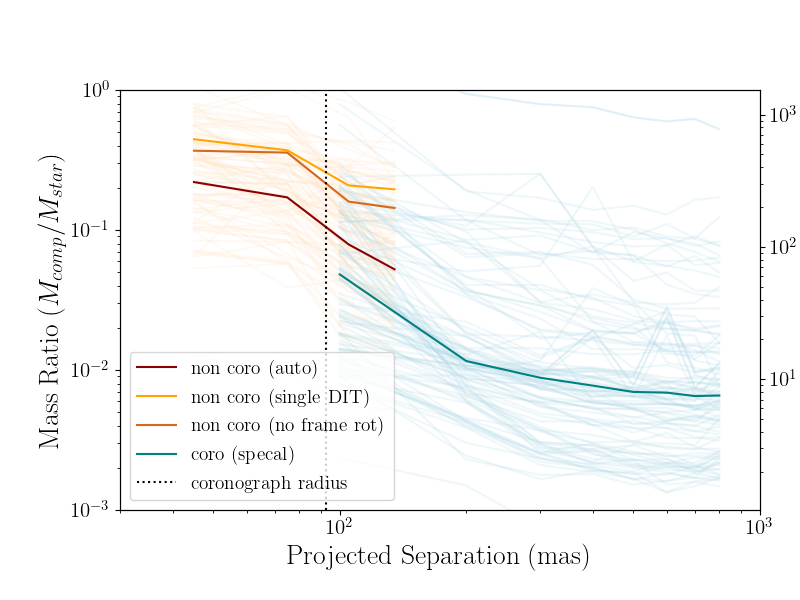}
    \caption{Minimum mass (top panel) and mass ratio (bottom panel) vs. projected separation of companions detectable in the coronagraphic (blue lines) and non-coronagraphic (yellow lines) images, obtained using the COND evolutionary models \citep{baraffe2003} to convert the magnitude limits in Fig.~\ref{fig:cnt_lim}. The solid lines show the median contrast obtained with the various methods described in Sect.~\ref{sec:detlim}. The dashed vertical line marks the coronagraphic radius. The dotted-dashed line in the top panel marks the hydrogen burning limit. We note that objects with multiple epochs will appear more than once.}
    \label{fig:mass_lim}
\end{figure}

Contrast limits for the individual datasets are shown in Fig.~\ref{fig:cnt_lim} and reported in Table~\ref{tab:ifs_ccurves}. We only considered separation within 800 mas, thus within the IFS FoV, and shown separately the limits for non-coronagraphic and coronagraphic observations. Limiting contrasts at separations larger than 800 mas are expected to be at least as deep as those obtained at this separation.
The corresponding mass and mass ratio limits, obtained using the Cond evolutionary models \citep{baraffe2003} to convert the magnitude limits in Fig.~\ref{fig:cnt_lim}, are shown in Fig.~\ref{fig:mass_lim}. 

\begin{table*}
\caption{Summary of the characteristics of the observed systems. If  more than one SHINE epoch was available, only the separation ($\rho$) and position angle (PA) from the first epoch are reported; information on the single measurements for these objects can be found in Table~\ref{tab:astroph}. 
The values of the masses of the primaries ($M_A$) and companions ($M_B$) were derived as described in Sect.~\ref{sec:mstar}. Minimum and maximum values are provided for both (full table available through CDS). \label{tab:sys_char}}
    \begin{tabular}{rlllllll}
\hline
\hline 
ID  & $M_A$ $_{min}^{max}$  & $M_B$ $_{min}^{max}$  & q             & $\rho$    & PA    & N$_{comp}$ & Notes \\
    & ($M_{\odot}$)         & ($M_{\odot}$)         & ($M_B/M_A$)   & (mas)     & (deg) &            &       \\
\hline
HIP 2729~AB & 0.77$_{0.77}^{0.80}$ & 0.76$_{0.76}^{0.78}$ & 0.99 & 21.40 & 62.70 & 2 & N \\
AF Hor~Aab$^1$ & 0.57$_{0.54}^{0.56}$ & 0.55$_{0.51}^{0.56}$ & 0.96 & 50.40 & 302.20 & 4 & W \\
TYC 8491-0656-1~Aab$^1$ & 0.72$_{0.72}^{0.74}$ & 0.73$_{0.72}^{0.74}$ & 1.01 & 45.41 & 104.02 & 4 & W \\
TYC 8497-0995-1~AB & 0.68$_{0.67}^{0.68}$ & 0.46$_{0.48}^{0.42}$ & 0.69 & 62.88 & 348.06 & 2 & N \\
HIP 17157~AB & 0.73$_{0.73}^{0.74}$ & 0.19$_{0.20}^{0.16}$ & 0.25 & 1423.40 & 314.38 & 3 & N \\
HIP 17157~AC & 0.73$_{0.73}^{0.74}$ & 0.28$_{0.29}^{0.24}$ & 0.38 & 1764.91 & 266.27 & 3 &  \\
\hline
    \end{tabular}\\
{\footnotesize $^1$AF~Hor~Aab and TYC 8491-0656-1~Aab form a wide pair, separated by $\sim$20\arcsec, which is why are both reported as quadruple. {\bf Notes.} {\it A:} possible unresolved additional companion; {\it W:} Additional wide companion, see Table~\ref{tab:widebin}; {\it K:} previously known as binary but not resolved or with poor information on orbit. {\it SB:} previously known as spectroscopic binary;  {\it S:} suspected binarity (based on RV or dynamical signatures); {\it P:} dedicated SHINE publication; {\it T:} transit candidate detected in TESS data; {\it sub:} companion could be sub-stellar if star very young; {\it G:} no parallax in EDR3, mass estimates done using the DR2 values.}
\end{table*}

\section{Results}
\label{sec:results}

\begin{figure*}[htbp]
   \centering
   \includegraphics[width=18.truecm]{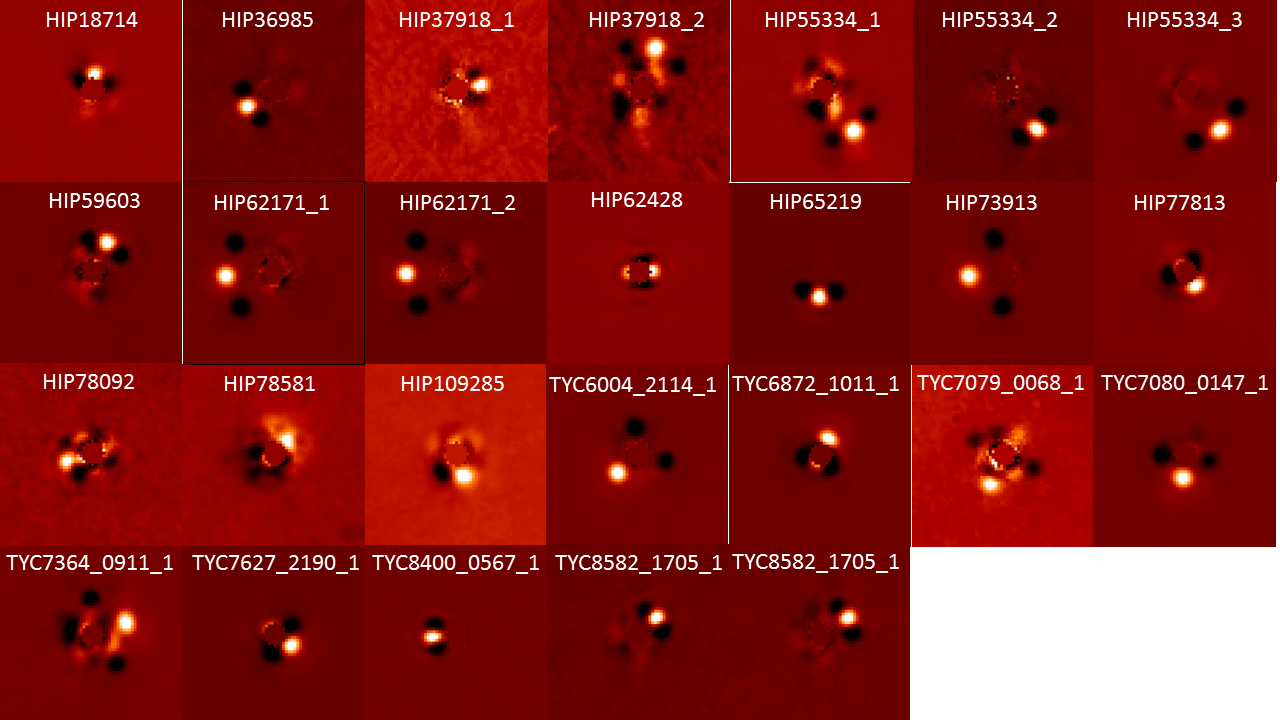}
   \includegraphics[width=18.truecm]{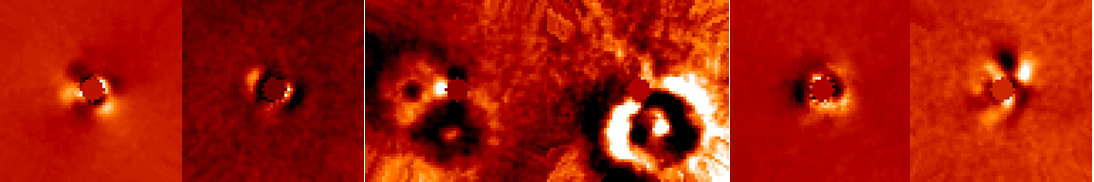}
   \caption{Gallery of confirmed (top 4 rows) and rejected (bottomw row) companions retrieved with the automatic procedure in the non-coronagraphic images.  All images are in the $J-$band. In all images, N is to the top and E to the left, and the central region within 3 pixels of the centre (22 mas) is masked. The images are square with sides of 64 pixels=477 mas}
    \label{fig:gallery}%
\end{figure*}

\begin{figure}
    \centering
    \includegraphics[width=9cm]{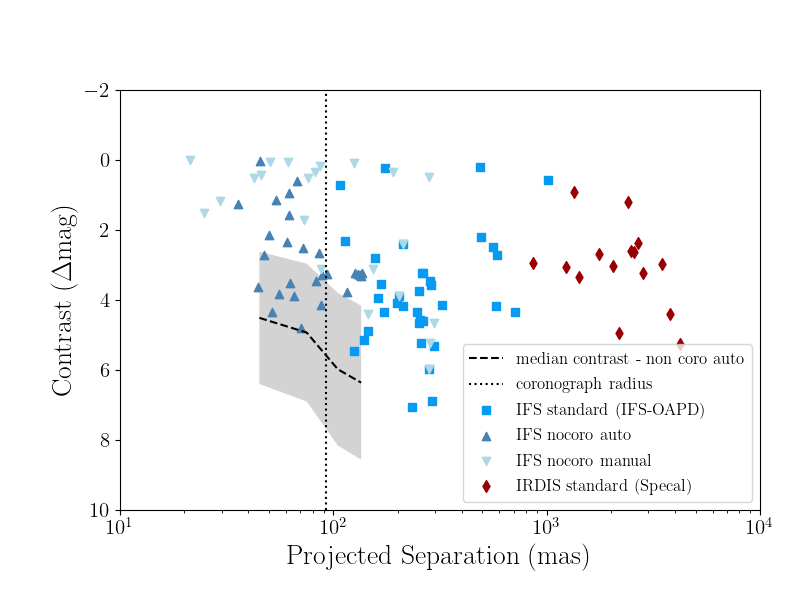}
    \caption{Contrast (in $\Delta$mag) vs. projected separation of all the detected companions. The different plot symbols reflect the various reduction methods described in Sect.~\ref{sec:data_red}. The dotted line marks the position of the edge of the coronagraph. The average contrast obtained for the automatic detections on non-coronagraphic observations is shown for comparison (black dashed line; see Sect.~\ref{sec:non_coro_auto} for details). The shaded area marks a 1$\sigma$ boundary around it. The corresponding limit for the standard reduction would be below the plot limits. We note that objects with multiple epochs will appear more than once if different reduction methods were used. }
    \label{fig:detections}
\end{figure}

Binaries with separation $<5$ arcsec known at the epoch of compilation of the original list (Summer 2014) were removed from the SHINE sample. Nevertheless, we found 78 out of the 463 stars observed so far as part of the statistical sample of the SHINE survey have companions within this separation range, 56 of which are new discoveries. Twenty-one of these systems have three or more components. Figure~\ref{fig:gallery} shows some examples of detections in the IFS FoV. 
The main characteristics of the systems in our sample are listed in Table~\ref{tab:sys_char}. 
As shown in Fig.~\ref{fig:detections} a significant fraction of the companions lays below the inner working angle limit imposed by the coronagraph and were detected using the non-coronagraphic PSF sequence, as described Sects. \ref{sec:non_coro_manual} and ~\ref{sec:non_coro_auto}. 

\begin{table*}
\caption{SPHERE astrometry and photometry for all the stars in the sample. Each epoch is reported separately. The last column shows the reduction method used, as described in Sect.~\ref{sec:data_red}  (full table available through CDS). \label{tab:astroph}}
\resizebox{\linewidth}{!}{        \begin{tabular}{r|l|ll|l|l|l|l|l|l|l|l}
\hline \hline
ID  & Obs. Date & rho & PA & \multicolumn{2}{c|}{IFS Phot.} & \multicolumn{4}{c|}{IRDIS Phot.}&  Prob bkg & Red.Meth \\
    & (MJD) & (mas) & (\degree) &  \multicolumn{1}{|c}{$\Delta Y$} & \multicolumn{1}{c|}{$\Delta J$}  & \multicolumn{1}{c}{$\Delta H_2$} & \multicolumn{1}{c}{$\Delta H_3$} & \multicolumn{1}{c}{$\Delta K_1$} & $\Delta K_2$ &  &  \\
\hline
HIP 2729~B    & 57357.01 & 25.51 $\pm$ 1.13 & 254.0 $\pm$ 3.9 &  & 0.03 &  &  &  &  & 1.52e-10 & NcP \\
              & 58378.19 & 26.75 $\pm$ 0.09 & 245.0 $\pm$ 1.2 & 0.08 & 0.08 &  &  &  &  &  & NcP \\
AF Hor~Ab      & 57323.19 & 50.40 $\pm$ 1.72 & 302.20 $\pm$ 0.93 & 0.13 & 0.06 &  &  &  &  & 7.16e-10 & NcM \\
TYC 8491-0656-1~Ab & 58088.09 & 45.41 $\pm$ 1.02 & 104.02 $\pm$ 2.24 & -0.04 & -0.02 &  &  &  &  &  & NcA \\
TYC 8497-0995-1~B & 57356.08 & 62.88 $\pm$ 0.85 & 348.06 $\pm$ 11.06 & 0.79 & 1.12 &  &  &  &  & 1.39e-09 & NcM \\
HIP 17157~B   & 57675.26 & 1423.40 $\pm$ 3.54 & 314.38 $\pm$ 0.18 &  &  & 3.38 & 3.34 &  &  & 2.51e-06 & NcM \\
HIP 17157~C   & 57675.26 & 1764.91 $\pm$ 4.01 & 266.27 $\pm$ 0.02 &  &  & 2.74 & 2.68 &  &  & 2.06e-06 & NcM \\
\hline
\end{tabular}}\\

\footnotesize{Reduction methods: \textbf{NcM:} non-coro manual (see Sect.~\ref{sec:non_coro_manual}); \textbf{NcA:} non-coro auto (see Sec.~\ref{sec:non_coro_auto}); \textbf{NcP:} non-coro partially resolved; \textbf{IFS-OAPD:} standard SHINE reduction for IFS data (see Sec.~\ref{sec:specal}); \textbf{Specal:} standard SHINE reduction for IRDIS data (see Sec.~\ref{sec:specal})}
\end{table*}

\subsection{SPHERE astrometry and photometry}
\label{sec:sphere_astroph}

The astrometry and photometry measurements from all our SHINE observations, are listed in Table~\ref{tab:astroph}. For each epoch we report projected separation and position angle, and the contrast (expressed as apparent magnitude difference) in the IFS $Y$ and $J$ filters, as well as the IRDIS $H_2$ and $H_3$ filters for the observations performed in IRDIFS mode, and $K_1$ and $K_2$ for the IRDIFS-EXT observations\footnote{See \url{www.eso.org/sci/facilities/paranal/instruments/sphere/inst/filters.html} for a full description of the SPHERE filters.}. The probability that the source is a background star, evaluated as described in Sect.~\ref{sec:common_pm}, is also listed. The last column specifies which method, among those described in Sect.~\ref{sec:data_red}, was used to obtain each measurement.

\begin{table*}
   \caption{Gaia astrometry and photometry of companions retrieved in EDR3 (Full Table available through CDS). We note that while we used the ID of the target to be consistent with the rest of the tables in the paper, the values of the parallax and proper motion reported are those retrieved in EDR3 for the companions. Although the companions of HIP 19183 and HIP 77388 were detected by Gaia, no astrometric solution was available in EDR3 (hence the blank fields). } \label{tab:known_bin}
    \begin{tabular}{rlllllll}
   \hline \hline
      $ID_A$            & parallax                      & \multicolumn{2}{c}{Proper Motion}           & $\Delta$mag  & separation   & PA    \\
                & (mas)                         & RA (mas/yr)                   & Dec. (mas/yr)    & Gaia G band           & (mas)         & (\degree) \\
      \hline\hline    

HIP 17157~C             &   39.082 $\pm$ 0.018  & 91.550 $\pm$ 0.018    & 102.454 $\pm$ 0.026     &  -3.597   &   1775.646        &   86.973 \\
HIP 19183~B                 &                                       &                                       &                                        &   8.057   &       4193.240         &  205.885 \\ 
TYC 7059 1111 1~Ab      &   15.839 $\pm$ 0.033  & 14.075 $\pm$ 0.030    & -32.569 $\pm$ 0.041     &  -0.121       &   1000.068    &   88.371 \\
\hline 
    \end{tabular}
\end{table*}

\subsection{Gaia astrometry and photometry}
\label{sec:gaia_astroph}
We checked the Gaia mission EDR3 archive \citep{gaia2016, gaia2018, GaiaEDR3} looking for detection of the secondaries for the programme systems. The majority of the systems were too close to yield separate entries in the Gaia EDR3 catalogue. The secondary was detected as a separate object by Gaia EDR3 for 16 binaries (see Table~\ref{tab:known_bin}); these are wide, low contrast systems. The comparison between Gaia EDR3 and SPHERE positions confirms the physical association between the two components in all cases except for a wide (separation of 4.8 arcsec) candidate in the field of HIP~75367, which is more compatible with a background star. We note that this is the candidate at the largest separation within our sample, and it is not included in the remaining tables; we found, however, a closer companion to HIP~75367 that appears to be physically linked to the star, which was then retained as a binary. These systems may be used to confirm the SPHERE astrometric calibration. For this purpose, we did not consider HIP~28036 and HIP~70833 because there is some evidence of additional close companions, and HIP~77388 because there is no proper motion of the secondary in Gaia EDR3. For the remaining 12 stars, we considered the relative proper motion between the two components between the Gaia EDR3 and SPHERE epochs using the Gaia EDR3 data. On average, the difference in the separation and position angles between Gaia EDR3 and SPHERE measurements for these systems is $1.6\pm 0.8$~mas (rms=2.8 mas) and $-0.12\pm 0.03$~degrees (rms=0.11 degrees), respectively. The small zero point offsets in scale and PA are well within the uncertainties of the SPHERE astrometric calibration \citep[see][]{Maire2016}. The comparison with Gaia indicates that the accuracy of SPHERE astrometry is better than 3 mas even at large separation. For the remaining systems, either the Gaia measurements are uncertain because of the large magnitude difference between the two components, or there may be significant orbital motion between Gaia and SPHERE observations. In all cases, Gaia data were used to provide a further epoch for each object.

Given the relatively small size of SPHERE's FoV our observations only allow for the detection of companions out to few hundred au. Hence a significant number of wider companions could have been missed by our observations. We therefore performed a search in Gaia EDR3 for additional common proper motion sources within 10000 au from the objects in our sample, using the method presented in \cite{Fontanive2019}.  We selected sources that were consistent with relative differences of less than 20\% in parallax and in at least one of the two proper motion components, with a maximum relative discrepancy of 50\% in the other proper motion component. The search returned 11 entries. The characteristics of these objects, together with the additional two known wide companions not retrieved in Gaia, are listed in Table~\ref{tab:widebin}.  A similar search for wide companions to all stars in the F150 sample \citep{vigan2020} returned a total of 24 sources.

\begin{table*}
\caption{Additional companions outside the SPHERE FoV. Separations and position angles were derived using the positions from Gaia EDR3, when available. Further details about the known systems (marked with $W$ in Table~\ref{tab:sys_char}) can be found in Appendix~\ref{app:targets} (full table available through CDS). }\label{tab:widebin} 
 \begin{tabular}{rlllllll} \hline \hline
      ID                & parallax                      & \multicolumn{2}{c}{Proper Motion}           & $\Delta$mag  & separation   & PA    \\
                & (mas)                         & RA (mas/yr)                   & Dec. (mas/yr)    & Gaia G band           & (arcsec)              & (\degree) \\
      \hline\hline    
 AF Hor~B$^1$           & 22.937 $\pm$ 0.046    & 97.940 $\pm$ 0.047    & -14.129 $\pm$ 0.055     &  -1.331       &   22.190  &   190.879 \\
 TYC 8491-0656-1~B$^1$  & 23.075 $\pm$ 0.044    & 93.852 $\pm$ 0.050    & -11.790 $\pm$ 0.056     &   1.331       &   22.190  &    10.881 \\
 HIP 17797~B                    & 18.798 $\pm$ 0.058    & 63.372 $\pm$ 0.055         &  -8.121 $\pm$ 0.065   &   0.598       &    8.370  &   216.621 \\
 HIP 17797~C                    & 18.799 $\pm$ 0.018    & 74.352 $\pm$ 0.017         &  -4.885 $\pm$ 0.023   &   7.496       &   86.207  &   141.388 \\
\hline
\end{tabular}\\

{\textbf{Notes:} $^1$ TYC 8491-0656-1 Aab; $^2$ AF Hor~Aab; $^3$ HIP~25436, also a close binary; $^4$ companion not retrieved in EDR3 because of very bright magnitude, separation estimated from WDS; $^5$ UY~Pic;  (see Appendix~\ref{app:targets} for details about all these objects).}
\end{table*}

\subsection{Common proper motion confirmation}
\label{sec:common_pm}

\begin{figure}
    \centering
    \includegraphics[width=0.9\linewidth]{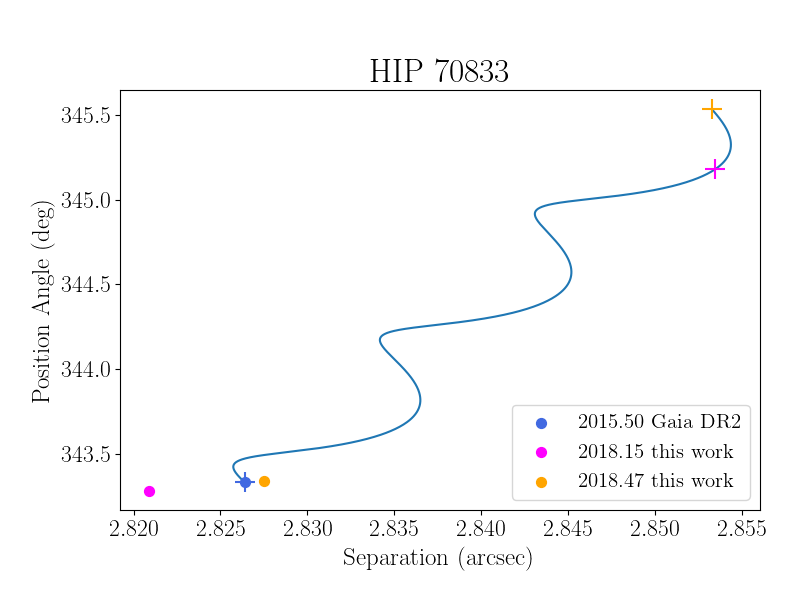}
    \includegraphics[width=0.9\linewidth]{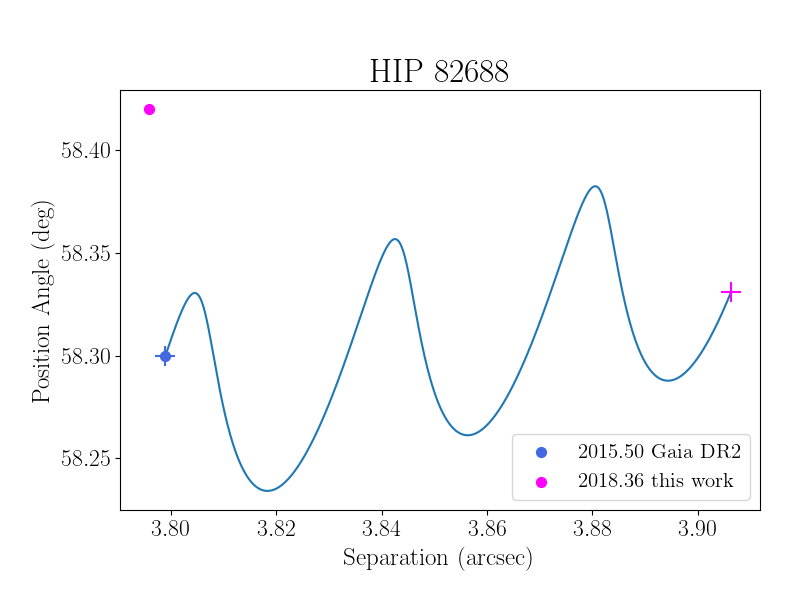}
    \caption{Common proper motion analysis of HIP~70833 (top) and HIP~86288 (bottom). In both panels the filled circles mark the measured separation (in arcsec) and position angle (in degrees) at the epochs listed in the legend. The corresponding expected values for a background source (assumed to have proper motion equal to zero) are marked with plus symbols of the same colours.}
    \label{fig:cpm}
\end{figure}

Multiple SHINE epochs, confirming the co-moving nature of our candidates, are available for about half of the programme stars (40 out of 78). The remaining objects are bright companions at very small separation and are then very likely physically related as the probability of having such bright background stars at these separations is very low. To confirm this, we used the code described in Section 5.2 of \citet{Chauvin2015} and adapted it to the SHINE results to estimate the probability of finding a background contaminant at the given separation and contrast as a function of galactic coordinates by comparison with the prediction of the Besan\c{c}on galactic model \citep{Robin2012}. These probabilities are listed in the \textit{Prob bkg} column of Table~\ref{tab:astroph}; they are below 1E-4 for all targets but the companions of TWA~24, HIP~64322, HIP~70833, and HIP~82688. In these four cases, the physical link is confirmed by the common proper motion, as shown in Fig.~\ref{fig:cpm} or by Gaia data.

As further confirmation we were able to retrieve additional epochs from other surveys, catalogues (including Gaia) or papers dedicated to specific objects for all but ten of our targets. The complete list of astrometric measurements for all our systems is presented in Table~\ref{tab:multi_epoch}, together with the references used for each entry. 

\begin{table*}
\caption{Complete list of all the astrometric data available for our systems (full table available through CDS). \label{tab:multi_epoch}}    
    \begin{tabular}{r|l|ll|ll|l}
\hline \hline
ID & Obs.Date       & rho   & $e_{rho}$ & PA        & $e_{PA}$  & Ref. \\
   & (MJD-245000)   & (mas) & (mas)     & (\degree) & (\degree) & Ref. \\

\hline
HIP 2729~B      & 57357.01 & 25.51 & 1.13 & 254.0 & 3.9 & this paper \\
                & 58378.19 & 26.75 & 0.09 & 245.0 & 1.2 & this paper \\
AF Hor~Ab       & 56993.0 & 75.0 & 1.0 & 174.5 & 0.36 & S07 \\
                & 57323.19 & 50.4 & 1.72 & 302.2 & 0.93 & this paper \\
TYC 8491-0656-1~Ab & 57292.516 & 34.3 & 1.0 & 97.9 & 1.5 & T16 \\
                & 57322.3 & 57.7 &  & 90.8 &  & this paper \\
                & 57738.262 & 44.5 & 1.2 & 51.8 & 2.5 & T18 \\
                & 58089.8 & 45.41 & 1.02 & 104.02 & 2.24 & this paper \\
                & 58144.915 & 35.9 & 1.7 & 99.7 & 1.7 & T19 \\
                & 58324.167 & 41.7 & 0.9 & 123.5 & 0.9 & T19 \\
HIP 17157~C     & 57205.79 & 1782.2 &  & 267.31 &  & GDR2 \\
                & 57675.26 & 1764.91 & 4.01 & 266.27 & 0.02 & this paper \\
\hline 
\end{tabular}\\ 
\footnotesize{\textbf{References:} \textit{S07:} \cite{Shan2017}; \textit{T16}: \cite{Tokovinin2016}; \textit{T18}: \cite{Tokovinin2018}; \textit{T19}: \cite{Tokovinin2019}; \textit{J12}: \cite{Janson2012}; \textit{G16}: \cite{Galicher2016}; \textit{K19}: \cite{kammerer2019}; \textit{J17}: \cite{Janson2017}; \textit{J130}: \cite{Janson2013}; \textit{B14}: \cite{Brandt2014}; \textit{T20}: \cite{tokovinin2020}; \textit{K05}: \cite{kouwenhoven2005}; \textit{K07}: \cite{Kouwenhoven2007}; \textit{L14}: \cite{Lafreniere2014}; \textit{R13}: \cite{Rameau2013}; \textit{R19}: \cite{ruane2019}; \textit{AT19}: \cite{asensiotorres2019}; \textit{H15}: \cite{hinkley2015}; \textit{S09}: \cite{shkolnik2009}; \textit{M19}: \cite{metchev2009}; \textit{S21}:\cite{Steiger2021}}
\end{table*}

\subsection{Constraints on the binary orbits}
\label{sec:orbits}

We performed an accurate literature search to retrieve as much information as possible about the systems considered in this paper, including not only relative astrometry, but also absolute astrometry and RVs. 
This information was then used to constrain orbital parameters for 25 of our systems. The orbital parameters were derived with two distinct approaches, depending on the amount of information available: a direct orbit determination or a Monte Carlo approach. Methods and results are discussed in the rest of this subsection. Table~\ref{tab:orbits} summarises the orbital parameters obtained, with a clear specification of the class of dataset considered and of the method used to derive constraints on the orbital parameters.

\subsubsection{Orbital fitting}
\label{sec:orbital_fit} 

When combined with literature astrometric and RV data, our SPHERE astrometry allows the (relative) orbits of a fraction of our targets to be constrained. To this purpose, we used the code {\it Orbit} by \citet{Tokovinin2016b}\footnote{\url{https://zenodo.org/record/61119}} that is based on a Levenberg-Marquard optimisation algorithm. This code allows us to combine astrometric and RV data. We were able to obtain full (relative) orbital solutions for four systems (TYC~6820-223-1, HIP~95149, HIP~107948AB, HIP~113201). Useful constraints on the orbits were obtained for 14 additional systems 
by assuming masses for the components as given by the analysis of the photometry described above. Table~\ref{tab:orbits} lists the orbital parameters we derived for 18 systems using this method. Detailed discussions for each individual case are given in Appendix~\ref{app:targets}. 

The systems for which orbital solutions were obtained mainly have intermediate separation because not enough data are generally available for very close systems (mostly unknown before our survey), and a tiny fraction of the orbit was covered for wide systems. The median values for the periods and semi-major axes are $\sim$20~yr and $\sim$8~au, and the ranges are $3-1000$~yr and $2-100$~au, with those with full orbit determination being closer systems than those for which solutions were found assuming the masses of the components. This bias should be taken into account when discussing our results.

\begin{figure}
    \centering
    \includegraphics[width=0.9\linewidth]{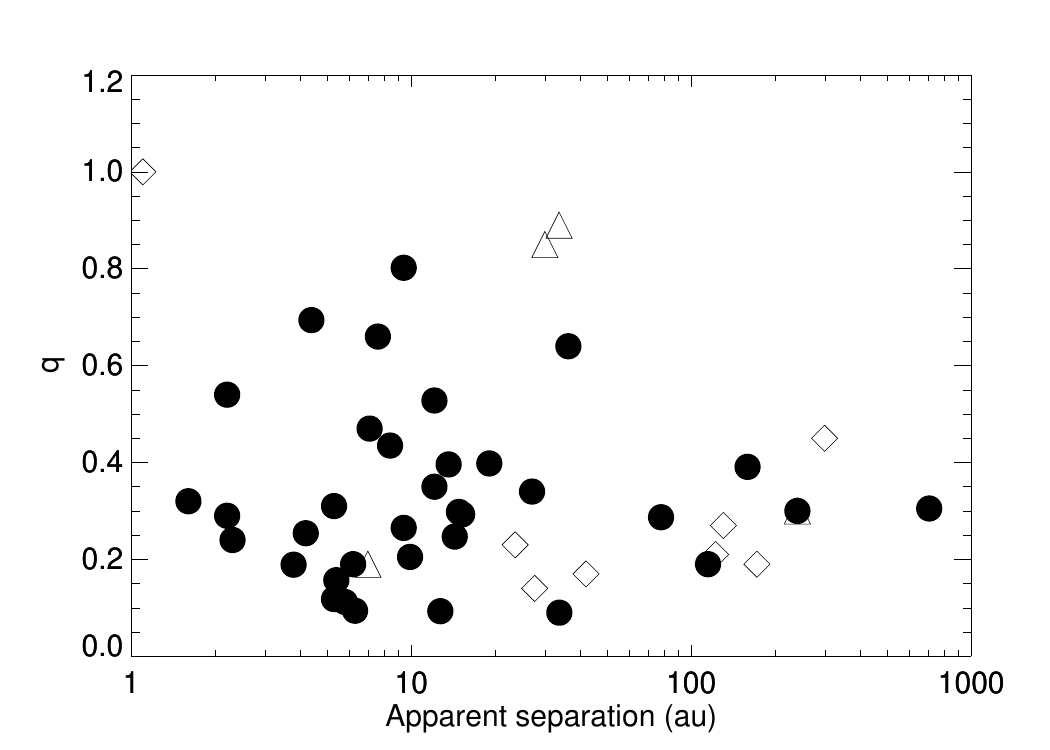}
    \caption{Mass ratio $q$ (see Sect. 5.2) vs. apparent separation (in au) for Hipparcos stars in our sample. Filled circles are binaries included in the catalogue by \citet{Kervella2019} and showing PMAs; empty diamonds are binaries that are included in the catalogue but not classified as PMAs; open triangles are stars that do not appear in the catalogue. }
    \label{fig:det_pma}
\end{figure}

 \begin{figure}
   \centering
   \includegraphics[width=8.8truecm]{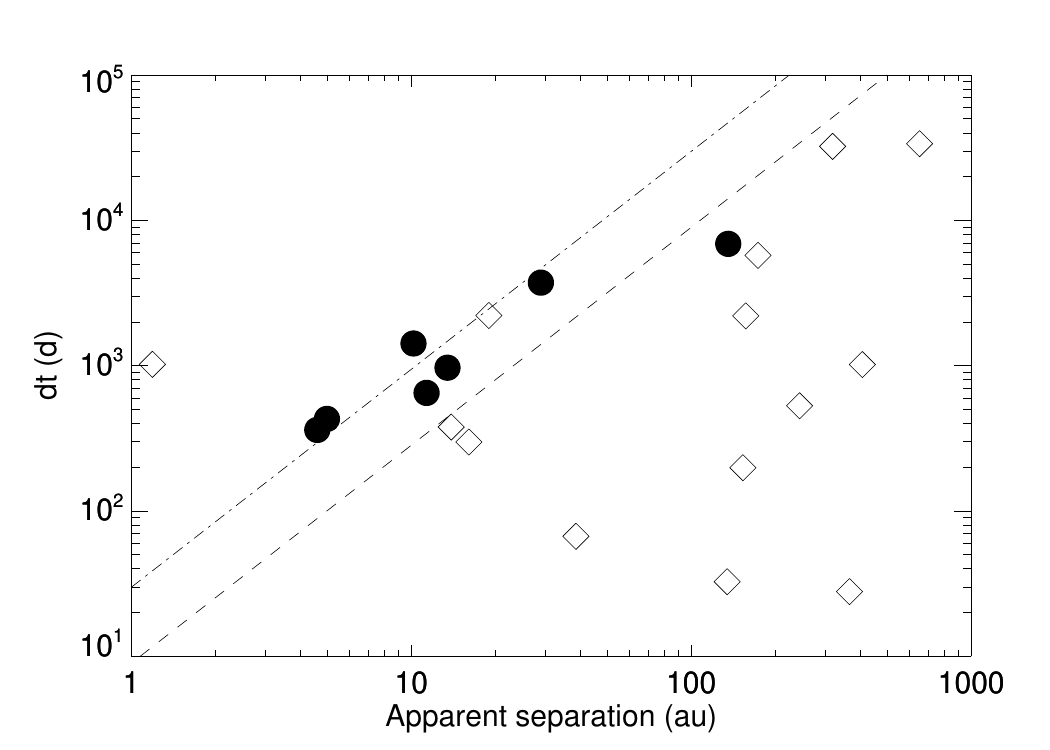}
   \caption{Time span of the astrometric observations $dt$ vs. apparent separation for systems with multiple epochs and not analysed using the code {\it Orbit}. Filled circles are systems for which the Monte Carlo analysis provided some constraints on the orbits; open diamonds are systems for which no useful result could be obtained. The dashed (dot-dashed) line marks a 0.03 (0.10) coverage of the orbit assuming that the semi-major axis is equal to the apparent separation and that the system mass is 1.5~$M_\odot$ }
\label{fig:orb_dt}%
    \end{figure}

\subsubsection{Targets with proper motion anomalies}
\label{sec:PMA}

\citet{Kervella2019} evaluated the PMAs (the motion of the photocentre with respect to a straight uniform motion) at the epochs 1991.25 and 2015.5 for stars that are present in both the Hipparcos and Gaia catalogues; the anomalies are then relative to a straight motion fit through epochs 1991.25 (Hipparcos) and 2015.5 (Gaia DR2), which, however, is not exactly the motion of the barycentre. Taking this into account, we can compare these motions with those predicted for the primaries using the relative orbits and mass ratios we determined in this paper. We defer a full analysis to a future paper, including simultaneous fitting of the relative positions of the components and of the PMA at the two epochs. In this paper we simply compare the results obtained by \citet{Kervella2019} with the orbits or family of orbits that we determined from our data alone. This comparison allows us to validate results obtained with two completely different approaches based on independent datasets. 

In this respect, it is interesting to remark the very high level of overlap between the two catalogues (see Fig.~\ref{fig:det_pma}). There are 329 Hipparcos stars observed within our statistical sample, and we found a stellar physical companion within 5 arcsec for 48 of them. Of these, 36 are also identified as binaries by \citet{Kervella2019}, who, by construction, only included Hipparcos stars. Four stars (HIP~19183, HIP~26369, HIP~70350, and HIP~107948) are missing from their catalogue because they have poor astrometric solutions either in Hipparcos or in Gaia DR2 (at least in some case this can be explained by confusion due to the secondary). Eight of our binaries are not identified as binaries on the basis of the PMA: Most of them are objects at large separation and with small mass ratios that likely have a very small PMA. In addition, HIP2796 -- which is the object with the shortest projected separation and likely has a period $<1$~yr -- was not detected as binary by \citet{Kervella2019}. We note that for some system the object responsible for the PMA may be a closer companion that is undetected in our observations, rather than the one we detected; for example, this is likely the case for HIP~70833 that is the object with the largest projected separation in Fig.~\ref{fig:det_pma} (see also Appendix A).

On the other hand, a cross-match with the full SHINE statistical sample showed that 70\% of the binaries present in both samples were detected. 
Objects with detected PMA from \citet{Kervella2019} with no detections in SHINE will likely have companions at very small separation and/or very low mass ratio (that is, they likely are at lower left corner of Fig.~\ref{fig:det_pma}) and will be the subject of a separate study. 

Considering PMAs, it is possible to better constrain the orbit and to estimate the uncertainties existing in the mass of the secondaries for some binary. 
For this purpose, we may consider two groups of systems:\\

\noindent {\it 1) Systems with one much fainter component}\\
    In this case the contribution of the secondary to the position of the photocentre is small enough that the estimates by \citet{Kervella2019} essentially coincide with the motion of the primary, with at most small corrections -- which we, however, considered in the following discussion. This makes the comparison more robust. For the purpose of illustrating the potential of this comparison, we consider here the relative orbits for five such targets:\\
    \textbf{HIP37918:} In this case we have position measures at three epochs and RVs for two. While the number of measurements is small, the relative orbit is rather well fixed once we assume masses from the photometry. The very small variation in RV indicates that the orbit is seen close to face on; we then assumed $i=0$. The best solution has a semi-major axis of 78.3 mas and period of 3.958 yr. In this case, \citet{Kervella2019} obtained motion anomalies of -12.97 mas/yr in RA and -3.78 mas/yr in Dec. For our best orbit we obtain motion anomalies of -14.20 mas/yr in RA and -5.13 mas/yr in Dec. for the epoch 2015.5. Deviations are significant at about 5.9~$\sigma$ if only the errors given by \citet{Kervella2019} are considered. This residual difference may be eliminated assuming an orbit with a slightly larger semi-major axis (79.9 mas) and period (4.079 yr) that also matches very well the PMA measured by \citet{Kervella2019} for the Hipparcos epoch (1991.25), which is very sensitive to the adopted period. We obtain a total reduced $\chi^2=2.08$ once we combine the contribution of the residuals in the orbital fit with those on the PMA. A fully integrated optimisation may further refine this orbital solution.\\
    \textbf{HIP~79124:} This is a triple system, but the outer companion is so far and faint that we can neglect it in this analysis, so we focused on the inner binary (HIP~79124AaAb). Given the large mass ratio, the contribution of the secondary to the photocentre is negligible. The orbit analysis yields a family of possible solutions. Since the portion of the orbit covered by the observations is small and the S/N of the PMA obtained by \citet{Kervella2019} is rather low (S/N=2.8), the period is only constrained to be $>50$~yr. However, independent of the period, we found that the mass of the secondary should be $0.10\pm 0.03~M_\odot$ to satisfy all the astrometric constraints, in good agreement with the results obtained from the photometric analysis (see also \citealt{asensiotorres2019}).\\
    \textbf{HIP~93580:} With four position measures and RVs at four epochs, we can set a family of relative orbits for this system depending on period when we assume masses from photometry, with preference for orbits longer than 15 years. However, only orbits with a period of about 30 yr match reasonably well the motion anomalies for the epoch 1991.25 and 2015.5 obtained by \citet{Kervella2019}. The best solution is for an orbit with a period of 28.1~yr that produces PMAs of 6.4 mas/yr in RA and 2.0 mas/yr in declination at the epoch 1991.25; and of 0.9 mas/yr and 5.0 mas/yr for the epoch 2015.5. \citet{Kervella2019} obtained values of $6.3\pm 0.3$ and $3.2\pm 0.3$ mas/yr at 1991.25 and of $1.4\pm 0.3$ and $6.3\pm 0.3$ at 2015.5 in RA and declination, respectively. Since errors are small, we are still formally out by a few $\sigma$. The remaining difference between the two results might be explained by either small errors in our orbit (not accounted for in our procedure) or by optimising the masses of the two components.\\
    \textbf{HIP~95149:} Three astrometric epochs and a quite long RV sequence allow the relative orbit to be derived by assuming masses from photometry. This orbit yields a PMA of 29.36 mas/yr in RA and 0.41 mas/yr in Dec. at 2015.5, while the values obtained by \citet{Kervella2019} are 25.79 mas/yr in RA and 0.46 mas/yr in Dec. Given their small errors, the discrepancy is at 3.0~$\sigma$; this might be solved by reducing the secondary mass from the nominal value given by the photometric analysis ($0.26^{0.02}_{-0.048}~M_\odot$) to 0.225~$M_\odot$, which is well within the error bars. We conclude that in this case our orbit matches well the PMA measured by \citet{Kervella2019}.\\
    \textbf{HIP~113201:} We obtained a full relative orbit solution for this star. Assuming the mass ratio given by the photometric analysis, the motion anomaly predicted for 2015.5 is -5.86 mas/yr in RA and 23.84 mas/yr in Dec. The values listed by \citet{Kervella2019} are -6.41 mas/yr in RA and 21.90 mas/yr in Dec. Since errors are very small for this system, the two results are formally discrepant at 8.6~$\sigma$. This difference would be minimised by assuming a mass of 0.093~$M_\odot$ (rather than $0.099^{0.001}_{-0.001}~M_\odot$ as given by the photometry) for the secondary. Given the uncertainties existing in the masses of low-mass stars, we conclude that there is good agreement between our analysis and that of \citet{Kervella2019}.\\
    
\noindent {\it 2) Systems with components of similar brightness} \\
For these objects the correction required to obtain the motion of the primary from that of the photocentre
is large and strongly depends on the luminosity ratio between the two stars in the Gaia photometric band and on the mass ratio between the two components. These quantities are not directly measured and should be inferred from the photometry in the NIR. 

The only nearly equal-mass system with information on the orbit in common between our sample and that of \citet{Kervella2019} is HIP~65219. In this case we estimated that the photocentre PMA is a factor of 2.22 smaller than the PMA in the primary orbit. Once corrected for this factor, the motion anomaly for the primary given by the \citet{Kervella2019} measurements is 3.10 mas/yr in RA and 3.86 mas/yr in Dec. at 2015.5, with a large error because their detection of the PMA has a low S/N, 3.78 (the S/N is even much lower for the 1991.25 epoch). The PMA predicted from the relative orbit depends on the assumed period (which is not well determined); since our observations were obtained not far from the reference epoch, the main effect is the contribution of the orbital motion to the estimate of the long-term trend by \citet{Kervella2019}. The orbital solution that gives the best agreement with \citet{Kervella2019} results is for a semi-major axis of 86 mas (=11.0 au) and a period of 20.3 yr, longer though within the error of the orbit fitting our astrometry alone. In this case the values we obtain for the primary motion anomaly at this epoch are about 5.88 mas/yr in RA and 4.33 mas/yr in declination. While the direction of motion is well reproduced, we would still expect a larger motion anomaly (by a factor of $\sim 1.5$) for the system photocentre than observed by \citet{Kervella2019}. This difference might be due to an underestimate by us of the correction from the photocentre to primary motion anomaly, for example. because the luminosity difference between the two components in the Gaia photometric system is smaller than we assumed ($\sim 0.7$~mag rather than $\sim 0.9$~mag).
On the whole, we conclude that these comparisons support the orbit determination presented in this paper.

  \begin{figure*}[ht]
   \centering
   \includegraphics[width=18truecm]{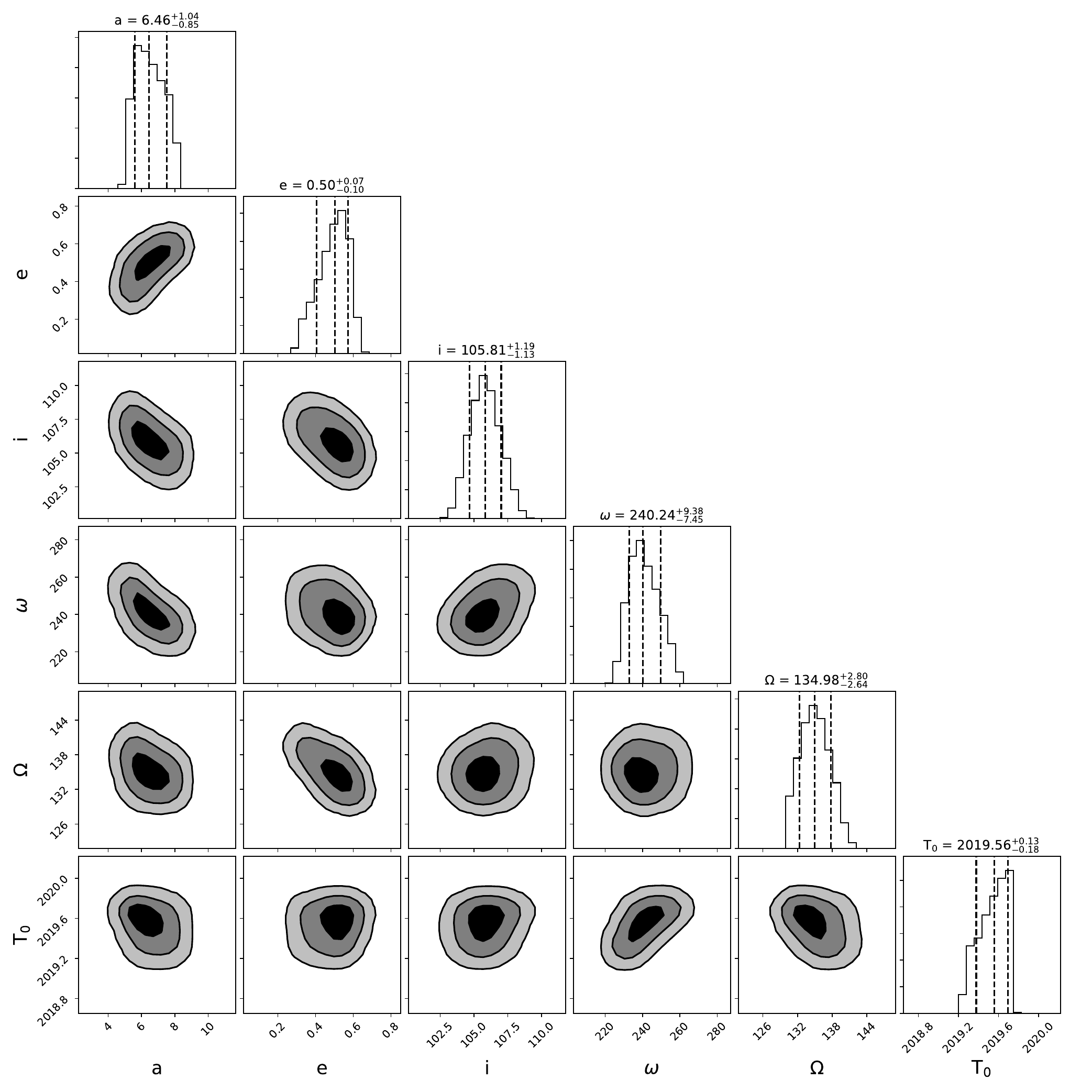}
     \caption{Posterior distribution of the orbital parameters (from left to right: semi-major axis, eccentricity, inclination, longitude of node, longitude of periastron, time of periastron passage) for TYC~6872-1011-1 obtained with the Monte Carlo method. }
\label{fig:orb_TYC68}%
    \end{figure*}

\subsubsection{Statistical constraints on orbits for systems with few observations}
\label{sec:MCorbit}
For the binary systems with multiple relative astrometric observations covering only a small fraction of the orbital motion, and therefore unsuitable for a derivation of orbital parameters using {\it Orbit}, we ran a Monte Carlo simulation to explore the possible families of orbits allowed. The procedure (hereafter MC) follows \cite{2018MNRAS.480...35Z}: we created $2*10^7$ orbits with random orbital parameters and selected only the ones that fitted the astrometric points. The fitting procedure is based on the visual binaries constants of Thiele-Innes. In this way we can understand whether the posterior distributions of the orbital parameters for a given target are uniform, or whether certain orbits are preferred. Figure~\ref{fig:orb_dt} shows that useful constraints were obtained for seven systems where roughly 10\% of the orbit is covered using this approach, though in favourable conditions even a shorter coverage may give some hints. An example of constraints used in this procedure is shown in Fig.~\ref{fig:orb_TYC68}.
The results of the MC for the orbits selected are listed in Table~\ref{tab:orbits}. To estimate the error bars of the parameters summarised in that table, we calculate the 0.16 and 0.84 quantiles of each posterior distribution. The median value (0.5 quantile) is assumed as the most probable value. We note that some posterior distributions are more stringent, while others permit a wide range of values for the parameters. That is normal for a very small coverage of the orbit.  

\subsection{Sensitivity to additional companions} 
\label{sec:mess_sim}

\begin{figure}[h]
    \includegraphics[width=9cm]{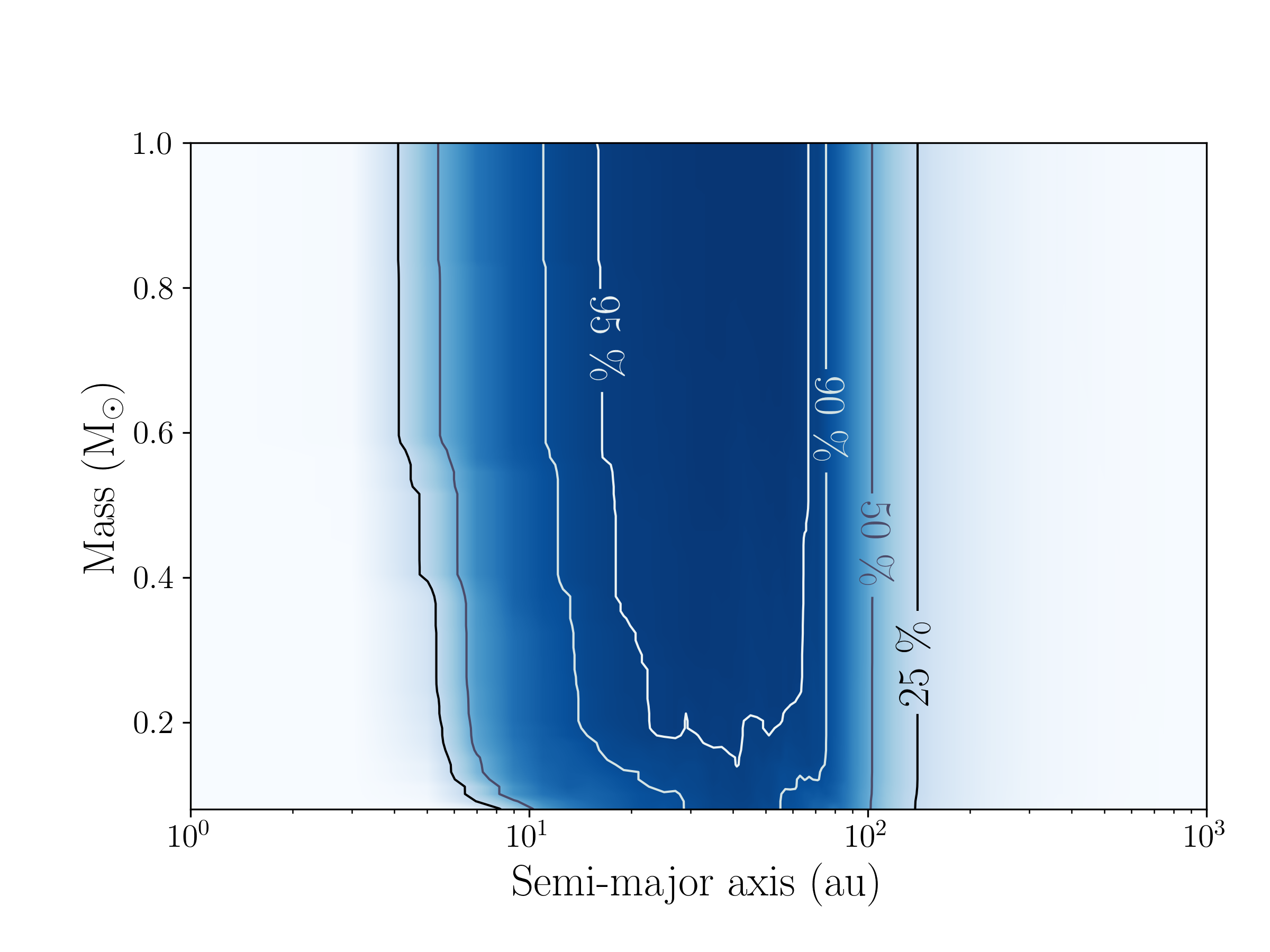}
    \caption{Probability of detecting additional companions around the stars in our sample, calculated using the Exoplanet Detection Map Calculator (Exo-DMC; see \citealt{Bonavita:2020ascl} for details) and the median mass limits shown in Fig.~\ref{fig:mass_lim}.} 
    \label{fig:det_prob}
\end{figure}

We used the Exoplanet Detection Map Calculator \citep[Exo-DMC;][]{Bonavita:2020ascl}\footnote{\url{https://ascl.net/2010.008}} to obtain a first estimate of the completeness of our sample in terms of additional stellar companions within the IFS FoV. 
The Exo-DMC is the latest (and for the first time in Python) rendition of the MESS \citep[Multi-purpose Exoplanet Simulation System][]{Bonavita2012}, a Monte Carlo tool for the statistical analysis of direct imaging survey results. 
In a similar fashion to its predecessors, the DMC combines the information on the target stars with the instrument detection limits to estimate the probability of detection of companions in a given mass and semi-major axis range, ultimately generating detection probability maps. 

For each star in the sample the DMC produces a grid of masses and physical separations of synthetic companions, then estimates the probability of detection given the provided detection limits. 
In order to account for the chances of each synthetic companion to be in the instrument's FoV, a set of orbital parameters is generated for each point in the grid, which allows an estimation of the range of possible projected separations corresponding to each value of semi-major axis.
The default setup uses a flat distribution in log space for both the mass and semi-major axis with all the orbital parameters uniformly distributed except for the eccentricity, which is generated using a Gaussian eccentricity distribution with $\mu =0$ and $\sigma = 0.3$, following the approach by \cite{Hogg2010} \citep[see][for details]{Bonavita2013}. 
The detection probability at a given semi-major axis is then calculated as the fraction of orbital sets that, for a given mass, allows for the companion to be detected. 
 
Figure~\ref{fig:det_prob} shows the median detection probability for companions with masses over 70~$M_{\rm Jup}$ and separations between 1 and 100 au, obtained using the limits shown in Fig.~\ref{fig:mass_lim}. Instead of the default setup, for this specific case we used the mass and semi-major axis distributions from \cite{Raghavan2010}.
Separate runs were performed for the targets for which both coronagraphic and non-coronagraphic images were available, then combined considering the best performance at each point in the grid.

\section{Discussion}
\label{sec:discussion}

\subsection{Survey completeness and binary frequency}
\label{sec:bin_freq}

\begin{figure}
    \centering
    \includegraphics[width=0.5\textwidth]{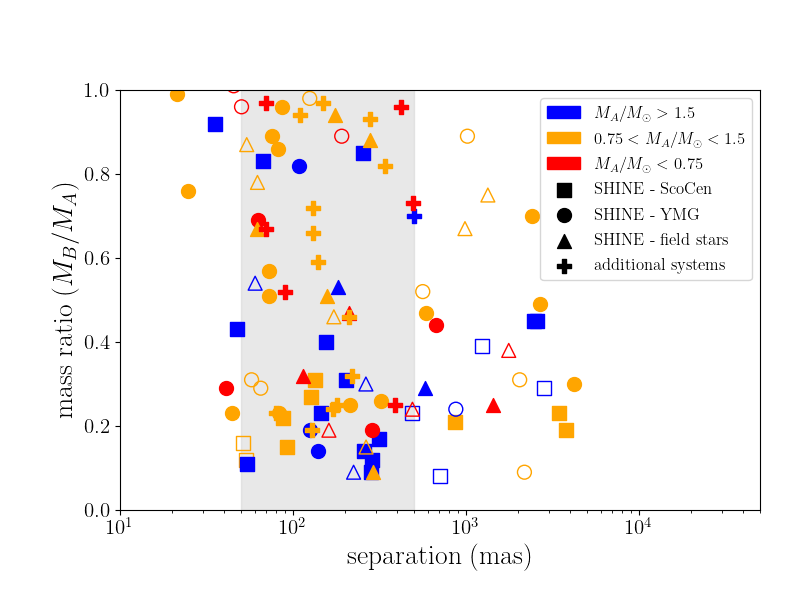}
    \caption{Mass ratio ($M_B/M_A$) vs. separation for the binaries in our sample, marked with circles for the young moving group members, triangles for the young field objects, and squares for the members of the Scorpius Centaurus region (US, LCC, and UCL in Table~\ref{tab:age_per}). The new systems discovered in this paper are marked by filled symbols. The plus signs show the position of the additional objects listed in Table~\ref{tab:known_bin}. Different colours represent different ranges of primary masses. }
    \label{fig:sepq}%
\end{figure}

We assume our sample to be reasonably complete at separations between 0.05$\arcsec$ and 0.5$\arcsec$, thus including systems that would have been too close to be resolved in past observations (but see later discussion), but still sufficiently wide that any stellar companion should have been detected in our data. This is confirmed by the limits in Figs.~\ref{fig:mass_lim} and \ref{fig:det_prob}, which show that our sensitivity in this separation range is well below the hydrogen burning limits for most of our targets.
As a consequence, even at the lowest value of the mass ratio compatible with a stellar secondary, we expect that only a small fraction of the stellar secondaries should have been missed in this range of separations.
Given the high level of completeness, this reduced sample could then be used to draw some preliminary conclusion on the impact of our results on the frequency and properties of young binaries.

In order to properly do so, however, it is first necessary to correct for the effect of the initial selection biases of the SHINE Survey. As previously mentioned, in fact, we removed any objects with known visual companions within the FoV at the time of selection. Several additional systems were also removed mid-way through the survey, following the publication of dedicated works characterising the binarity of stars in young associations \citep[see e.g.][]{Elliott2016}.

We were able to retrieve the information about the objects that were part of the original list of members of nearby moving groups ($\beta$~Pic=BPIC, Tucana=TUC, Columba=COL, Carina=CAR, AB Doradus=ABDO, Argus=ARG; see Table~\ref{tab:age_per} for a full description and age references.), and excluded from the final target list because of their binarity. This list includes 62 systems, of which 20 have $0.05\arcsec < \rho < 0.5\arcsec$ and are listed in Table~\ref{tab:known_bin}. 
We note that this list also includes a handful of objects that were observed within SHINE with special status (TWA~5) or as part of the binary filler programme (HIP~25647 and GJ 2060). 
Without the bias against binaries all these objects would have all been part of the SHINE statistical sample, but not all of them would have been observed and therefore included in our sample of new binaries, because not all the original targets of the survey were actually observed. To take this into account, we derived for each of these rejected objects the value of the merit function described in \cite{Desidera2021} that was used to assign each target to the four priority bins considered in the survey; and then derived their probability of being observed, based on the fraction of objects observed within SHINE for each of those bins (75.5\% for P1, 36.5\% for P2, 26.5\% for P3 and 18.5\% for P4). Most of the excluded systems were originally classified as P1 (53 out of 62), and only six, one, and two were marked as P2, P3, and P4, respectively. They would therefore count as 42.84 additional detected binaries, of which 13.91 have $0.05\arcsec < \rho < 0.5\arcsec$. 

The SHINE statistical sample also includes a number of objects belonging to the Upper Scorpius (USco), Lower Centaurus Crux (LCC), and Upper Centaurus-Lupus (UCL) regions (ScoCen). A lot more information about the multiplicity of these objects would have been available at the time of the target selection, making the bias against binarity a lot more effective. For example, most of the early-type stars in this region have Hipparcos observations, and therefore a higher sensitivity to similar-luminosity binaries down to small separation, compared to late-type objects in this and other regions. Several dedicated surveys for multiplicity were also performed
\citep{kouwenhoven2005,Janson2013,Lafreniere2014}.
However, given that a fixed number of stars from ScoCen were added to the SHINE initial sample (40 for each priority bin, see \citealt{Desidera2021}), evaluating the probability that any of the excluded binaries would have been observed is extremely difficult.  
We therefore chose to not re-include those in our sample, and we also exclude all the ScoCen systems (squares in Fig.~\ref{fig:sepq}) from the following analysis.

Finally, there are a number of young field objects included in the SHINE sample, which cannot be associated with any of the moving groups mentioned above. The information on the binarity in this case is rather incomplete and coming from scattered sources, and evaluating the impact on the initial bias would be quite complicated. Even if we could retrieve the full list of excluded binaries, to estimate the expected priority we would need to simulate the information available at the time of the selection, in particular regarding the age of the system. Then a full re-assessment of all the systems characteristics would be needed to properly consider them in our analysis. As this would be well beyond the scope of this paper, and will be presented in the final statistical analysis of the SHINE survey, we decided to also exclude the field binaries from the following discussion. 

Taking into account all the caveats discussed above, we then construct a \emph{complete moving group sample} limiting the census to targets belonging to young moving groups and re-introducing the excluded objects mentioned above. This sample includes a total of 231.84 stars (189 stars observed within SHINE and 42.84 that would have been observed but actually were not because known binaries), of which 32.91 are binaries with a separation of $0.05\arcsec < \rho < 0.5\arcsec$. This leads to a first estimate of the binary frequency in this separation range of $14.2 \pm 2.9\%$ (the error being given by Poisson statistics).
This value appears to be slightly higher than what was reported in previous studies \citep{Duquennoy1991, Raghavan2010} that suggested a frequency of companions with periods comparable to those in our statistical sample close to $\sim 11\%$. This difference is significant only at slightly more than 1$\sigma$, and it may then be an artefact of low number statistics. If confirmed by more data, it would resemble the case of the Taurus SFR that shows a slightly higher-multiplicity frequency with respect to field objects \citep{Kraus2011}. This is attributed to the young age and the fact that these are low-density environments.

   \begin{figure}
   \centering
   \includegraphics[width=8.8truecm]{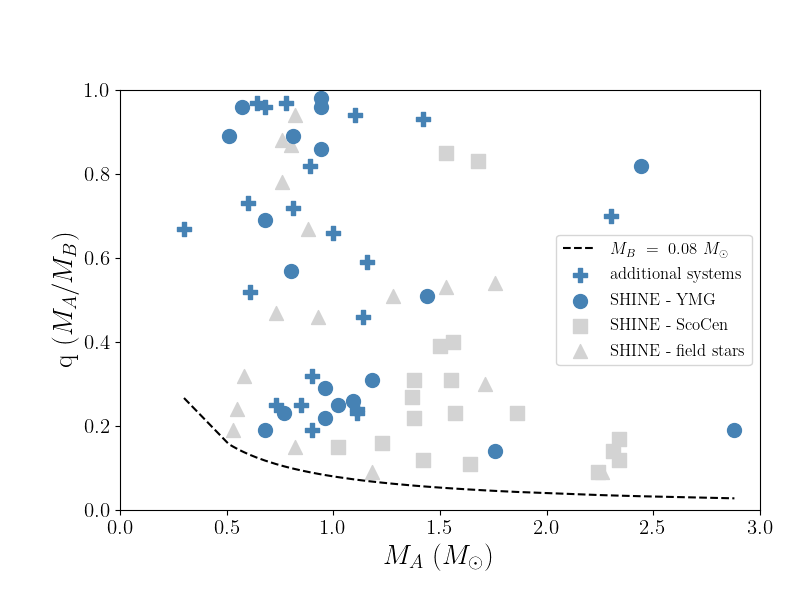}
   \caption{Primary mass ($M_A$) vs. mass ratio ($M_B/M_A$) for binaries in our reduced sample. The light grey squares and triangles show the position of the binaries in ScoCen and in the field excluded from the reduced sample, respectively. The plus signs show the position of the additional systems from Table~\ref{tab:known_bin}. The dashed line  shows the position of the hydrogen burning limit ($M_B = 0.08 M_{\odot}$).}
\label{fig:mA_q}
    \end{figure}

   \begin{figure}
   \centering
   \includegraphics[width=8.8truecm]{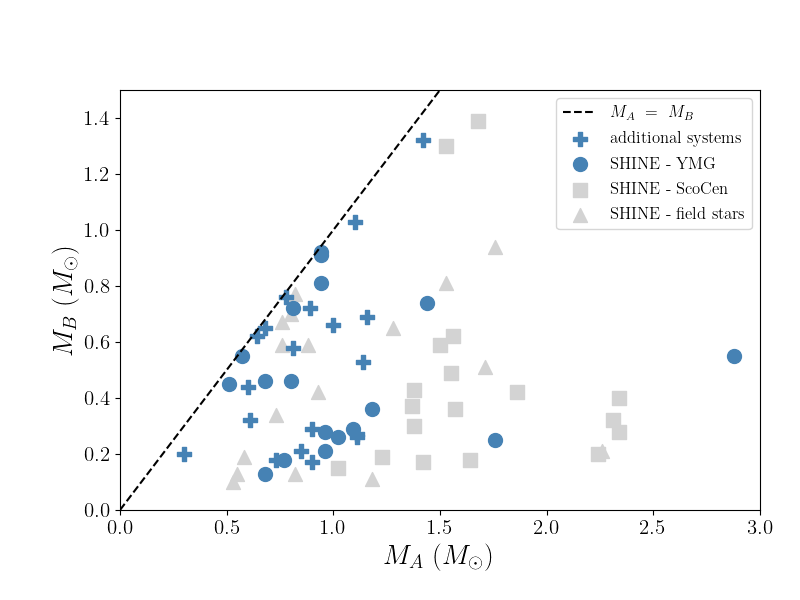}
   \caption{Primary mass ($M_A$) vs. secondary mass ($M_B$) for binaries with separation between 0.05 and 0.5 arcsec in our sample. The light grey squares and triangles show the position of the binaries in ScoCen and in the field excluded from the reduced sample, respectively. The plus signs show the position of the additional systems from Table~\ref{tab:known_bin}. The dashed line indicates mass equality}
\label{fig:mA_Mb}%
    \end{figure}

\subsection{Mass ratio distribution}
\label{sec:mratio}

Using the reduced sample defined in Sect.~\ref{sec:bin_freq}, we performed a tentative analysis of the mass ratio distribution of the binaries in the SHINE statistical sample. 
In the following discussion we consider two entries for triple (hierarchical) systems, one with the mass ratio obtained by summing the masses of the closer binary, and another with the mass ratio within the closer binary, that is, our $q$ values are $q_{\rm sys}$ according to the notation of \citet{Tokovinin2014b}. However, at variance from the definition used there, we forced $q$ to be $\leq 1$ for triple systems: that is, in cases where the mass ratio for triple systems obtained with the optically brighter primary would be larger than one, our $q$ value is the reciprocal of it.
Figures~\ref{fig:mA_q} and \ref{fig:mA_Mb} show the mass ratio $q=M_B/M_A$ and the secondary mass $M_B$ as a function of primary mass $M_A$ for the objects with companion in the reduced sample (once again with the additional binaries from Table~\ref{tab:known_bin} shown with a different symbol). 

As shown in Fig.~\ref{fig:mA_q}, there seems to be no clear trend with the primary mass, except that equal mass binaries seem to be slightly more common among the stars with $M_A<1~M_\odot$.
While our data are not robust enough to warrant any solid conclusion in this respect, we note that a similar trend has been obtained independently by \citet{Moe2017} (see their Fig. 35), extending over a much wider mass range but with a smaller statistics in this particular mass range; and by \citet{El-Badry2019} with a much larger statistics but wider separations using Gaia data (similar result but with smaller statistics was also previously obtained by \citealt{Soderhjelm2007} using Hipparcos data). This seems then a consolidated effect and might be related to, for example, differences in the migration efficiency within disks as a function of stellar mass, so that in massive systems equal mass binaries end up closer to the star than the region we are considering (see \citealt{Pinsonneault2006, Moe2017, TokovininMoe2020}).

   \begin{figure}
   \centering
   \includegraphics[width=8.8truecm]{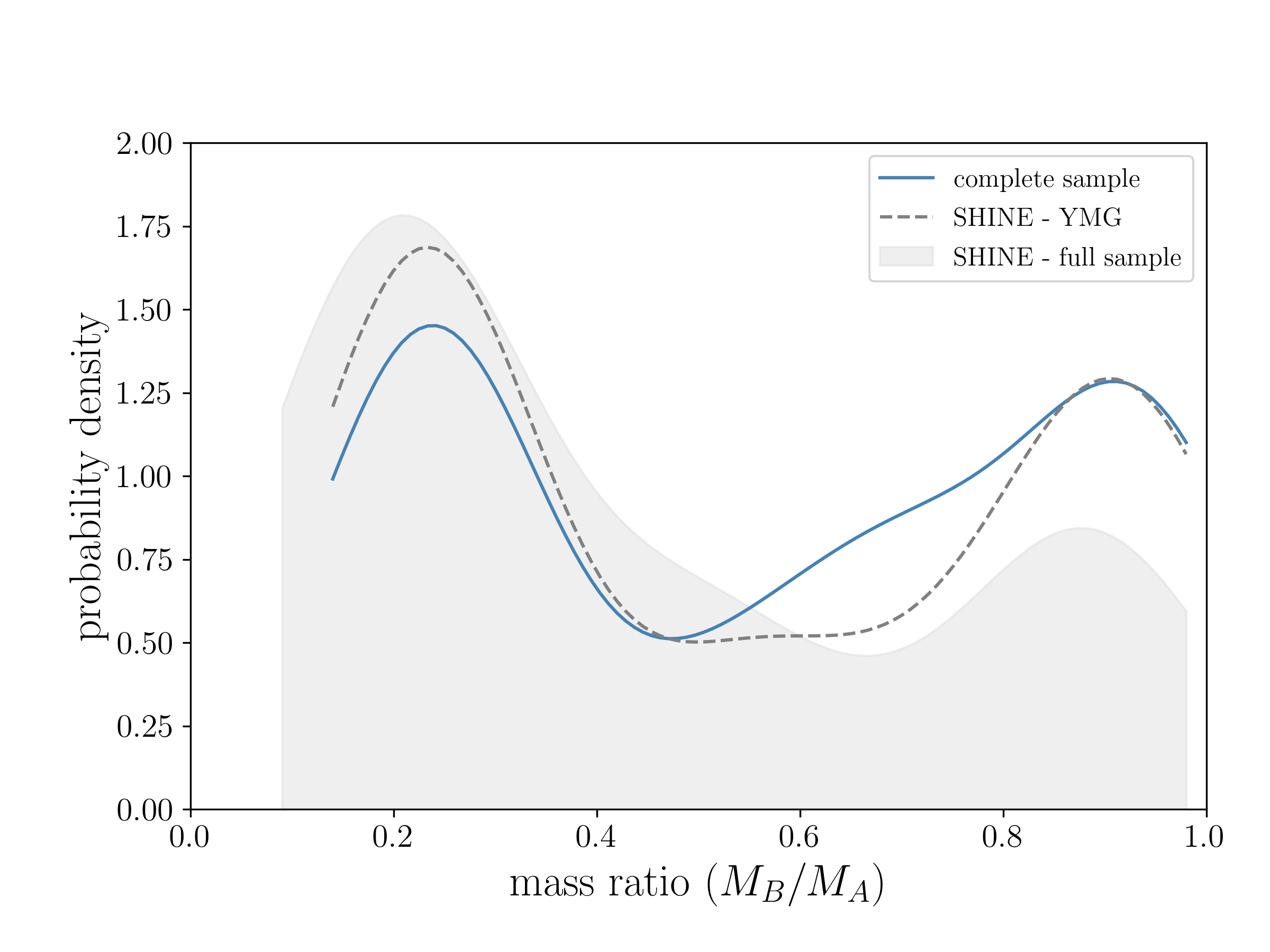}
   \caption{Mass ratio distribution obtained using the kernel density estimate (KDE) method \citep[see e.g.][]{Silverman1986} and a Gaussian kernel with $\sigma=0.1$. The solid blue line shows the result obtained for the systems in the complete sample described in Sect.~\ref{sec:bin_freq} including the additional objects from Table~\ref{tab:known_bin}, weighted according to their probability of being observed (see Sect.~\ref{sec:bin_freq} for details). 
   The grey shaded area shows the distribution obtained using the full sample of SHINE binaries from this work, while the dashed grey line shows the distribution obtained using only the SHINE young moving group systems included in the complete sample.}
\label{fig:distq}%
    \end{figure}

   \begin{figure}
   \centering
   \includegraphics[width=8.8truecm]{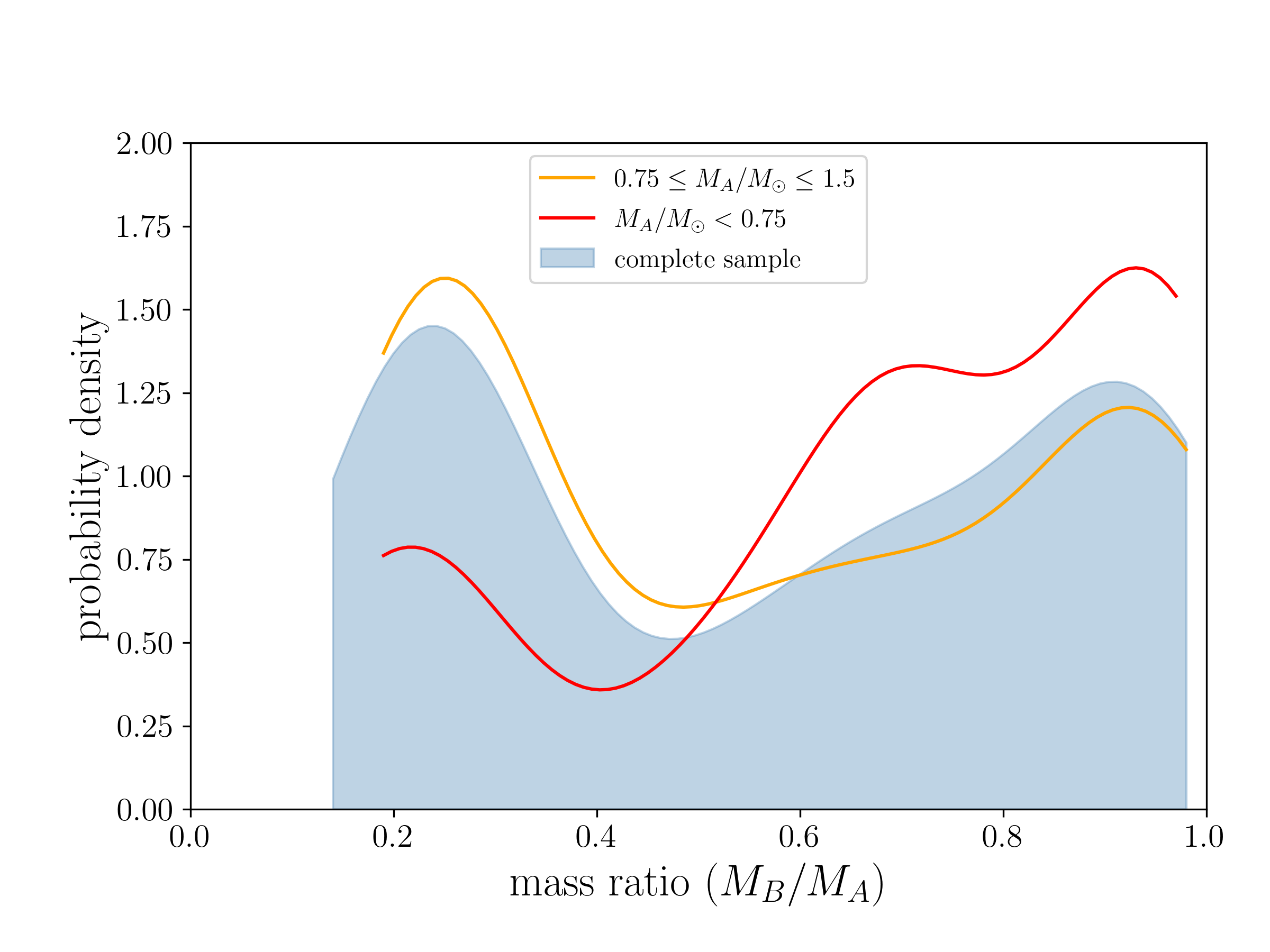}
   \caption{Mass ratio distribution obtained using the KDE method \citep[see e.g.][]{Silverman1986} and a Gaussian kernel with $\sigma=0.1$. 
   The blue shaded area corresponds to the solid blue line in Fig.~\ref{fig:distq}, while the coloured solid lines show the distributions for solar-type (gold) and low-mass (red) primaries. }
\label{fig:distq_mbins}%
    \end{figure}

Figure~\ref{fig:distq} shows the distribution of the values of $q$ obtained with the kernel density estimate (KDE) method \citep[see e.g.][]{Silverman1986} and a Gaussian kernel with $\sigma=0.1$.
Given the wide range of primary masses in our sample it is difficult to properly assess our completeness in a fixed mass ratio range, as the upper and lower bounds can correspond to very different companion masses. 
Figure~\ref{fig:distq_mbins} shows the mass ratio distributions obtained considering subsamples selected according to the primary mass, separating between solar-mass stars ($0.75 < M_A/M_{\odot}< 1.5$) and low-mass stars ($M/M_{\odot} < 0.75$). Given that the exclusion of the ScoCen members (shown with light grey squares in Fig.~\ref{fig:mA_q} and \ref{fig:mA_Mb}) effectively removed most of the systems with high-mass primaries from the reduced sample, we did not consider primaries with $M_A/M_{\odot} > 1.5$.
While the resulting subsamples are too small to draw any conclusion on the single distributions (the low-mass primaries bin only includes ten systems), this should at least clarify how each group contributes to the different peaks in the full distribution shown in Fig.~\ref{fig:distq}.

Our distribution show two distinct peaks, one including nearly equal mass systems, and the second peaking at $q\sim 0.21$. While the overall shape of the distribution apparently contradicts earlier results showing flat distributions \citep[see e.g.][]{Moe2017, El-Badry2019}, it is difficult to say whether such a difference can be explained by the difference in terms of completeness between our sample and those used to obtain such results.

\subsection{Dynamical masses}\label{sec:dyn_mass}
\begin{figure}
   \centering
   \includegraphics[width=8.8truecm]{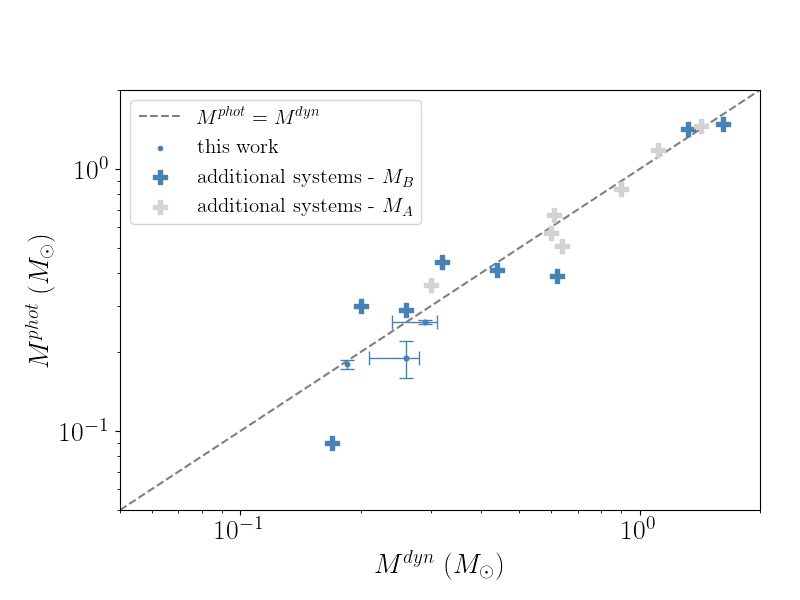}
   \caption{Dynamical mass ($M^{dyn}$) vs. photometric mass ($M^{phot}$) for the companions from our sample for which we were able to obtain an estimate of $M^{dyn}$ (blue dots; see Table~\ref{tab:orbits}), as well as both components of the additional systems from Table~\ref{tab:known_bin} for which an orbital solution was available in \cite{tokovinin_multip2018}.}
\label{fig:dynmass}%
    \end{figure}

The values of the masses used so far are derived from a comparison of the position of the stars in the colour-magnitude diagrams with isochrones, assuming their ages (evolutionary masses; see Sect.~\ref{sec:mstar}). However, for a few objects we are also able to estimate the dynamical mass using the amplitude of the observed VC curve $K$\footnote{When using {\it Orbit}, $K$ is a free parameter that fits the RV curve. Hence, we could use it to derive the masses of the secondaries despite the fact that we usually assumed the total mass of the system as derived from evolutionary models.} and assuming it as the velocity of the primary. This is true for systems with a large contrast between the components, where and the secondary is very faint in the optical. For these systems, we derived the mass of the secondaries $M_B$\ under the assumption that the total system mass of the system $M_A+M_B$ is the one obtained from evolutionary considerations. The relation between $K$ and $M_B$\ requires knowledge of the orbit inclination; this is very poorly determined for systems seen close to face-on, so we did not consider such cases. We have only three systems for which all the needed requisites are satisfied (HIP~36985, HIP~95149, and HIP~97255). Results are given in Table~\ref{tab:orbits}. 
Figure \ref{fig:dynmass} shows the comparison between the photometric mass $M_B^{phot}$ and the dynamical mass ($M_B^{dyn}$) for the 3 systems for which we performed the analysis (blue dots).
We found that dynamical and evolutionary masses agree within their errors for all of our three objects.
This is also true for both components of the additional systems from Table~\ref{tab:known_bin} for which an orbital solution was available in \cite{tokovinin_multip2018} (shown in Fig.~\ref{fig:dynmass} as grey and blue crosses).
This also further supports the goodness of the orbit derivations discussed in Sects.~\ref{sec:orbital_fit} and \ref{sec:MCorbit}.

\begin{table*}
\caption{Orbital parameters for all the systems in our sample for which an orbital fit was possible. For the few systems for which it is available, we report the dynamical mass of the secondary ($M_B^{dyn}$). The mass obtained as described in Sect.~\ref{sec:mstar} ($M_B^{phot}$) is reported for all objects.  The last two columns list the code (O={\it Orbits}; MC=MonteCarlo) and the kind of datasets used (AS=relative astrometry; RV=radial velocities, PMA=proper motion anomaly in Hipparcos and Gaia DR2 from \citealt{Kervella2019}) to obtain the orbital solution (full table available through CDS). }    \label{tab:orbits}
\resizebox{\linewidth}{!}{\begin{tabular}{r|llllllll|cc|cc|l|l}
\hline\hline
ID & $\rho$  & a  & P  & T0  & e & $\Omega$  & $\omega$  & i  & $K_A$  & $V_0$  & $M_B^{dyn}$ & $M_B^{phot}$  & Method & Data \\
  & (\arcsec) & (au) & (yr) & (yr) & & ($\degree$) & ($\degree$) & ($\degree$) & (km/s) & (km/s) & ($M_{\odot}$) & ($M_{\odot}$) & & \\
\hline
AF Hor~Aab & 0.11 $\pm$ 0.02 & 4.8 $\pm$ 0.8 & 10 $pm$ 2 & 2015.694 & 0.62$^{+0.05}_{-0.14}$ & 332.1 & 302.3 & 22.5 $\pm$ 22.5 &  &  &  &  & O & AS \\
TYC 8491-0656-1~Aaab & 0.0499 $\pm$ 0.0008 & 2.2 & 2.31 $\pm$ 0.12 & 2014.803 & 0.27 $\pm$ 0.01 & 109 & 261.8 & 127.0 $\pm$ 1.7 &  &  &  &  & O & AS \\
HIP 17157~Aab & >1.09 & >29 & >160 &  & 0.74 $\pm$ 0.04 &  &  & 157 $\pm$ 22 &  &  &  &  & O & AS+RV+PMA \\
2MASS J05195513~AB & 0.53 $\pm$ 0.16 & 30.7 $\pm$ 9 & 198 $\pm$ 64 & 2007 $\pm$ 52 & 0.40$^{+0.37}_{-0.27}$ & 1 $\pm$ 110 & 3 $\pm$ 120 & 135 $\pm$ 27 &  &  &  &  & MC & AS \\
TYC 7059-1111-1~Aab & >1.7 & >106 & >900 &  & 0.2 $\pm$ 0.1 & 268.4 &  & 89.95 &  &  &  &  & O & AS \\
HIP 36985~AB & 0.482 & 6.78 & 20.2 $\pm$ 0.3 & 2019.211 & 0.147 $\pm$ 0.008 & 134.24  $\pm$ 0.22 & 79.02 $\pm$ 3.5 & 92.84 $\pm$ 0.29 & 2.40 $\pm$ 0.09 & -2.34 $\pm$ 0.08 & 0.180 $\pm$ 0.007 & 0.185 & O & AS+RV+PMA \\
HIP 37918~Aab & 0.0784 $\pm$ 0.013 & 2.7 & 3.78 $\pm$ 0.09 & 2013.927 $\pm$ 0.018 & 0.546 $\pm$ 0.015 & 68 $\pm$ 136 & 106 $\pm$ 136 & 2.4 &  &  &  &  & O & AS+RV+PMA \\
TYC 8582-1705-1~AB & 0.057$^{+0.032}_{-0.028}$ & 3.2$^{+1.8}_{-1.6}$ & 6$^{+6}_{-4}$ & 2020 $\pm$ 3 & 0.86$^{+0.04}_{-0.38}$ & 335 $\pm$ 110 & 330 $\pm$ 110 & 55$^{14}_{-31}$ &  &  &  &  & MC & AS \\
TYC 8944 1516 1~AB & 0.030$^{+0.020}_{-0.015}$ & 3.6$^{+2.4}_{-1.8}$ & 7$^{+8}_{-5}$ & 2018 $\pm$ 3 & 0.84$^{+0.04}_{-0.45}$ & 300 $\pm$ 90 & 317 $\pm$ 90 & 53$^{+17}_{-29}$ &  &  &  &  & MC & AS \\
GSC 08584-01898~AB & 0.53 & 70.3 & 700 & 1770 & 0.56 & 137 & 284 & 109 &  &  &  &  & O & AS+RV \\
TYC 7191-007-1~AB & 1.4 $\pm$ 0.3 & 175 $\pm$ 35 & 1600$\pm$500 & 1920 $\pm$ 380 & 0.25 $\pm$ 0.10 & 123.14$\pm$ 0.3 & 160$\pm$ 20 & 82 &  &  &  &  & O & AS \\
HIP 55334~AB &  &  &  &  &  &  &  & 80 $\pm$ 6 &  &  &  &  & O & AS+PMA \\
HIP 65219 & 0.069$^{+0.021}_{-0.013}$ & 8.8$^{+2.8}_{-1.6}$ & 14.5$^{+7.4}_{-3.8}$ & 2020.3 $\pm$ 1.0 & 0.33$^{+0.14}_{-0.21}$ & 1 $\pm$ 110 & 337 $\pm$ 120 & 32$^{+15}_{-18}$ &  &  &  &  & MC & AS \\
HIP 79124~Aab & >0.12 & >16 & >50 &  & <0.5 & 65 $\pm$ 2 &  & 63 $\pm$ 6 &  &  &  &  & O & AS+PMA \\
HIP 79156~Aab & 0.96 $\pm$ 0.13 & 144 $\pm$ 20 & 1180 $\pm$ 250 & 2115 $\pm$ 300 & 0.08 $\pm$ 0.04 &  &  & 158 $\pm$ 10 &  &  &  &  & MC & AS \\
TYC 6820-223-1~AB & 0.117 & 9.73 $pm$ 0.15 & 20.78 $\pm$ 0.84 & 2009.35 $\pm$ 0.24 & 0.264 $\pm$ 0.026 & 35.3 $\pm$ 4.8 & 12.8 $\pm$ 8.8 & 147.2 $\pm$ 5.9 &  &  &  &  & O & AS \\
TYC 6872-1011-1~AB & 0.090$\pm$0.014 & 6.68 $\pm$ 1.0 & 19 $\pm$ 4 & 2019.56 $\pm$ 0.16 & 0.52$^{+0.06}_{-0.09}$ & 134.6 $\pm$ 2.5 & 239 $\pm$ 8 & 105.8 $\pm$ 1.0 &  &  &  &  & MC & AS \\
HIP 87386~AB & 0.168$^{+0.058}_{-0.042}$ & 10.6$^{+3.6}_{-2.7}$ & 23$^{+13}_{-8}$ & 2020 $\pm$ 8 & 0.48$^{+0.28}_{-0.34}$ & 350 $\pm$ 100 & 10 $\pm$ 110 & 122 $\pm$ 16 &  &  &  &  & MC & AS \\
HIP 93580~AB & 0.215 & 17.4 & 28.1 & 2000.4 $\pm$ 4.4 & 0.27 $\pm$ 0.09 & 118 $\pm$ 11 & 156 $\pm$ 52 & 59.6 $\pm$ 2.4 & 3.58 $\pm$ 2.16 & -23.5 $\pm$ 0.5 &  &  & O & AS+RV+PMA \\
HIP 95149~AB & 0.208 $\pm$ 0.003 & 4.16 $\pm$ 0.18 & 8.702 $\pm$ 0.010 & 2016.29 $\pm$ 0.36 & 0.289 $\pm$ 0.033 & 39.3 $\pm$ 5.1 & 194 $\pm$ 15 & 44.1 $\pm$ 2.4 & 1.79 $\pm$ 0.31 & 0.05 $\pm$ 0.06 & 0.19 $\pm$ 0.03 & 0.26$_{-0.05}^{+0.02}$ & O & AS+RV+PMA \\
HIP 97255~Aab & 0.275 $\pm$ 0.005 & 8.02 $\pm$ 0.15 & 19.36 $\pm$ 0.53 & 2008.00 $\pm$ 0.15 & 0.235 $\pm$ 0.006 & 14.2 $\pm$ 0.5 & 8.9 $\pm$ 2.1 & 83.6 $\pm$ 1.9 & 2.41 $\pm$ 0.05 & -8.05 $\pm$ 0.02 & 0.261 $\pm$ 0.005 & 0.29$_{-0.05}^{+0.02}$ & O & AS+RV+PMA \\
HIP 107948~Aab & 0.152 $\pm$ 0.004 & 4.6 $\pm$ 0.1 & 9.55 $\pm$ 0.09 & 2019.33 $\pm$ 0.19 & 0.41 $\pm$ 0.03 & 48.4 $\pm$ 9.9 & 106 $\pm$ 10 & 28 $\pm$ 4 &  &  &  &  & O & AS+PMA \\
HIP 107948~AB & 0.98 & 29.9 & 165 & 2010.4 $\pm$ 0.7 & 0.398 $\pm$ 0.015 & 216 $\pm$ 5 & 132.8 $\pm$ 2.1 & 38.6 $\pm$ 1.2 &  &  &  &  & O & AS \\
HIP 109427~AB & 0.235 & 6.65 & 10.896 & 2014.85 $\pm$ 0.14 & 0.53 $\pm$ 0.09 & 193 $\pm$ 11 & 61 $\pm$ 11 & 52.5 $\pm$ 0.5 &  &  &  &  & O & AS+PMA\\
HIP 113201~AB & 0.262 & 6.2 & 19.835 & 2015.465 $\pm$ 0.012 & 0.507 $\pm$ 0.005 & 275.3 $\pm$ 0.9 & 22.7 $\pm$ 0.7 & 179.9 &  &  &  &  & O & AS+RV+PMA \\
\hline\hline
\end{tabular}}
\end{table*}

\subsection{Orbital parameters} \label{sec:orbit_par}

   \begin{figure}
   \centering
   \includegraphics[width=8.8truecm]{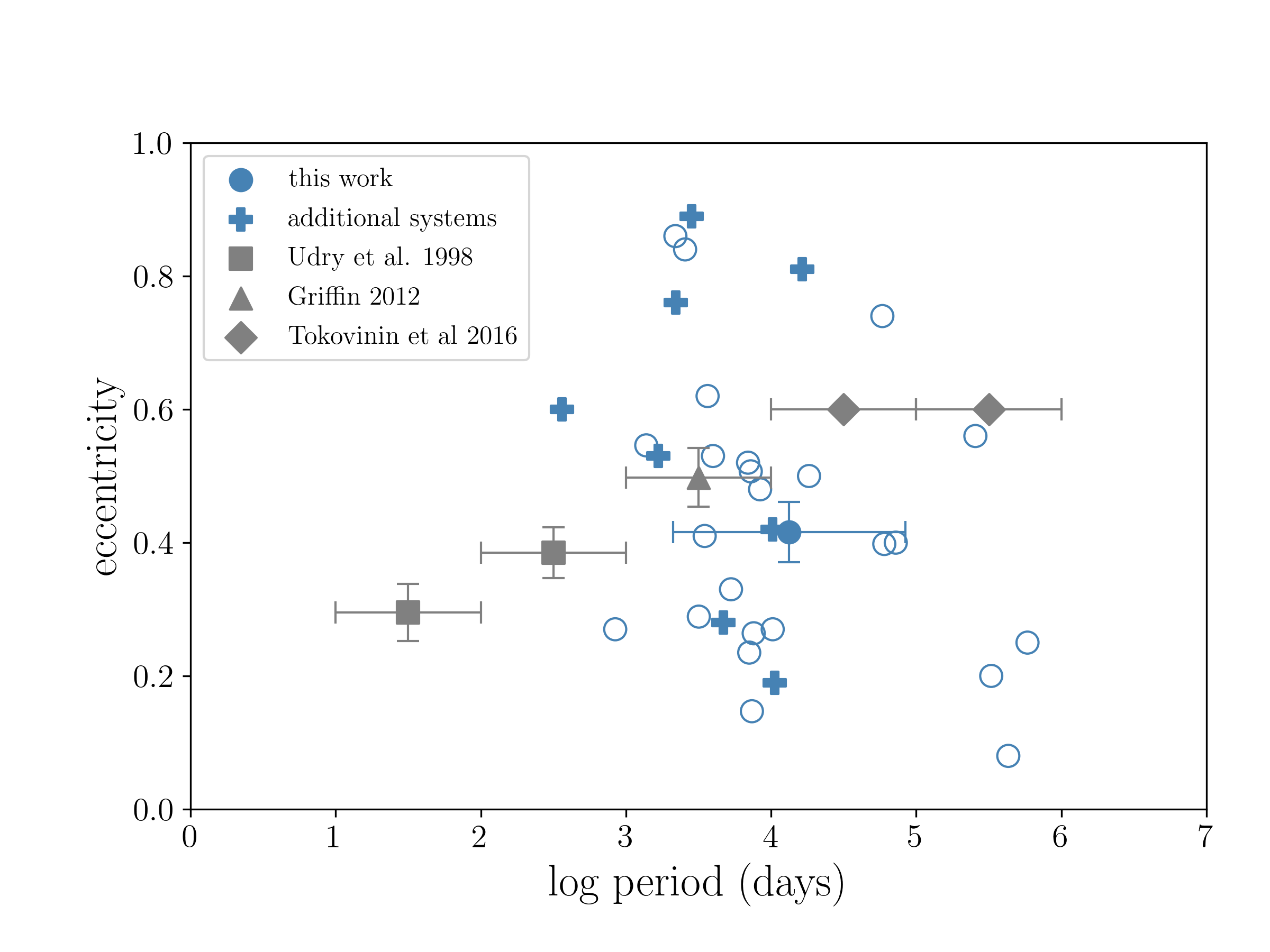}
   \caption{Eccentricity vs. orbital period for the systems in our sample with orbit determination (blue open circles). The blue filled circle is the average value ($e=0.416\pm 0.043$). The values for the additional objects from Table~\ref{tab:known_bin} are shown with plus signs. The grey squares and triangles are the average values for spectroscopic binaries obtained by \citet{Udry1998} and \citet{Griffin2012}, respectively. The grey diamonds are estimates of average values for long period visual binaries by \citet{Tokovinin2016c}. }
\label{fig:ecc}%
    \end{figure}
    
In Sect.~\ref{sec:orbits} we presented the orbital analysis for 25 of the systems presented in this paper. The same kind of analysis was not possible for systems with very long periods, for which the orbital coverage was too small even when multiple epochs were available, nor for those with very short periods because of the lack of observations (only the SHINE epochs were available). 
This obviously introduces several biases in the derived period and eccentricity distributions, 
which are difficult to assess and correct for, also due to the difference in accuracy of our orbital determinations discussed in Sect.~\ref{sec:orbits}.
Therefore, while we could identify some interesting trends emerging in our sample, the uncertainties and biases that affect our orbit determinations do not allow for an accurate analysis of the orbital parameters distributions of our targets, making our results only tentative. 

Figure~\ref{fig:ecc} shows the eccentricity versus orbital period for the 25 SHINE targets with orbital solutions (blue open circles) and the additional targets from Table~\ref{tab:known_bin} (plus signs), as well as the mean values from several previous works (grey symbols).  
The mean eccentricity of our targets ($e=0.416\pm 0.043$, with an r.m.s. of 0.21; marked as a blue filled circle in Fig.~\ref{fig:ecc}) appears to be lower than the typical value of $e=0.498\pm 0.044$ for spectroscopic binaries in the Hyades studied by \citet{Griffin2012} (blue square in Fig.~\ref{fig:ecc}) and usually considered as the reference in this range of periods (see e.g. \citealt{Tokovinin2016c}). Our value seems also very close to the average value obtained for shorter period field spectroscopic binaries by \citet{Duquennoy1991} and \citet{Udry1998}.

The orbital elements of visual binaries have a preference to lower eccentricities owing to observational limitations \citep{Finsen1936}.
For the same reason, we also expect a trend for the time of passage at periastron ($T0$) to be within the range of the observed epochs; this is not obvious in our data. However, given that our errors on the eccentricity estimates vary quite strongly due to the partial (and in some case very poor) orbital coverage, the amplitude of this effect on our sample is not trivial to assess. 

Keeping in mind the caveats discussed above, it is also worth pointing out that our data seem to show a lower mean eccentricity for higher-order systems (about half of those discussed here) compared to the simple binaries.  
Confirming this result (moderate eccentricities that are fairly independent of semi-major axis up to a few tens of au) with a larger sample would be important to clarify the origin of these systems as we expect that disk fragmentation would favour lower-eccentricity orbits with respect to core fragmentation (see discussion in \citealt{Tokovinin2016c}) that might be responsible for the formation of wider binaries. More observations and analysis is required to clarify this issue.

   \begin{figure}
   \centering
   \includegraphics[width=8.8truecm]{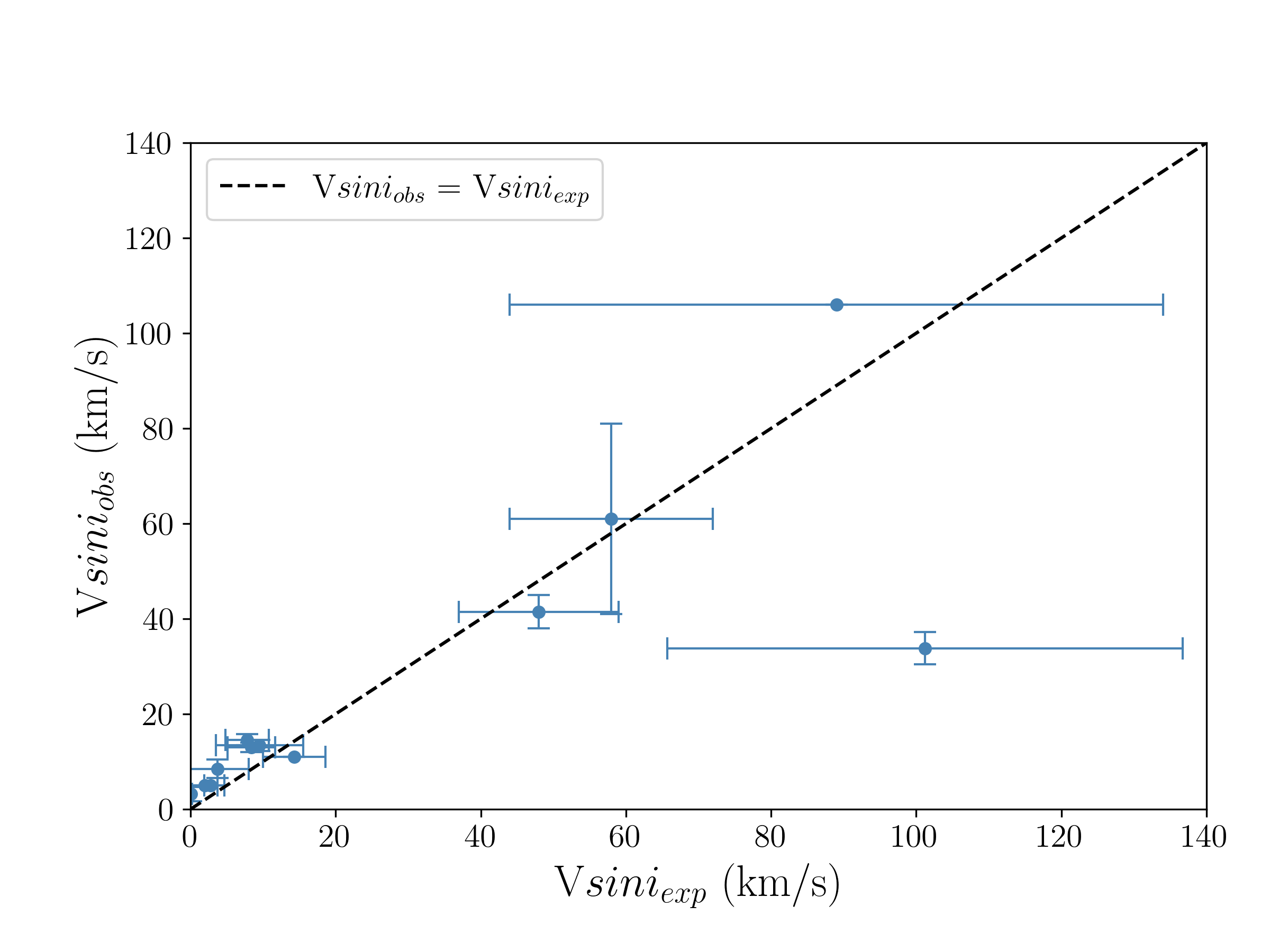}
   \caption{Comparison between the values of the stellar rotational velocity expected under the assumption that the spin axis of the primary is aligned with the orbital axis of the binary ($V~\sin{i}_{\rm exp}$) and the values measured from the spectra ($V~\sin{i}_{\rm obs}$). The error bars are estimated as described in Sect.~\ref{sec:spin-orbit}. All the values are reported in Table~\ref{tab:vsini}. The dashed black line denotes equality.}
\label{fig:vsini}%
    \end{figure}

\begin{table*}   
    \caption{Observed and predicted values of the stellar rotational velocities $(V~\sin{i})_{\rm obs}$ and $(V~\sin{i})_{\rm exp}$, respectively, in the hypothesis of spin-orbit alignment (full table available through CDS).}     \label{tab:vsini}
    \begin{tabular}{lcccccc}
\hline \hline
System          & $P_{\rm rot}$ & $R$         & $i_{\rm orbit}$ & $(V~\sin{i})_{\rm exp}$  & $(V~\sin{i})_{\rm obs}$ & Ref \\
                & (days)        & ($R_\odot$) &     (deg)       & (Km/s)                    & (Km/s) & \\
\hline
AF Hor          & $3.39\pm 0.12$   & $0.65\pm 0.08$ & $22.5\pm 22.5$ & $3.7\pm 4.3$     & $8.5\pm 1.9$  & K14, T06      \\
TYC 8491-0656-1 & $0.520\pm 0.003$ & $0.74\pm 0.16$ & $127\pm 1.7$   & $58\pm 14$       & $61\pm 20$    & K14, T06      \\
HIP~17157       & $8.0\pm 0.3$     & $0.81\pm 0.20$ & $157\pm 22$    & $2.0\pm 2.7$     &  5.0          & G20           \\
TYC 7059-1111-1 & $0.697\pm 0.007$ & $0.66\pm 0.15$ & $89.95\pm 11$  & $48\pm 11$       & $41.5\pm 3.5$ & T06           \\
HIP~36985       & $11.1\pm 2.6$    & $0.61\pm 0.06$ & $92.8\pm 0.3$  & $2.8\pm 0.9$     &  5.0          & G20           \\
HIP~37918       & $24\pm 8$        & $0.89\pm 0.26$ & $2.4\pm 2.4$   & $0.1 \pm 0.1$    &  $3.2\pm 1.5$ & D06           \\
TYC 8582-1705-1 & $3.03\pm 0.18$   & $0.69\pm 0.15$ & $55\pm 23$     & $9.5\pm 6.0$     & $13.4\pm 1.2$ & T06           \\
HIP 55334       &  $0.797\pm 0.01$ & $1.42\pm 0.66$ & $80\pm 6$      & $89\pm 45$       & 106           & C11           \\
TYC 6820-223-1  & 2.857            & $0.81\pm 0.18$ & $147.2\pm 5.9$ & $7.8\pm 3.0$     & $14.6\pm 1.2$ & T06           \\
TYC 6872-1011-1 & $0.504\pm 0.005$ & $1.04\pm 0.35$ & $105.8\pm 1.0$ & $101.2\pm 35.5$  & $33.8\pm 3.4$ & T06           \\
HIP 95149       & $3.77\pm 0.29$   & $0.89\pm 0.24$ & $44.7\pm 2.4$  & $8.4\pm 3.3$     & $13.0\pm 1.0$ & T06           \\
HIP~97255       & 3.51             & $0.99\pm 0.30$ & $83.6\pm 1.9$  & $14.3\pm 4.3$    & 11.0          & G20           \\
\hline \hline
    \end{tabular}\\
\footnotesize{\textbf{References:} \textit{K14:} \citet{kraus2014}; \textit{T06:} \citet{Torres2006};, \textit{G20:} \citet{Grandjean2020}; \textit{D06:} \citet{desidera2006a}; \textit{C01:} \citet{chen2011}} 
\end{table*}    

\subsection{Spin-orbit alignment}
\label{sec:spin-orbit}
If confirmed, the behaviour of the mass ratios and eccentricities discussed in the previous sections would suggest that a significant fraction of the systems with projected separation below a few tens of au ($<0.5$~arcsec in our sample) actually formed through disk fragmentation, rather than core fragmentation followed by capture. An additional expectation of such a scenario is that the spin axis of the star should be roughly parallel to  the direction of the orbital angular momentum, though as shown by, for example, \citet{Bate2018}, late accretion can considerably affect this prediction possibly leading sometimes to misaligned disks. On the other hand, the situation is less clear in case of binaries forming by core fragmentation along filaments, depending on the relative role of magnetic fields (leading to alignment: \citealt{Galli1993, PlanckXXXII2016,PlanckXXXV2016} and turbulent fragmentation (leading to misalignment: \citealt{Offner2016}), with a quite confuse picture emerging from observations with a predominance of alignments over random distribution \citep{Kumar2011, Kong2019, Menard2004, Davis2009, Lee2016, Stephens2017, Ansdell2016, Ansdell2018}. Spin-orbit alignment in binaries has been studied over the last 50 years (\citealt{Weis1974, Hale1994, Harding2013} and references therein). In general, it has been found that the majority of systems with separation $<40$~au are reasonably well aligned \citep{Hale1994}, although this claim have been recently challenged by \cite{Justesen2020}, whose results suggest that the reported trend that binaries with separations below 30~au are preferentially aligned could be spurious. Furthermore, the fraction of systems with aligned spin-orbit seems to decrease with separation \citep{Jensen2004, Monin2006}.  \\
\indent Most of these past studies are based on statistical arguments or make use of the Rossiter-McLaughlin effect for eclipsing binaries. Our sample is too small for a statistical study and the binaries are too wide to be eclipsing. To verify the spin-orbit alignment for the binary of our sample we would need measures of the orbital inclination, of the projected rotational velocity of the star $(V~\sin{i}_{\rm obs})$ from spectroscopy and of the rotation period from photometry; we note that any result given by this comparison is still statistical, because having the same value of $\sin{i}$ does not necessarily imply alignment of the axes. The required quantities are actually available only for 12 systems (see Table~\ref{tab:vsini}). The comparison we considered is that between the observed value of $(V~\sin{i}_{\rm obs})$, and the one expected $(V~\sin{i}_{\rm exp})$ given the radius of the star, the rotational period measured from photometry, and the orbital inclination, assuming that the latter represents the inclination of the stellar rotation axis. To estimate the radius (of the primaries) we compared its absolute $J$ magnitude - corrected for the contribution of the secondary - to that of main sequence stars of the same spectral type (a proxy for the temperature) in the tables by Pecaut \& Mamajek \footnote{\url{http://www.pas.rochester.edu/~emamajek/EEM_dwarf_UBVIJHK_colors_Teff.txt}}. The result of this comparison is shown in Fig.~\ref{fig:vsini}. While the sample is clearly very limited, we obtain a reasonably good agreement between the predicted and observed value for $V~\sin{i}$, supporting the spin-orbit alignment for the systems considered in this analysis. The most discrepant case is TYC~6872-1011-1, which is a very fast rotator with a photometric period of only 0.503 d but has only a moderately high value of $V~\sin{i}=33.8$~km/s \citep{Torres2006}; the discrepancy would be largely resolved if the real rotational period is a multiple of the photometric period, but there is no clear evidence of this in the light curve. Obviously more data are required to confirm this result with an extended sample.

\section{Conclusions and final remarks}
\label{sec:conclusions}

In this paper we present 78 multiple systems with separation $<5$ arcsec observed within the context of the SHINE survey (including 15 triple systems and 6 quadruple), 56 of which are new discoveries.  
Given the extremely heterogeneous nature of the datasets used, several different methods were employed to detect these companions in the high contrast imaging data provided by SPHERE.
In particular, newly developed dedicated routines were used to deal with very close systems. 

We combined the SHINE dataset with literature and archival data, trying to better characterise these systems. In particular, we used TESS data to extract stellar rotation periods for a large part of our sample; we carefully derived ages of all the systems following the selection criteria of the SHINE survey, finding that they are generally young. Finally, masses for all components were derived from a comparison with model isochrones.

Given the strong selection bias against binarity applied while selecting the SHINE sample \citep[excluding any object with stellar companions within the SPHERE FoV at the time of selection; see][for details]{Desidera2021}, it was not possible to draw any definitive conclusion regarding the impact of our results on the overall frequency of binaries. 
In an attempt to compensate for the original bias while estimating the completeness of our sample, we limited our analysis to the young moving group members included in the original SHINE list, which allowed us to retrieve and re-introduce the excluded targets, also taking into account their probability of being observed, based on the SHINE priority bins. 
This extended sample of SHINE young moving group members counts 231.84 targets including 76.84 multiple systems (34 of which are among the 189 stars in moving groups observed within SHINE and 42.84 were originally excluded from the selection because they were already known to be binaries). We note that the census of binaries in this sample of young moving stars is still incomplete because it does not include systems with very small or very wide separation.

Restricting the analysis to companions with $0.05\arcsec < \rho < 0.5"$ -- where, given the quality of our observations,  our sample is more likely to be complete -- we find a binary frequency of $14.2 \pm 2.9\%$. 
This indicates a possible excess of binaries in our sample with respect to that with similar periods in \citet{Duquennoy1991} and \citet{Raghavan2010}, that is, about 11\%; this difference is, however, only marginally significant.

A full discussion of all the properties of these systems was beyond the purposes of this paper; however, we note some very interesting trends.
A very tentative discussion of the mass ratio distribution of the binaries with $0.05\arcsec < \rho < 0.5"$ highlighted two distinct peaks, one that includes nearly equal mass systems and a second that peaks at $q\sim 0.21$. While this result seems to disagree with previous claims of smooth mono-modal distributions from previous studies, such as \citet{Moe2017} or \citet{El-Badry2019}, it is not clear whether such a difference can be explained as a result of selection effects and possible differences in the samples. 

We derived orbital parameters for 25 systems by combining our data with literature data. To this purpose, we not only considered high contrast imaging and speckle interferometry data, but also separate detections of the components on the Gaia DR2 archive, the derivation of the PMA at the epochs of Hipparcos and Gaia by \citet{Kervella2019}, and RVs whenever available. While additional constraints might possibly be derived in the future for additional targets (for instance, with a more systematic combination of PMA with high contrast imaging), there are some interesting preliminary results. For instance, we derived generally moderate eccentricities for most of our targets, but once again it is unclear if this result is actually robust against selection effects.
Exploiting the knowledge of the rotation period of the stars, we could compare the observed values for the projected rotational velocity of the primaries ($V \sin{i}$) with that expected if the stellar and orbital spins are aligned for 12 systems, in general finding a good agreement (but with one possible exception) and confirming, via a star-by-star comparison, the results found using statistical arguments by \citet{Hale1994}. 

Overall, the properties of the binaries in our sample seem to favour a disk fragmentation scenario \citep{Bate2002} for the formation of most of the systems we studied.
However, we should once more stress that these conclusions are highly speculative, given the biases affecting our sample. 

An example of the possible constraints that could be obtained using systems such as those discovered in our survey is provided by HIP~107948 \citep{Elliott2015}. 
This is the only triple system for which we were able to derive constraints on the orbits of all components, thanks to the combination of our data with a rich literature\footnote{As mentioned in Appendix~\ref{app:targets}, an orbital solution for both components of this system was already available in \cite{Tokovinin2020a}, showing values in good agreement with the ones presented in Table~\ref{tab:orbits}}. Very intriguingly, we found that the orbits of components Aa and B  (the latter around the barycentre of the Aab system) are both prograde, are roughly co-planar (within $16\pm 13$ degrees), have a very similar eccentricity ($e\sim 0.4$), and have a similar longitude of the periastron. This suggests a common origin within a disk and important dynamical interactions; this is similar to the majority of (but not all) cases of triples with both orbits determined studied by \citet{Tokovinin2021}. 

We conclude by stressing the importance of combining different techniques when tackling complex problems such as the formation of binaries. Even restricting ourselves to a limited range in separation, such as the one considered here, the most exciting results from our study required the combination of high contrast imaging, speckle interferometry, space astrometry and photometry, high precision RVs, and high resolution spectra. While our analysis is essentially serendipitous, because SHINE was focused on sub-stellar companions around single stars, we demonstrate how large samples are often useful for purposes different from those originally considered. 





\begin{acknowledgements}
The authors would like to thank the referee, Dr. Tokovinin, for providing very insightful feedback that significantly improved the clarity of the paper.
SPHERE is an instrument designed and built by a consortium consisting of IPAG (Grenoble, France), MPIA (Heidelberg, Germany), LAM (Marseille, France), LESIA (Paris, France), Laboratoire Lagrange (Nice, France), INAF - Osservatorio di Padova (Italy), Observatoire de Gen\`eve (Switzerland), ETH Z\"urich (Switzerland), NOVA (Netherlands), ONERA (France) and ASTRON (Netherlands) in collaboration with ESO. SPHERE was funded by ESO, with additional contributions from CNRS (France), MPIA (Germany), INAF (Italy), FINES (Switzerland) and NOVA (Netherlands). SPHERE also received funding from the European Commission Sixth and Seventh Framework Programmes as part of the Optical Infrared Coordination Network for Astronomy (OPTICON) under grant number RII3-Ct-2004-001566 for FP6 (2004--2008), grant number 226604 for FP7 (2009--2012) and grant number 312430 for FP7 (2013--2016). 
This research has made use of the SIMBAD database and Vizier services, operated at CDS, Strasbourg, France and of the Washington Double Star Catalog maintained at the U.S. Naval Observatory. 
This work has made use of data from the European Space Agency (ESA) mission
{\it Gaia} (\url{https://www.cosmos.esa.int/gaia}), processed by the {\it Gaia}
Data Processing and Analysis Consortium (DPAC,
\url{https://www.cosmos.esa.int/web/gaia/dpac/consortium}). Funding for the DPAC
has been provided by national institutions, in particular the institutions
participating in the {\it Gaia} Multilateral Agreement.
This paper includes data collected with the TESS mission, obtained from the MAST data archive at the Space Telescope Science Institute (STScI). Funding for the TESS mission is provided by the NASA Explorer Program. STScI is operated by the Association of Universities for Research in Astronomy, Inc., under NASA contract NAS 5–26555.
This paper has made use of data products available in ESO archive. 
This work has been supported by the project PRIN INAF 2016 The Craddle of Life - GENESIS-SKA (General Conditions in Early Planetary Systems for the rise of life with SKA), by the "Progetti Premiali" funding scheme of the Italian Ministry of Education, University, and Research
and by the ASI-INAF agreement n.2018-16-HH.0.
M.B. acknowledges funding by the UK Science and Technology Facilities Council (STFC) grant no. ST/M001229/1.
AV acknowledges funding from the European Research Council (ERC) under the European Union's Horizon 2020 research and innovation programme (grant agreement No.~757561).
C.\,P. acknowledge financial support from Fondecyt (grant 3190691) and financial support from the ICM (Iniciativa Cient\'ifica Milenio) via the N\'ucleo Milenio de Formaci\'on Planetaria grant, from the Universidad de Valpara\'iso. \par
For the purpose of open access, the authors have applied a Creative Commons Attribution (CC BY) licence to any Author Accepted Manuscript version arising from this submission.

\end{acknowledgements}

\bibliographystyle{aa}
\bibliography{biblio}


\begin{appendix}
\section{Photometric variability and rotation periods}
\label{app:prot}

We used data from the TESS mission \citep{tess2015} in order to determine their ages through gyrochronology, and to look for evidence of transiting planets. To robustly calibrate the age-period-colour relation, we also considered the (single) stars in the F150 sample \citep[][]{Desidera2021}. We downloaded data from the Mikulski Archive for Space Telescopes (MAST) portal (\url{https://archive.stsci.edu/tess/}). We considered here only light curves and the standard analysis for transits produced by the Science Processing Operations Center (SPOC; \citealt{Jenkins2016}) available on the archive on July 20, 2020. The transit analysis includes a measure of the shift of the photocentre during the transit, pre-whitening of the light curve, and extraction of transit parameters. These data are clearly preliminary; in particular, transit signals should be further checked to confirm them. Search areas were kept at $<10$ arcsec (half a TESS pixel) around the nominal star position to avoid misinterpretation of data. We determined periods from peaks in the Scargle periodogram \citep{Scargle1982} extracted from the light curves. Error bars are the half width half maximum of the peak values.

\begin{figure*}[hbt]
  \centering
  \begin{tabular}{cc}
  \includegraphics[width=0.45\textwidth]{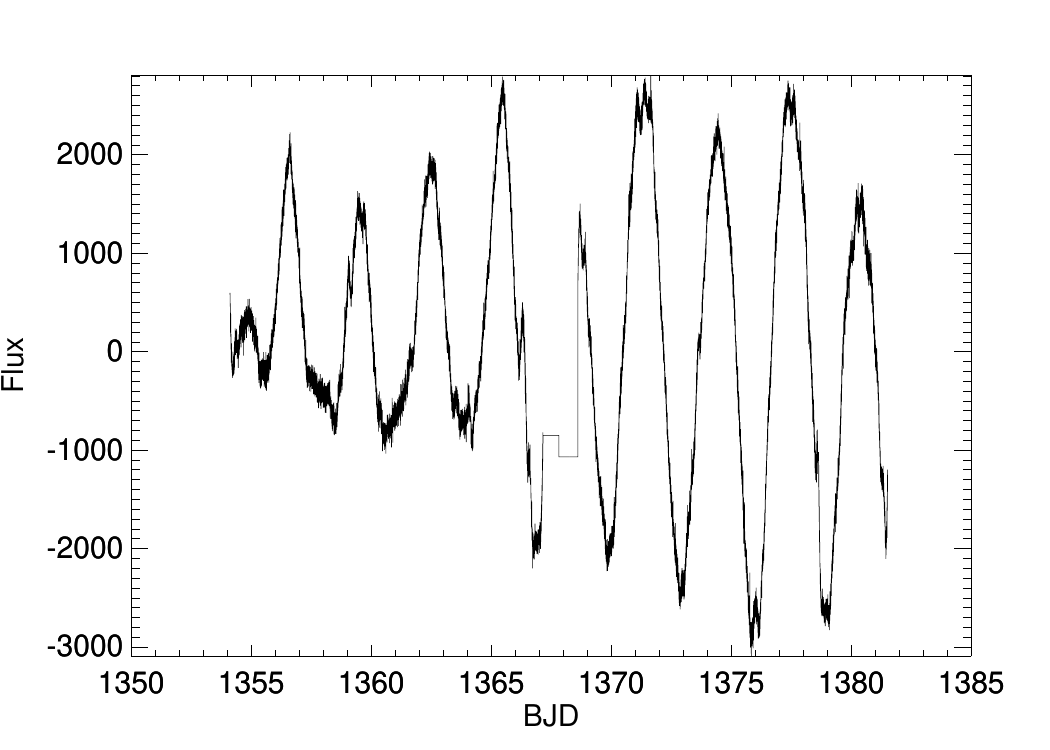}   & 
  \includegraphics[width=0.45\textwidth]{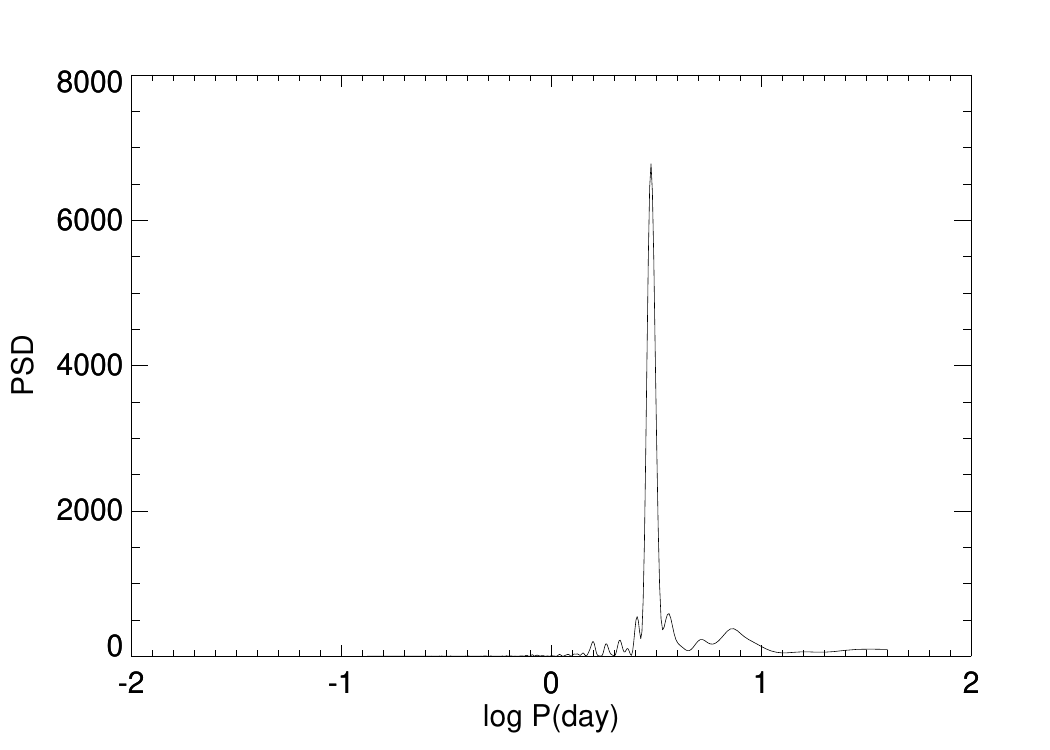}  \\
  \includegraphics[width=0.45\textwidth]{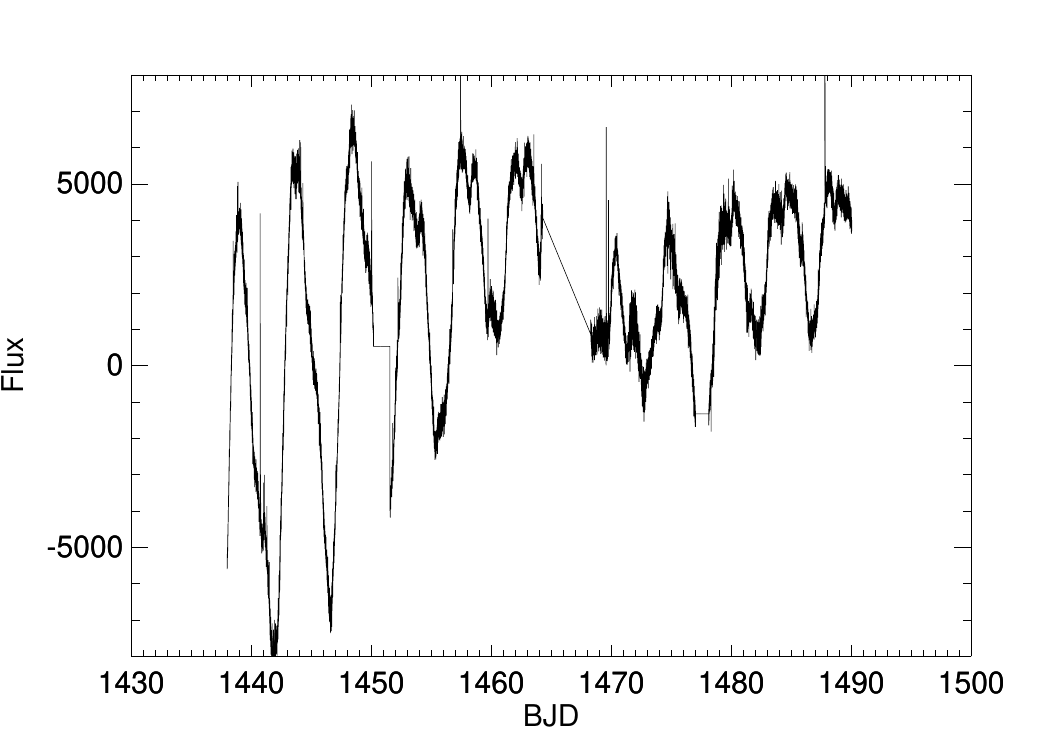}   &
  \includegraphics[width=0.45\textwidth]{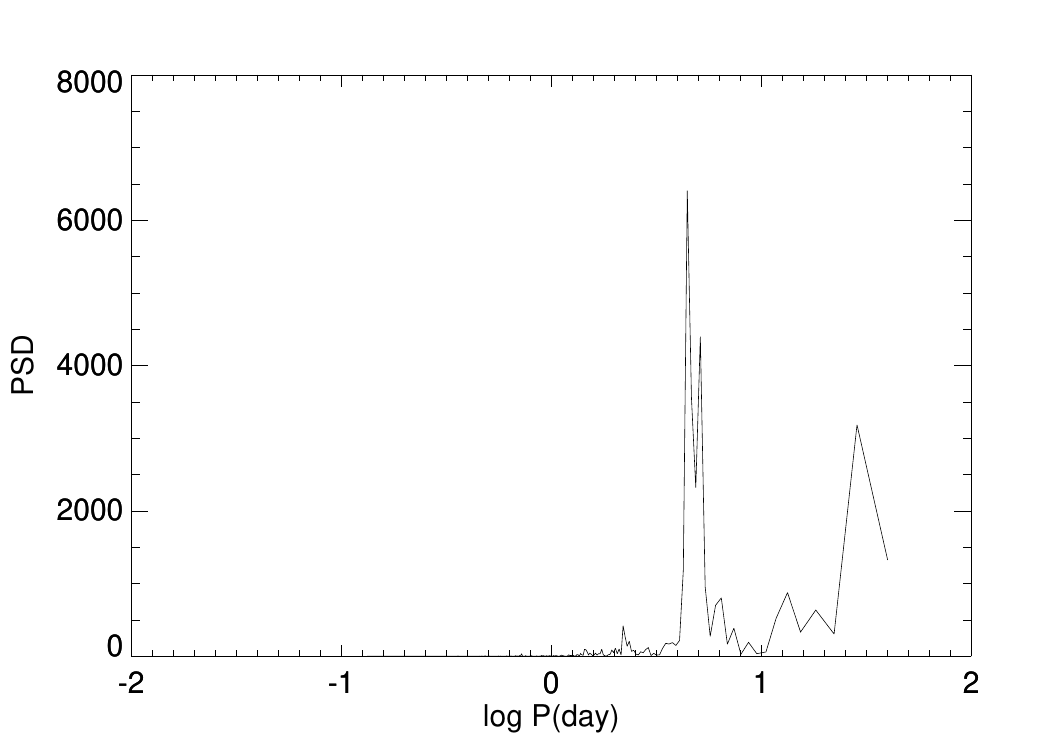} \\
  \includegraphics[width=0.45\textwidth]{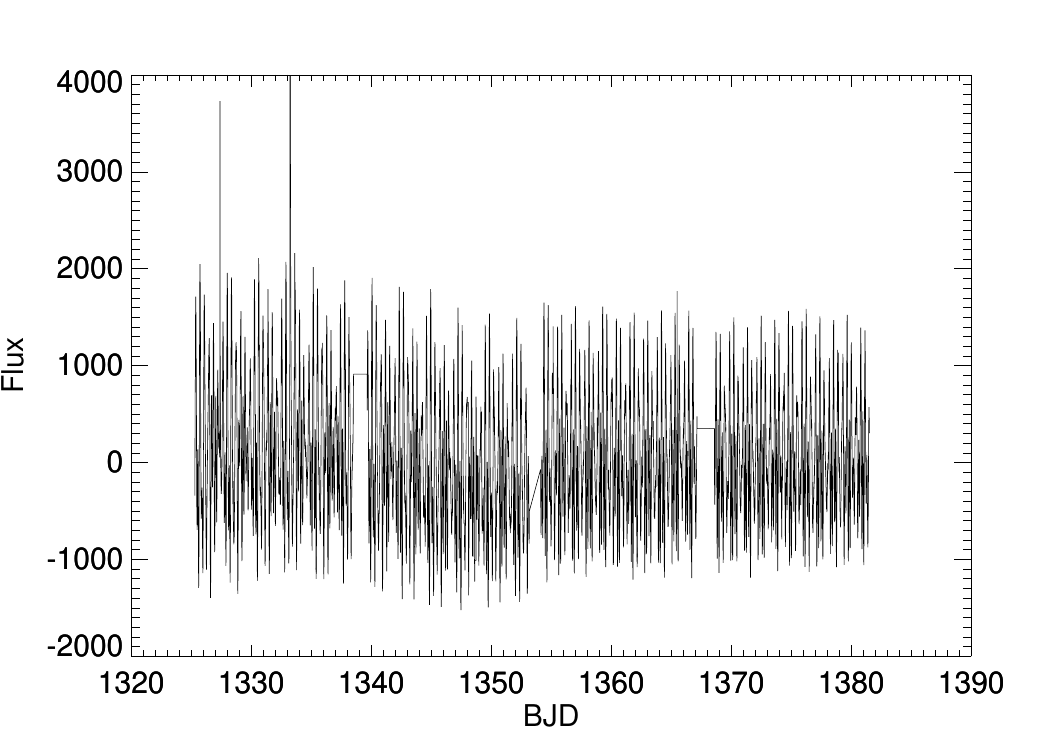}   & 
  \includegraphics[width=0.45\textwidth]{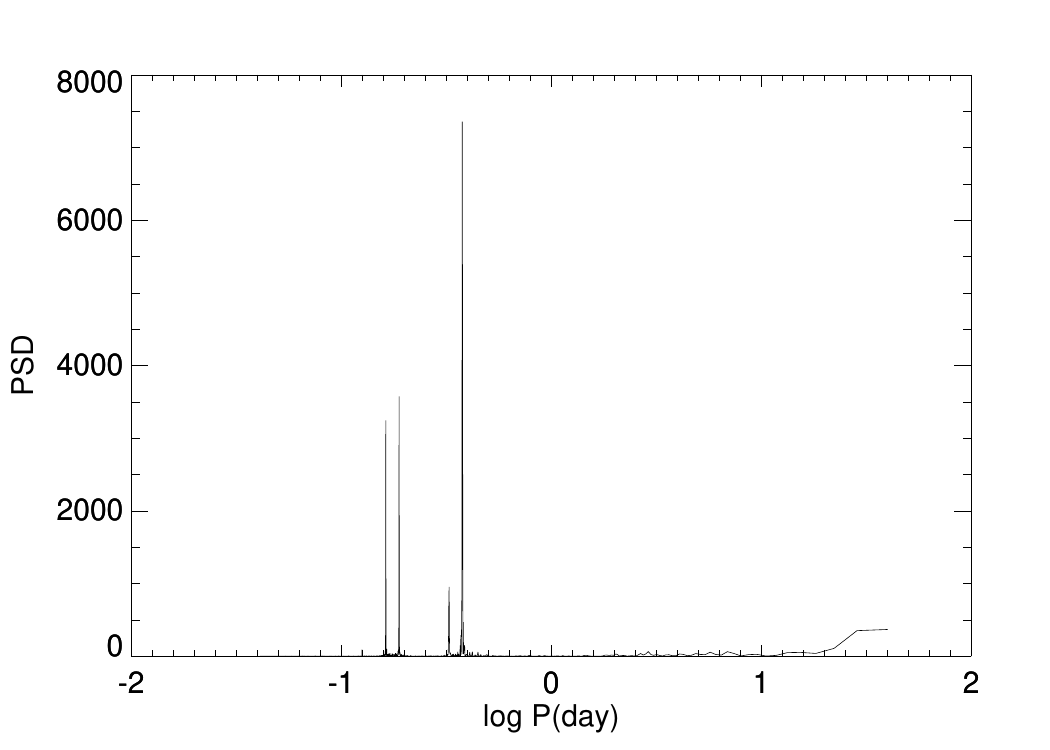}  \\
  \end{tabular}
  \caption{Examples of TESS light curve (left) and Lomb-Scargle periodograms (right). Upper row: HIP~490, a single star from the F150 sample with a single peak in the periodogram. Middle row: HIP~26369, a normally rotating binary with two distinguishable peaks in the periodogram that can be attributed to the two components. Lower row: HIP~2729, a fast rotator binary with two distinguishable peaks in the periodogram that can be attributed to the two components; the other peaks seen in the periodogram are due to the harmonics of these two periods. }
  \label{fig:tess}
\end{figure*}

  \begin{figure}[ht]
   \centering
   \includegraphics[width=9truecm]{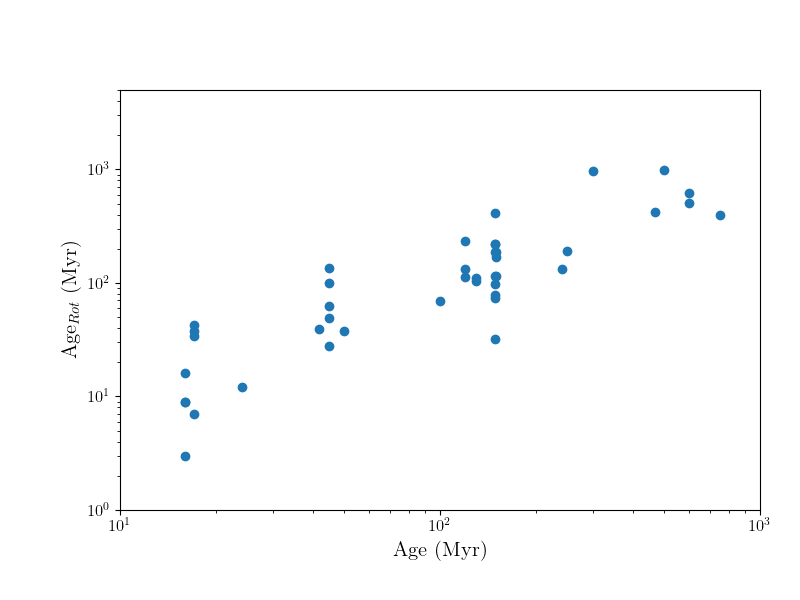}
   \caption{Comparison between the ages obtained by the gyro-chronological calibration using rotation periods from TESS (Age$_{Rot}$) and the adopted age (see Table \ref{tab:age_per} for details).} 
    \label{fig:age_cal}
    \end{figure}

On July 20, 2020, SPOC TESS light curves were available for 48 stars of the binary sample, and for 96 stars in the SHINE F150 sample; in both cases, they are about 60\% of the total. All stars show strong and highly significant peaks in their periodograms. Since the programme stars are typically young, the highest peak is likely related to stellar rotation for the late-type stars, while the typically very short periods (with quite small amplitudes) obtained for early-type stars (spectral type earlier than F3) are most likely related to pulsation. In most cases the strongest peak can be attributed to the primary (see upper row of Fig.~\ref{fig:tess}; this concern does not exist for the single stars in the F150 sample); however, in a few cases additional signals may be related to the secondaries (see middle row of Fig.~\ref{fig:tess}), though there is some ambiguity in these attributions. Long periods for systems with an early-type primary and a late-type secondary were attributed to the secondary. In addition, depending on the distribution of active regions on the stellar surfaces, the strongest peak in the periodogram may be a harmonic of the real rotational period.

With these caveats, we used these periods (complemented by a few data points from the literature) to obtain an estimate of the stellar ages for stars with spectral type later than F7. We did not consider here fast rotating late-type stars ($P<1$~day) because they bias the calibration of gyrochonology for very young stars \citep{messina2017, Messina2019}; they are further discussed below. Once these are eliminated, we do not find any systematic difference between close binaries (projected separation $<100$~au) on one side and wide binaries and bona fide single stars on the other. We used here the colour-period-age calibration by \citet{Angus2019}, which is appropriate for stars of spectral type later than F7. This calibration uses Praesepe and the Sun as benchmarks; since most of the programme stars are much younger than Praesepe (age of about 650 Myr), we corrected the ages obtained in this way to match those obtained for young associations as compiled by \cite{Desidera2021}. When we compare the ages obtained by this TESS gyrochronology with those obtained using the approach by \cite{Desidera2021}, we obtain a very high Persson linear correlation coefficient of r=0.75 over 90 stars, and the rms scatter around the mean relation is 0.26 dex, indicating that the accuracy of these ages is about 81\%. The errors are much larger than those expected from the uncertainties in the periods alone. 
The result of this calibration is shown in Fig.~\ref{fig:age_cal}. 
If we compare these gyro-chronological ages ($t_{\rm us}$, in Myr) with those obtained using the calibration by \citet{Barnes2007} ($t_{\rm Barnes}$, in Myr), we found the relation $\log{t_{\rm us}}=1.37~\log{t_{\rm Barnes}}-0.69$, with a very small scatter. The small scatter means that the colour terms are handled in a similar way in the two calibrations. The two age estimates agree very well for ages around $\sim 100$~Myr (that is the median value in our sample) but our ages are lower for younger objects and higher for older ones. Since the SHINE survey focuses more on the youngest objects, we comment these ones here. On this respect, we notice that the youngest calibrating objects considered by \citet{Barnes2007} have an age of 30 Myr, while we also used younger ones (age $\sim$ 10 Myr); this makes his calibration less adequate for the youngest objects in our sample. For instance, the median age of ScoCen members derived using our calibration is 17~Myr, in good agreement with that obtained by \citet{Pecaut2012}, while the value obtained using the calibration by \citet{Barnes2007} is 24~Myr.

A few of the programme stars of spectral type from late-K to early-M are fast rotators (periods $<$1 day) that do not match the colour-period-age calibration in a reasonable way. A sequence of fast rotators is typically observed when examining young associations (see e.g. \citealt{Meibom2009}). \citet{Rappaport2014} found several similar examples in Kepler data, where they constitute a few per cent of all M stars. This fraction is lower than observed in young clusters, likely because of an age dependence. In a fraction of the cases, \citet{Rappaport2014} found two measurable short periods, which may be attributed to individual components in binary systems (see lower row of Fig.~\ref{fig:tess}). These fast-rotations are attributed to an early disruption of the circumstellar disk that is expected to keep the rotation of contracting stars slow through disk-locking \citep{Bouvier2020, Lamm2005, Bouwman2006, Jayawardhana2006, Fallscheer2006, Weise2010}. We could then expect that fast rotators are related to moderately close binaries (projected separation of a few au). Indeed, \citet{tokovinin_multip2018}
found that late-type multi-periodic variables are typically close visual binaries. Furthermore, in these systems both components could be fast rotators - and if the mass ratio is not too large, two periods are observed in the light curve, related to the two components (as also observed by \citealt{Rappaport2014} for a fraction of the Kepler short period stars). We find that the observed light curves for a few systems (HIP~2729, AF~Hor, TYC~7627-2190-1, TYC~6872-1011-1) may indeed be explained in this framework; the high fraction of fast rotators within our sample may be explained by the combination of a young age and the fact that we are specifically observing binaries. On the other hand, there are a couple of fast rotators that are in quite well separated binaries (2MASS J05195513, TYC~7059-1111-1) though we notice that 2MASS J05195513 is in a quadruple system and there is a hint that the primary of TYC~7059-1111-1 is itself a close binary. In addition, nine of the bona fide single stars in the F150 sample are also fast rotators; while of course some of them could be close binaries so far undetected, this explanation seems unlikely for the full set of objects. Finally, there are several close binaries (projected separation of a few au) with the classical longer periods expected on the basis of the colour-period-age calibration. We notice that small projected separation may be obtained by chance alignment along an actually (relatively) wide orbit and that the orbits may be highly eccentric. Further analysis is thus needed to clarify this issue.

In addition to the rotational period, SPOC analysis of TESS photometry provided candidate transits among a dozen of the systems in the binary sample. 
More details about these results are given in the context of the individual objects in Appendix~\ref{app:targets}.\\

\section{Notes on individual targets}
\label{app:targets}
\medskip
\noindent{\bf HIP 2729 = HD 3221}
This target was included in various surveys looking for faint companions (SEEDS: \citealt{Janson2013}; NICI: \citealt{Biller2013}; VLT-NACO LP: \citealt{Vigan2017}; GPIES: \citealt{Nielsen2019};  WEIRD: \citealt{Baron2018}) but it was not mentioned as binary. Our data indicate that this is a close system with two components having nearly the same luminosity.
Kinematics indicates membership to the Tuc-Hor group. \citet{Grandjean2020} found a large range for RVs ($RV_{\rm amp}=4.79$~km/s), suggesting that this is a spectroscopic binary. Since apparent separation of the binary is about 1 au, the RV variability may well be due to the newly detected companion.
TESS periodogram shows two very significant short period peaks consistent with both the components being fast rotators.

\medskip
\noindent{\bf AF Hor = GSC 08491-01194}
It shares proper motion with TYC~8491~656~1 ($\rho>20$~arcsec); both systems are member of the Tuc-Hor group.  
Binary nature was discovered by \citet{Shan2017}. The target was included in the NICI \citep{Biller2013}, SEEDS \citep{Janson2013}, and WEIRD \citep{Baron2018} surveys but no companion was detected. 
Our data indicate that this is a close system with two components having nearly the same luminosity. TESS reveals a period of 3.39 days that, if considered the rotation of one of the two components, agrees with the expectation for a member of the Tuc-Hor group.

A number of constraints can be considered for the orbit of AF Hor, in addition to the relative astrometric position measured by \citet{Shan2017} and by us. These include (i) the lack of RV variations \citep{Torres2006, kraus2014}, (ii) the difference in the proper motion measured by Gaia DR2 with respect to historical values (SPM4 and UCAC4), from which we can extract a PMA at epoch 2015.5 (which yields values of 1.0 and -7.8 mas/yr in RA and Dec., respectively), and (iii) the lack of detection in the observations with the The Near Infrared Camera and Multi-Object Spectrometer mounted on the Hubble Space Telescope \citep[NICMOS@HST][]{nicmos}. on October 24, 2004, and August 2, 2005 (proposal ID-10176, PI Song) that we retrieved from the MAST archive\footnote{\url{https://archive.stsci.edu/hst/search_retrieve.html}}, which implies that separation was $<0.15$~arcsec at the epochs. The constraints from RV and PMA should take into account the dilution effect due to the contribution by the secondary that is large; for instance for RV we estimate an upper limit of 5~km/s on the amplitude of RV variations for the primary. With these data, we used {\it Orbit} to explore the range of possible solutions assuming the masses from photometry. We found $P=10\pm 2$~yr (that implies $a=110\pm 20$~mas), $e=0.62^{+0.05}_{-0.14}$, and $0<i<45$~degree.

\medskip
\noindent{\bf TYC 8491-0656-1}
It is a wide companion of AF Hor whose kinematics indicate membership in the Tuc-Hor association. No companion is detected in SEEDS \citep{Janson2013}, NICI \citep{Biller2013}, or \citet{Janson2017}. It was previously resolved by Speckle \citep{Tokovinin2016,Tokovinin2018,Tokovinin2019}.
Our data indicate that this is a close,  nearly equal mass binary. Thus, this is a quadruple system. 
It shows significant $\Delta\mu$ between Tycho, Gaia DR1, and Gaia DR2. 
The TESS periodogram shows two very significant short period peaks consistent with both the components being fast rotators. Similar periods were also reported by \citet{kiraga2012} and \citet{oelkers2018}.

Combining our astrometry with literature data suggests a binary period of about 2.3 yr and semi-major axis of about 45-50 mas, yielding a total mass reasonably consistent with that determined from the photometric analysis of the two components. When speckle data are taken at face value, residuals from the best orbit are always very large; the best orbital solution has a very large eccentricity and a quite large inclination. This might perhaps explain why the binary is resolved only in some of the Speckle observations. However, it is possible that the two stars are exchanged in some of the speckle dataset. This opens the possibility to an orbital solution with very small residuals; this orbit would have low eccentricity (e=$0.27\pm 0.01$) and an intermediate inclination ($i=127.0\pm 1.7$~degree). A more careful examination of existing data and/or acquisition of new data might clarify this issue. We note that interferometric observations might be needed for a full orbital solution.

\medskip
\noindent{\bf TYC 8497-0995-1 = GSC 08497-00995}
Our data indicate that this is a close system with a small luminosity difference between the two components. 
Kinematics indicate membership in the Tuc-Hor association. No companion was detected by \citet{Janson2013}, \citet{Biller2013},  \citet{Shan2017}, or \citet{Janson2017}. 
The star has a constant RV from the SACY catalogue \citep[Search for Associations Containing Young Stars][]{Elliott2014}. 
TESS reveals a period of 7.408 days that, if considered the rotation of one of the two components, agrees fairly well (it is actually a bit longer) with the expectation for a member of Tuc-Hor.
A similar period was reported by \citet{messina2010} and \citet{kiraga2012}.

\medskip
\noindent{\bf HIP 17157 = CD-48 1042}
This is a binary system with both components in Gaia DR2, with $\Delta$G=3.44. They have similar parallaxes and proper motion although with formally significant differences. There is also a significant Gaia-Dr1 versus Tycho2 $\Delta \mu$. Notably, our observation reveals a third object not much fainter than the secondary. Since this further object is only detected in our observation, it is not sure that it is a physical companion to the other two objects; however, given the large distance of this star from the galactic plane, this is highly probable. This third object is far enough from the two other components so that the Gaia astrometry is unlikely to be affected significantly; however, the impact of this source on the dynamics of the system is unclear.

The star is not associated with known moving groups, but all the indicators consistently indicate a young age. 
Periodogram of TESS data indicate a period of 7.694 d that nicely overlaps the sequence of the Pleiades.
Lithium and activity are consistent with an age similar or slightly older then the Pleiades.

Gaia DR2 gives two separate entries for the components of this system with full astrometric solution. If we combine these data with our position and the RVs from \citet{Trifonov2020} (who used the same observations as described in \citealt{Grandjean2020}), and if we assume the masses from photometry, we derive a period of $>160$~yr (semi-major axis $a>1.09$~arcsec, that is, $a>29$~au), an eccentricity between 0.70 and 0.78, and an orbit seen with inclination $i>135$~degree, under the assumption that this result is not significantly affected by the third component. If we add the constraints on the PMA by \citet{Kervella2019}, which are, however, not well matched (perhaps because of the third component), the best result is at the lower edge of the period range ($P=160$~yr, $e=0.78$\ and $i=167$~degree, that is, an orbit seen almost face-on). Given the complex and unclear dynamics of this system, we do not use it in the statistical discussions of Section 5.

\medskip
\noindent{\bf HIP 17797}
This is a triple system member of the Tuc-Hor association. The secondary at about 8 arcsec is HD 24071, also an early-type star. 
Both components are in Gaia DR2, with similar parallax but with large offsets in proper motion (16.6 mas/yr in RA and 3.6 mas/yr in Dec.) between the two components in Gaia DR2. 
We found that the primary is a tight binary, possibly explaining the offset in proper motion. Nothing found at large separation by WEIRD \citep{Baron2018}. 
The TESS periodogram has a peak at 6.897 days that is consistent with the expected rotation of the companion revealed by our data. 

\medskip
\noindent{\bf HIP 18714 = HD 25402A}
This star has a wide companion at 8.6 arcsec \citep{Tokovinin2014} confirmed by Gaia DR2 parallax (HD 25402B). There are large offsets in proper motion (2.6 mas/yr in RA and 6.6 mas/yr in Dec.) and in RV (3.7 km/s) between the two components in Gaia DR2. The primary was shown to be a tight binary by our SPHERE observation; this might explain the offset in proper motion with the wide companion. This is a new discovery from this study \citep[the star was previously observed by ][with nothing found]{Baron2018}.
The star was classified as a member of the Tuc-Hor association by \citet{kiss2011} and classified as bona fide member in several successive studies.  BANYAN kinematic analysis with Gaia parameters yield 22.0\% membership probability on Tuc-Hor for the primary and 99.9\% for the secondary. Independent of the kinematics, the age indicators are somewhat ambiguous: the Li EW by \citet{kiss2011} is compatible with Tuc-Hor but also with the locus of older systems as the Pleiades and AB Dor MG, while clearly above the Hyades. 
No rotation period was detected by \citet{messina2011} and the star has no X-ray emission from ROSAT. However, analysis of TESS light curve shows a photometric modulation with possible period of 3.58 days, slightly slow for a Tuc-Hor member. The position on the CMD of the secondary is well above the Zero Age Main Sequence, indicating a young age of few tens of megayears, unless it is itself a binary with a companion contributing significantly to the integrated flux.
\citet{nordstrom2004} noted significant RV variability (rms 2.8 km/s from 2 epochs over 829 days). Considering the projected separation of 44 mas = 2.3 au, the newly detected companion can be the responsible for the observed variability.

\medskip
\noindent{\bf HD 25284A}
This multiple system is thought to be a member of the Tuc-Hor association. There is a wide companion (HD25284B) at 11 arcsec. 
HD 25284B was flagged as 0.5" binary in the Washington Double Star Catalogue \citep[WDS][]{wds}.  from Tycho, but this companion was not retrieved in our SPHERE observations \citep{Langlois2021} and then the target was kept in the SHINE statistical analysis as an isolated object. Nothing found at large separation by WEIRD \citep{Baron2018}. The companion we found around the primary of this system is a new discovery. 
The TESS photometry reveals both a short (0.312 d) and a long (4.652 d) period. Considering the $V~\sin{i}$ of the targets \citep[69.8 and 6.9 km/s for A and B, respectively, ][]{kraus2014}, it is plausible that the two periods are both real and belong to the two components.

\medskip
\noindent{\bf HIP 19183 = HD 25953}
This is an F6 member of AB Doradus. The secondary is in Gaia DR2 without parallax and proper motion, likely due to the large $\Delta G=7.75$. SPHERE second epoch confirms that the object is bound. The estimated mass for the companion is $0.14$~ $M_\odot$. The target is in GPIES \citep{Nielsen2019} and WEIRD \citep{Baron2018},  with no special comment.
The period from the TESS periodogram (2.007 d) agrees with expectation for a member of AB Dor.

\medskip
\noindent{\bf 2MASS J05195513-0723399 = UCAC4 414-008464}
Our detection confirms that the companion detected with Astralux \citep{Janson2012} is physically bound to the primary. There is another star at 15 arcsec (2MASS J05195412-0723359), which is itself a close
visual binary, which has similar proper motion, albeit with significant differences, likely due to orbital motion. This is then a quadruple system of 4 M-type objects:
the AB pair (observed with SHINE) with a separation of 0.5 arcsec (28 au);
and the CD pair (not observed with SHINE) with a  separation 0.8 arcsec (46 au).
Spectral type M1 was reported for A component and M4 for the C component.
2MASS J05195412-0723359 was flagged as a Columba member by \citet{malo2013} and \citet{gagne2015}, while
2MASS J05195513-0723399 was flagged as Tuc-Hor member by Schlieder (priv. comm.).
With Gaia DR2 parameters of the A component, there is not
association with known groups. The C component is in Gaia DR2 but without astrometric solution, likely
because of the effect its close companion. Astrometric solution for the A component has large errors for the same reason.

We observed the primary with FEROS  \footnote{Prog. ID 090.A-9010(A), PI J. Schlieder} on February 22 and 25, 2013. We classified the star as M1 and we measured RV 27.2$\pm$1.0 km/s, and $V~\sin{i}$=4 km/s. The RV is fully consistent with that measured by \citet{schneider2019} ($27.36\pm0.41$ km/s) and slightly discrepant with respect to Gaia DR2 (24.78$\pm$0.44). Indications of high level of magnetic activity were detected in our spectra, supporting the youth of the system (age $< 300 Myr$). 
We did not detect Lithium in the spectrum of the primary, which implies an age older than about 30-40 Myr. The TESS periodogram gives a short (P=0.610 d) and a long (P=6.452 d) period. If the long period is due to the rotation of the most massive (and brightest) object in the system, the age would be as low as 38~Myr (with an uncertainty of 50\%), which is in agreement with membership to Tuc-Hor.  
The position on the color-magnitude diagram (CMD) of the primary after correction for the flux of the companion is above main sequence, slightly below but compatible within errors with the locus of Tuc-Hor members.
We then adopt an age of 50 Myr, with minimum limit at 30 Myr from Lithium and maximum at 150 Myr.
Improved astrometric parameters with forthcoming Gaia releases will allow improved placement on CMD and
kinematic analysis and then to refine the system age.
For an age of 50 Myr and the Gaia distance to the primary, the masses of the components result of
0.60, 0.15, 0.20, 0.13 $M_\odot$, respectively.
Some constraints on the orbit are obtained using the MC method (see Sect. 4.5.3 and Table~\ref{tab:orbits}).

\medskip
\noindent{\bf HIP 25434}
This is a quadruple system, formed by two pairs (HIP 25434 and HIP 25436) separated by 12.0".
The close companion to HIP 25434 is a new discovery from the present paper, HIP 25436 was shown to be an SB by \citet{desidera2015}. The star has a large PMA from Hipparcos and Gaia DR2 \citep{Kervella2019}. The system is member of the Columba association \citep{moor2013,desidera2015}, confirmed after Gaia DR2. The stellar rotation period of 1.36 d obtained from TESS data agrees with membership in Columba association, it if belongs to HIP 25434.

\medskip
\noindent{\bf TYC 7059-1111-1 = UX Col}
The binarity of the object was originally discovered in 1930 (WDS J05290-3328AB), but in WDS there are no report of observation after 1942.
Beside our recent SPHERE observations, both components are in Gaia DR2 with similar parallax, some differences in proper motion likely due to orbital motion, and small magnitude difference ($\Delta G = 0.14$~mag). The larger magnitude difference in the NIR suggests the possibility of an additional component around the primary.

The star was flagged as an AB Dor MG member by \citet{torres2008}.
Our kinematic analysis yields probabilities larger than 95\% for the individual values of each component, adopting Gaia DR2 results. The indirect  age indicators are also compatible with membership. TESS shows a large variability (PTV $\sim$0.07 mag). The periodogram shows two strong peaks (at 0.6947 and 0.9712 d), plus a third one at 0.347 d, which is half the period of the highest peak. Periods similar to these ones were reported by \citet{messina2011} (0.694 days) and \citet{kiraga2012} (0.970 days). It is likely that these two periods are both real and correspond to the rotation periods of two components in the system. 
Large differences in proper motion between available measurement are reported.
There is also a quite large difference between the Gaia and SPHERE astrometry; the system was also observed optically in the thirties, at a much smaller separation than observed by Gaia and SPHERE (0.3 arcsec with respect to 1 arcsec at present epoch). Attempts to derive orbits adopting the masses derived from the photometric analysis suggest an orbit seen almost edge-on ($i\sim 89.95$ and $\Omega=268.45$) with a small or moderate eccentricity ($e<0.4$). The semi-major axis and period cannot be determined from existing observations alone; the only constrain is that semi-major axis is $>1.7$~arcsec and the period is $>900$~yr. 

An additional object with similar parallax and proper motion was identified by Gaia DR2: Gaia DR2 4823262708795280384 = 2MASS J05284233-3326596     at 193" $\sim$ 12000 au projected separation. It was flagged as a possible AB Dor member by \citet{gagne2018}, with an expected M3 spectral type. This star is also likely the counterpart of the X-ray source 1RXS J052843.5-332646 (nominal separation 21", well within ROSAT positional errors). The system is then at least triple.

\medskip
\noindent{\bf HIP 26369 = UY Pic B = CD-48 1893}
This is a triple system formed by the isolated K0 star HIP 26373 = UY Pic at 18" from the newly discovered
close binary HIP 26369 = UY Pic B. The system is a bona fide member of the AB Dor MG. There is a 2.2 km/s RV difference between Gaia DR2 and SACY \citep{Elliott2014}. Nothing was detected by  NICI \citep{Biller2013} and WEIRD  \citep{Baron2018}.
TESS photometry yields a period of P=4.445 d. This is very close to the measured period of
HIP 26373 by \citet{messina2010} and \citet{kiraga2012} (4.52 and 4.54 days, respectively).
It is then plausible that it belongs to HIP 26373, which is the brightest component in the system.

\medskip
\noindent{\bf HIP 28036 = HD 40216}
This is a member of Columba MG. The secondary is in Gaia DR2 but without parallax and proper motion, and $\Delta$G=5.37. It was previously detected by \citet{Galicher2016} and  \citet{Song2003}. Small RV difference between SACY (\citealt{Elliott2014}: 24.38$\pm$0.09~km/s) and Gaia DR2 (25.02$\pm$0.45~km/s). Nothing detected at large separation in WEIRD \citep{Baron2018}.
TESS photometry yields a period of P=0.948 d, consistent with the expected rotation of the primary if member of the Columba moving group.

\medskip
\noindent{\bf TYC 7079-0068-1 = CD-34 2676}
This is a new triple system, with a first component at 100 mas with $\Delta H$=5.1 and a second one at 560 mas with $\Delta H$=4.2 (physical association not confirmed but very likely). The primary is a G9V star, identified as a member of AB Dor MG by \citet{torres2008}. The membership probability results 52\%, when adopting \citep{Elliott2014} RV. The TESS light curve is not available but rotation period is available from \citet{messina2010}. Considering the fully consistent indirect age indicators, we adopt AB Dor membership and age. The components are not resolved in Gaia but there is a significant $\Delta \mu$ (3.5 $\sigma$ between Gaia DR2 and Tycho2).

\medskip
\noindent{\bf TYC 7080-0147-1 = CD-35 2749}
This star was identified as a member of AB Dor MG by \citet{Elliott2014}. The star was flagged as a possible SB in \citet{Torres2006}, from the RV difference in two epochs. The close companion detected with SPHERE (projected separation 61 mas = 4.9 au) is likely the responsible for the RV variability and the significant difference of the astrometric parameters between Gaia DR1 and DR2 and other astrometric catalogues. The kinematic analysis with the Gaia DR2 astrometric parameters yields a very low membership probability in the AB Dor MG (3.7\%), while when adopting Gaia DR1 the probability jumps to 36.8\%. Li EW is slightly larger than the mean locus of Pleiades and AB Dor MG members but well within the observed distributions. The rotation period by \citet{kiraga2012} indicates that the star is part of the sequence of fast rotators. TESS data are not available. Pending a full orbital solution and assessment of the impact of the binarity on the Gaia astrometric solution, we consider the system as a probable member of AB Dor MG and we adopt the group age.

\medskip
\noindent{\bf TYC 7627-2190-1   }
The star is classified as a member of AB Dor MG by \citet{torres2008}. However, the kinematic analysis with Gaia DR2 parameters yields 0\% membership probability.The trigonometric parallax indicates a distance significantly larger than the photometric one adopted in \citet{torres2008}, which was underestimated because of the unrecognised binarity. There are no indications of significant RV variability from sparse observation \citep[][, Gaia DR2, RAVE DR5]{Elliott2014}.
Our data indicate that this is a close system with a small luminosity difference between the two components. TESS periodogram shows two very significant short period peaks consistent with both the components being fast rotators. One of the periods is very similar to the one measured by \citet{messina2011}.
The Li EW from \citet{Torres2006} is larger than the observed values for Pleiades and AB Dor MG, suggesting a younger age, and similar to members of IC 2391/IC 2602. 
We then adopt an age of 50 Myr, with limits 30 to 100 Myr.

\medskip
\noindent{\bf UCAC2 06727592 = GSC 8544-1037}
This was flagged as a member of AB Dor MG in \citet{Torres2006}, while the membership was rejected in \citet{Elliott2014}. Our kinematic analysis also yield null membership probability.
The binary was originally discovered by \citet{Elliott2014}.
Both components are also in Gaia DR2 with $\Delta$ G=0.31. 
TESS photometry yields two periods of P=6.23 d and P=7.86 d, which are in agreement with the expected rotation periods of primary and secondary, respectively, if the system has an age close to that of AB Dor MG.
The Li Equivalent Width (EW) by \citet{Torres2006} is slightly above the mean loci of the Pleiades and AB Dor MG but compatible within the observed distributions.
Independent of the kinematic membership, we then adopted an age of 120 Myr.

\begin{figure}
\centering
\includegraphics[width=9truecm]{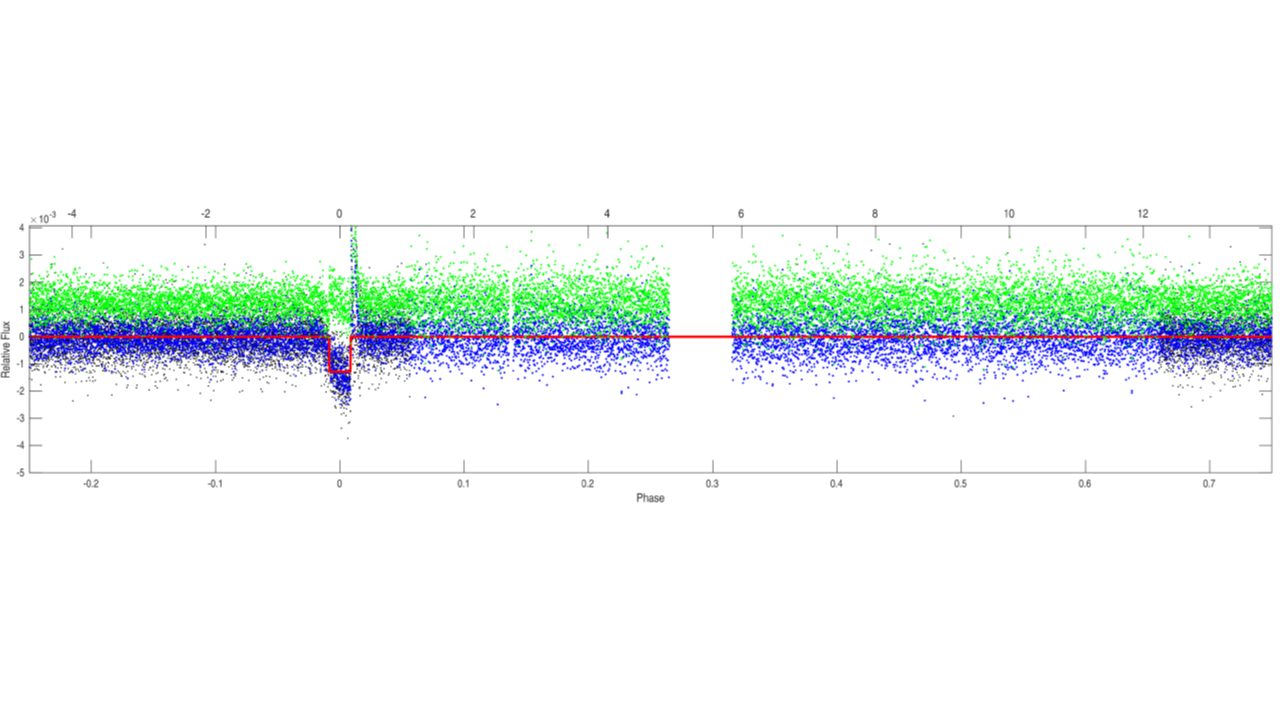}
\caption{Results from the SPOC analysis of TESS data for TYC~9493-0838-1. Black dots are the pre-whitened TESS photometry phased at a period of 18.558 d; blue dots are running median values. The red line is a trapezoidal transit model. The green points are residuals.}
\label{fig:tess_tyc_9493}
\end{figure}

\medskip
\noindent{\bf  TYC 9493-0838-1 = CD-84 80}
This is a member of AB Dor MG. The membership is confirmed by BANYAN analysis with the updated kinematic parameters. The star has constant RV according to SACY \citep{Elliott2014} and \citet{Brems2019}, with good agreement between these RV determinations.

The TESS data indicate a period of P=5.00 d, which agrees with expectation for membership to the AB Dor moving group. 
Similar periods were measured by \citet{messina2010,kiraga2012}.
In addition, the SPOC analysis of the TESS data detect a transit with S/N=12.2 and amplitude 2365~ppm (see Fig.~\ref{fig:tess_tyc_9493}). In Gaia DR2 catalogue there is no other star bright enough  ($G<16.24$) to be responsible for this signal within 30 arcsec from the primary. The stellar RV is stable within about a hundred m/s \citep{Brems2019}. The secondary has dJ=2.7 mag, sep=585 mas, and PA=135 degree, so that we cannot exclude that the transit occurs on the secondary. The in-transit vs. off-transit offset is of about $\sim$5~arcsec and PA=275~degree, which is roughly the expected accuracy for such a small transit. If the transit occurred on the secondary (that should be a star with an approximate spectral type of M2, and then a  radius of about 0.434~R$_\odot$ and a mass of 0.44~$M_\odot$) that should have a contrast of about 4 mag in visible light, it should have depth of about 94000 ppm. This requires an object of about  0.14~R$_\odot$. Given the estimated age of $149^{+31}_{-49}$~Myr, this radius would correspond to a $68^{+10}_{-16}$~M$_J$ BD or a low-mass star \citep{Baraffe2015}. On the other hand, if the transit is on the primary, the radius would be of only 0.38~R$_J$, and then a planet with a mass of the order of 5~$M_\oplus$ \citep{Linder2019}. This seems more probable than the other hypothesis, and it is clearly still consistent with a lack of detection from RVs.

\medskip
\noindent{\bf   HIP 36985 =     BD-02 2198 }
The orbit and dynamical masses coupling imaging, HARPS RV, and astrometry and comprehensive analysis of
the stellar properties will be presented in Biller et al. (in preparation). The star is considered here only for statistical purposes.

\medskip
\noindent{\bf  TYC 8911-2430-1 = CD-60 1850 = V838 Car}
This is a young object, classified as a possible member of Carina association by \citet{Elliott2016} but ruled out as
a member from our kinematic analysis. The lithium from \citet{Torres2006} is slightly larger than the mean loci
of Pleiades/AB Dor MG, and below Tuc-Hor one.
The rotation period from TESS is faster than the rotation sequence of Tuc-Hor and Argus associations, suggesting a younger age. More likely, the star is part of fast rotators subgroup. 
We adopt an age of 100 Myr with lower and upper limits of 50 and 150 Myr, respectively.
The primary is a solar-type star (mass of about 0.94 $M_\odot$), while the secondary has a mass of 0.37~$M_\odot$.

\begin{figure}
\centering
\includegraphics[width=9truecm]{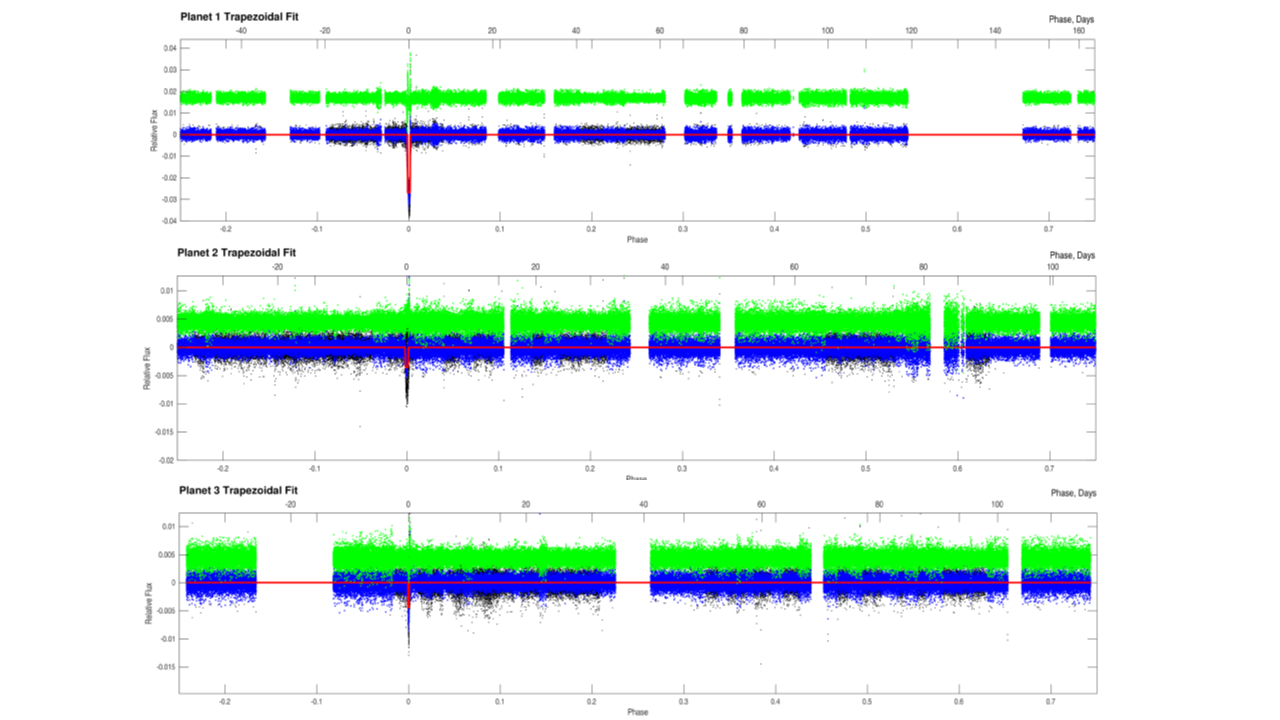}
\caption{Results from the SPOC analysis of TESS data for TYC~8911-2430-1. Black dots are the pre-whitened TESS photometry phased at a period of 18.558 d; blue dots are running median values. The red line is a trapezoidal transit model. The green points are residuals. The three panels are for the three candidates, with periods of 218.3 (planet 1, upper panel), 142.1 (planet 2, middle panel), and 155.8 days (planet 3, lower panel). }
\label{fig:tess_tyc_8911}
\end{figure}

Being close to south ecliptic pole, a very long TESS sequence is available for this binary (contrast of $\Delta J=2.71$ mag, separation of 172~mas, that is, a projected separation of $\sim$17~au, PA=172~degree, and age of $\sim$50-60~Myr). Gaia DR2 detected only a star slightly fainter then the limit requested to explain the candidate transits detected by TESS, at rather large separation of 20.8~arcsec and PA=5.2~degree. The SPOC analysis revealed three possible candidate planets (see Fig.~\ref{fig:tess_tyc_8911}). In transit-off transit offsets measured by SPOC analysis ($<4$~arcsec) are much smaller than this separation, are in a completely different direction, and are within the errors. Given the small separation of the binary, these offsets cannot be used to attribute the transits to each of the two components. The SPOC analysis detected three good candidates with periods of 218.3, 142.1 and 155.8 days, respectively, and an additional one with very short period that however does not look of good quality. It does not seem possible to attribute all these transits to the same component. If they are due to faint companion around the primary, they may be of planetary mass, while around the secondary they would be stellar. There should be serious concern about dynamic stability of a system with two close stellar companions. If these are all real transits, it seems reasonable to attribute two of the candidates to the primary and the third to the secondary. Still for dynamical stability, it seems then more likely that the primary has two planets with periods of 218.3 and 142.1 days (close to the 3:2 resonance) and the secondary has a stellar companion with period of 155.8 days. The planets would have radii of about 1.3 and 0.7 Jupiter radii, and should then be an external giant planet (mass of about 2 Jupiter mass) and a Neptunian inner one (mass in the range 10-20~M$_\oplus$, using the models by \citet{Linder2019}. The companion of the secondary would have a radius of 0.27 R$_\odot$, which, for the age of the system, would imply a mass of $\sim 0.2$~$M_\odot$, using the models by \citet{Baraffe2015}. TYC 8911-2430-1 is then potentially very interesting from a dynamical point of view because the binary looks quite close - though of course this may be a projection effect. We cannot test this because we have only a single SPHERE epoch.

\medskip
\noindent{\bf  HIP 37918 = HD 63581}
The star is a member of Carina-Near MG \citep{zuckerman2006}. It has an additional wide component at 23 arcsec, HIP 37923. The two components are very similar each other (same spectral type, K0, $T_{\rm eff}$ difference of 60-100 K, \citealt{desidera2006a}, $V$ magnitude difference 0.10 mag). The two stars have some differences in Li EW, $V~\sin{i}$ and possibly metallicity \citep{desidera2006a,desidera2006b,Torres2006}. Both components are metal enriched with respect to the Sun.
The TESS light curve yields a period of P=6.25 d, quite consistent with membership to Carina-Near MG.

Our two SPHERE observations revealed a companion around the star, at a projected separation of 60.5 mas (2.23 au) with a magnitude difference of $\Delta J=3.8$~mag. The companion was also detected by \citet{kammerer2019} with NaCo. 
The presence of the companion can also be inferred by the high $\Delta \mu$ signature and by some RV variability. In particular, the star is a RV variable by \cite{nordstrom2004}.

While we have only three astrometric epochs (that is, fewer than the seven parameters describing the orbit), we can set interesting constraints on the orbit of HIP~37918. In fact, the quite similar position found in \citet{kammerer2019} and in the second SHINE observation (but very different from the intermediate first SHINE observation) suggests a period of a little more than 3 yr. If we couple this with the small total mass of the system ($\sim 1.24$~$M_\odot$), the semi-major axis cannot exceed by far $\sim 70$~mas and the orbit should be fairly eccentric, with both these observations obtained rather close to apoastron. This constrains the orientation of the semi-major axis and the epoch of passage at periastron. On the other hand, the first SHINE observation has a separation not far from the expected semi-major axis, suggesting that also inclination is quite strongly constrained. This is further constrained by the very small differences between the RVs measured by \citet{desidera2006a}, GAIA DR2 and the catalogue value \citep{Gontcharov2006} (data from individual epochs by \citealt{nordstrom2004} are not published), which implies that the orbit is seen almost face on ($i < 3.6$~degree). With these consideration, we used the code {\it Orbit} by \citet{Tokovinin2016b} to find the best astrometric orbit assuming that the inclination at $i=2.4$~degree to fit a tentative orbit to the observed data. This preliminary orbit has a period of $3.78\pm 0.09$~yr, a semi-major axis of $a=78.4\pm 1.3$~mas, an epoch of T$_0=2017.707\pm 0.018$, and an eccentricity of $e=0.546\pm 0.015$. On the other hand, we cannot constrain significantly $\Omega$ ($68\pm 136$~degree) and $\omega$ ($106\pm 136$~degree). With this orbit, the sum of the masses of the two components is $M_A+M_B=1.38\pm 0.14$~$M_\odot$, in good agreement with the estimate based on photometry ($M_A+M_B=1.24\pm 0.02$~$M_\odot$). In Section 4.5.2 we showed that an orbit with parameters very close to this one matches very well the PMA measured by \citet{Kervella2019} from Hipparcos and Gaia measurements. Further observations are needed to confirm these values.

\medskip
\noindent{\bf  TYC 7657-1711-1 = CD-43 3604}
This is a member of the Argus group (98.1\% probability). It is a very close binary. There are significant RV difference (7.6 km/s)  between Gaia and SACY \citep{Elliott2014} RVs, but the star is indicated as constant RV by SACY \citep{Elliott2014}. This is qualitatively in agreement with the very small separation (40 mas = 3.4 au). There are two other sources within 5 arcsec in Gaia (delta mag 8-9 in G band). We retrieved them in our SHINE images, though at low S/N. While the time elapsed between the SHINE observation and Gaia epoch is short, both astrometry and position in the colour magnitude diagram indicates that they are background objects. No data are available yet from TESS. Very short rotation period is derived by \citet{messina2011,kiraga2012}.

\medskip
\noindent{\bf  TYC 7133-2511-1 = KWW 1637}
This is a binary first identified by \citet{Janson2017}. The star is also flagged as an SB2 by \citet{kim2005}, with an estimated RV difference between the components of 40-45 km/s and by \citet{Torres2006}. The estimated magnitude difference in the $V$ band is of 0.5 mag that is smaller than the observed $\Delta H$ (0.7 mag), suggesting a triple system.
The star was classified as a member of Columba by \citet{malo2014}, with a kinematic distance of 90 pc.
Gaia DR2 yields a very small parallax (1.42$\pm$0.38), which would imply a giant star.
Neglecting it (that is, assuming that Gaia DR2 astrometry is highly biased by binarity and that the star is a true young object close to main sequence), BANYAN analysis yields null membership probability for Columba and other known associations.
This is further supported by the extreme Li EW from \citet{Torres2006} (570 m\AA), which would suggest a younger object.
There are no data yet from TESS, but the rotation period was measured by \citet{kiraga2012} on All Sky Automated Survey \citep[ASAS][]{asas} time series.
Recently, \citet{yep2020} proposed the star to be a member of a small association near Cometary Globule CG 30 in the Gum Nebula, with an age of 0.5-1 Myr at a distance of 358.1$\pm$2.2 pc. 
The differences of the observed RV and proper motion of TYC 7133-2511-1 with respect to the group values
are rather small and compatible with the multiplicity of the object. The discrepancy of the Gaia DR2 parallax
with respect to the group distance becomes also smaller although still formally significant.
We consider this membership assignment as fully compatible with all the observational constraints and we adopt it.
The star has a prominent IR excess longwards of about 5.7 $\mu m$ from WISE \citep[Wide-field Infrared Survey Explorer][]{wise}, Spitzer \citep{spitzer}, and AKARI \citep{akari}

\medskip
\noindent{\bf  TYC 6004-2114-1 = HD 67945}
This is an F0 star that was proposed as a member of the Argus association by \citet{torres2008} but rejected as a member by \citet{zuckerman2019} because of the discrepant Gaia DR2 kinematics. Our kinematic analysis confirms the null probability for being member when adopting Gaia DR2 values. However, a 99\% membership probability was derived when adopting Gaia DR1 values, which differ by about 10 mas/yr in both coordinated of proper motion and 3.2 mas in parallax (30\%). The very close separation (72 mas = 6-8 au) of the newly detected companion is the likely explanation of the discrepancy in the astrometric parameters. The target was observed by \citet{Rameau2013} but with no detection of the companion identified with SPHERE. The star was also flagged as a possible SB2 by \citet{Torres2006}. A firm evaluation of the membership and age will require an astrometric solution that includes the presence of the companion. Considering the difficulty in the dating of early F type star, we tentatively adopt Argus membership, with upper limit as resulting from isochrone fitting.
No data are available yet from TESS.

\medskip
\noindent{\bf  TYC 8577-1672-1 = CD-57 2315}
This is a bona fide member of the Argus association, with constant RV from SACY \citep{Elliott2014} and Gaia DR2. 
No data are available yet from TESS. The photometric period by \citet{kiraga2012} (52.53 d) is not compatible with the available stellar properties.

\medskip
\noindent{\bf  TYC 8582-1705-1 = CD-52 2706}
This is a young star, not previously classified as a member of any known group. The kinematic analysis yields a 65.4\% membership probability in Carina-Near MG.
The TESS light curve yields a period of P=3.03 d, very close to the previously reported by \citet{kiraga2012}.
The Li EW \citep{Torres2006} and X-ray emission are close to the typical ones for Pleiades of similar colours. The rotation period is instead faster, being intermediate between the slow and fast rotators sequences in the Pleiades OC and similar to that of Tuc-Hor members if belonging to the sequence of slow rotators. Therefore, there is some tension between rotation (indicating an age of about 50 Myr of the star is on slow sequence) and lithium, which is only  marginally compatible with such an age and indicative of an age close the Pleiades. The possibility that the object is evolving from the fast rotators sequence might reconcile these indicators. This latter possibility may also be compatible with Carina-Near membership, although Li would be quite high in this case. Considering these results, we adopted an age of 120 Myr with limits from 40 to 300 Myr (this upper limit was adopted to allow possible Carina-Near membership).
The companion found with SPHERE is a new discovery. It is a very low-mass star (0.18 $M_\odot$ for the nominal age) at 70 mas (4 au) projected separation. Some constraints on the orbit are obtained using the MC method (see Sect. 4.5.3 and Table~\ref{tab:orbits}).

\medskip
\noindent{\bf  TYC 8944-1516-1 = CPD-62 1197}
This is a K star, identified as a member of the Argus association by \citet{torres2008} and confirmed in our
kinematic analysis.
No data are available from TESS but a fast rotation period (0.55 d) was derived by \citet{kiraga2012}
while \citet{messina2011} found a period of 1.26 d.
The first one is compatible with the large v sin i of the object \citep[84 km/s][]{Torres2006}.
The rotation and the other age indicators are compatible with Argus membership and age.
The companions detected in this work is a new discovery. Some constraints on the orbit are obtained using the MC method (see Sect. 4.5.3 and Table~\ref{tab:orbits}).

\medskip
\noindent{\bf  GSC 08584-01898}
The close companion (separation 264 mas) was previously detected by \citet{Shan2017}. The star was flagged as a member of Carina MG by \citet{moor2013}. Gaia DR2 and \citet{moor2013} show a marginally significant RV difference ($16.53\pm2.87$ and $22.6\pm0.8$, respectively). Independent of the adopted RV, BANYAN analysis classifies it as a field object when adopting Gaia astrometric parameters. The Li EW from \citet{moor2013} (315 m\AA) clearly indicates a young age (between beta Pic MG to Tuc-Hor/COL/Car MG; being consistent with both groups). The position on the CMD is also above the main sequence and consistent with the sequence of Tuc-Hor members. The young age is further supported by the X-ray emission detected by ROSAT. No data are available yet from TESS.
There are two other sources within 5 arcsec in Gaia DR2, both  background objects.

Since only a small fraction of the orbit has been covered since first detection, we can only derive quite loose constraints on the orbit. Given the current apparent separation, the semi-major axis is likely larger than 40 au (0.31 arcsec); also it is probably smaller than 100 au (0.75 arcsec), unless we consider the improbable circumstance of having observed the companion just when being projected very close to the primary in a very wide orbit. If in addition we assume the total mass of the binary as given by photometry, then a family of best solutions is found with all quantities correlated with each other. In particular, the period is in the range $270<P<1130$~yr, the eccentricity in the range $0.47<e<0.64$, the inclination is $105<i<113$~degree, $104<\Omega<170$~degree, $-98<\omega<-53$, $1740<T_0<1800$, and $21<V_0<25$~km/s. All these solutions provide a similar $\chi^2$ value when using {\it Orbit}, so they are equally acceptable. We notice that the selected period range is only a range of most probable values, and we cannot exclude longer periods.

\medskip
\noindent{\bf  TYC 4895-1137-1}
This is included in the target list as a possible member of Tuc-Hor (Schlieder, priv. comm.).
However, the kinematic analysis with Gaia parameters rejects the membership. Bright X-ray emission and a rotation period of 7.745 days have been reported by \citet{kiraga2012}, while TESS data are not yet available. We measured a Li EW of 92.8$\pm$3.4 m\AA\ on a FEROS  spectrum  \footnote{Prog. ID 090.A-9010(A), PI J. Schlieder}. Both the lithium and rotation period are close to the sequences of Pleiades members suggesting a similar or slightly older age.
The new companion is very bright ($\Delta J=0.49$ mag at 280 mas = 17.1 au projected separation).

\medskip
\noindent{\bf  TYC 8182-1315-1}
This star was identified as a member of AB Dor by \citet{torres2008}.
However, BANYAN analysis returns a very small membership probability (1\%) for this group.
The Li EW is compatible with Pleiades and AB Dor age (slighty below the mean value).
TESS photometry indicates a period of 3.126 d, somewhat short but not incompatible with membership to AB Dor. When coupled with the observed $V~\sin{i}$ \citep[6 km/s, ][]{Torres2006} this would
indicate a fairly pole-on observations. Alternatively, the true rotation period may be two times the observed one because of the configuration of the active regions. 
X-ray emission and $\log{R_{HK}}$ (measured by us on FEROS spectrum  \footnote{Prog. ID 083.A-9003(A), PI R. Launhardt} as in \citealt{desidera2015}, $\log{R_{HK}}$=-4.44) are slightly below the mean locus of Pleiades and AB Dor MG but well within the observed distributions.
Considering these ambiguities, we adopt an age of 150 Myr with limits 100 - 200 Myr.

\medskip
\noindent{\bf  HIP 49767 = HD 88201}
This is a field star. The TESS photometry indicates a period of P=5.00 d, close to the rotational sequences of the Hyades and Praesepe.
The Lithium EW we measured on FEROS spectrum  \footnote{Prog. ID 083.A-9003(A), PI R. Launhardt} (77.0$\pm$1.7 m\AA) is just smaller than that of the Hyades members of similar colours, while $\log{R_{HK}}$ and X-ray emission are slightly above Hyades loci.
The companion detected by SPHERE is thus a very low-mass star (0.11~$M_\odot$). 
There are good perspective for dynamical mass determination, as Hipparcos detected an astrometric acceleration and the star is flagged as RV variable by \citet{nordstrom2004} (see also the large PMA between Hipparcos and Gaia DR2 measured by \citealt{Kervella2019}).

\medskip
\noindent{\bf  TYC 7191-0707-1  HD 89326 = CI Ant}
This is the visual companion and was already known (Tycho, WDS).
The very short period (0.32d) photometric modulation seen in ASAS time series was interpreted as a contact eclipsing binary by \citet{bernhard2011,kiraga2012}, indicating a triple system. 
There are no data yet from TESS.
Kinematic analysis indicates a field object. 
Lithium EW \citep{Torres2006} is intermediate between Hyades and Pleiades.
                                                                
Existing data allow significant constraints to be put on the orbit of TYC 7191-0707-1. We can use the positions measured at three epochs over about 90 years. The intermediate measure (from the Tycho binaries catalogue) gives the widest separation. In addition, the three measures have very similar PA with variations of about 10\%  in separation. These facts essentially fix $\Omega$ and constrain $a$ (depending on $e$) and indicate that the orbit is seen at high inclination. If we now assume the masses given by the photometry and $i=82$ degree, we can find the best solution that has $a=1.4\pm 0.3$~arcsec (that is $a=175\pm 35$~au), $P=1600\pm 500$~yr, $T_0=1920\pm 380$~yr, $e=0.25\pm 0.10$, $\Omega=123.1\pm 0.3$~degree, and $\omega=160\pm 20$~degree.

\medskip
\noindent{\bf  HIP 54477 = HD 96819 = 10 Crt = TWA 43 = HR 4334}
This star was originally proposed as a possible member of the TWA association  by \citet{mamajek2005}. It was classified as a confirmed member and labelled as TWA 43 by \citet{gagne2017}. The updated kinematic analysis including Gaia DR2 astrometry fully supports this classification (membership probability 99.5\%). It was observed in several direct imaging surveys \citep{DeRosa2014,meshkat2015,Nielsen2019} without mention of the close companion identified with SPHERE.
The RV monitoring by \citet{lagrange2009} is not expected to have been sensitive to the new companion due to the short time baseline and the large RV errors due the high $V~\sin{i}$ of the star. The star has a $\Delta \mu$ in right ascension significant at  $>5 \sigma$ between Gaia DR2 and Hipparcos and $>4 \sigma$ between Gaia DR2 and Tycho2 (see also \citealt{Kervella2019}).

\medskip
\noindent{\bf  HIP 55334 = HD 98660}
This is an F2 star and a probable member of LCC (membership probability of 66.6\%). The star is moderately close on the sky (2.78 deg) and with similar kinematic parameter to the planet host HD 95086 \citep[see discussion in][]{Desidera2021}. It may be slightly older than the bulk of LCC, as also resulting from isochrone fitting.
Considering the somewhat ambiguous membership probability, we adopt LCC age with an upper limit from the isochrone fitting (pre-MS fit).
The companion at 133 mas is a new discovery. A background object at 3.3" was identified by \citet{Janson2013}. \citet{Kervella2019} measured a large PMA between Hipparcos and Gaia DR2. The motion and mass ratio of the companion we found makes it fully compatible with being responsible for the acceleration observed by \citet{Kervella2019}.

We tried to constrain the orbit of this system using our data. Unfortunately, only a limited portion of the orbit is available. The method we used, first derivation of best possible orbit and then application of the PMA data on the family of best orbits, is not adequate for this system, mainly because the fraction of the orbit covered is too limited. The only solid conclusion that we may draw with this approach is that the inclination at which we see the orbit is large (in the range 75-86 degree) and that PMA data do not agree well with highly eccentric orbit.

\medskip
\noindent{\bf  HIP 56128 = CD-33 7779 = TYC 7223-227-1}
The companion detected with SPHERE at 211 mas (7 au) is likely the responsible for the observed astrometric features (astrometric acceleration in Hipparcos, large Gaia/HIP/Tycho $\Delta \mu$, large RUWE in Gaia). Three RV measurements (RAVE, Gaia, and our own using FEROS \footnote{Prog. ID 083.A-9003(A), PI R. Launhardt}) are consistent within errors.
The object is not associated with known moving groups. 
The TESS photometry gives a period of 11.11 days, close to the rotational sequences
for the Hyades and Praesepe open clusters. 
The much shorter period (P=1.03 d) reported by \citet{oelkers2018} is likely an alias. 
The star has no lithium, indicating an age older than 300 Myr, and was not detected by ROSAT.
The $\log{R_{HK}}$ (re-derived from S index using B-V from \citealt{nascimbeni2016}, as the B-V from Hipparcos adopted in \citealt{gray2006} is highly discrepant
with respect to other photometric measurements and the spectral classification) is -4.71, lower than the locus of the Hyades.
We adopt an age of 600 Myr with limits of 300 and 2000 Myr.

\medskip
\noindent{\bf  HIP 56963 = HD 101523}
This is an A3V star, classified as an LCC member by \citet{dezeeuw1999} and \citet{rizzuto2011}. The kinematic analysis without including the RV (as the RV in SIMBAD is the astrometric one from \citealt{madsen2002}) yields a 51.5\% membership probability for LCC, 0.8\% for UCL, and 47.7\% for a field object. A slightly larger membership probability in LCC was found for Hipparcos astrometry (57.2\%). The star shows significant $\Delta \mu$ between various catalogues, most likely due to the companion at 0.2" and supporting its physical association. The companion detected in our study remained undetected by \citet{kouwenhoven2005}. Considering the somewhat ambiguous membership, we adopt the LCC age, but with the upper limit derived by isochrone fitting as for an isolate field object.

\medskip
\noindent{\bf  HIP 58859 = HD 104839} 
This B9 star is a confirmed member of LCC. It is a new close binary (separation 87 mas) from our observations.

\medskip
\noindent{\bf TWA 24 = CD-58 4411} 
TWA 24 was included among the TWA members by \citet{Song2003}.
The membership was however rejected by \citet{mamajek2005} and other studies, because of the distance larger than 100 pc, putting the star in the LCC association, in the background of TWA. The Gaia trigonometric parallax confirms this assessment and our kinematic analysis yields a membership probability to LCC of 99.9\% probability. Both components of the TWA 24 system are in Gaia DR2, confirming the physical association. The TESS periodogram yields two significant periods of 2.198 and 5.41 d, respectively; they are consistent with membership to the LCC association if they are interpreted as due to the rotation of the primary and secondary, respectively. \citet{messina2010} provided a shorter period of 0.68 days. The 2.198d period from TESS is compatible with the
observed $V~\sin{i}$ of the primary (17.1 km/s), the 5.41 is too long, while 0.68 would imply an unlikely orientation very close to pole-on.

Available NaCo archive imaging already allowed us to classify TWA24B as a physical companion before Gaia. The individual spectra gathered by \citet{mentuch2008}, with very strong Lithium line in both components, represent an additional confirmation of the physical link of the system. The RV difference between the components in Gaia DR2 (8.0$\pm$3.9 km/s) is only marginally significant due to the large errors. Archive HARPS RVs
 \footnote{Prog. ID 074.C-0037(A), 076.C-0010(A), 077.C-0012(A), 079.C-0046(A), PI E. Gunther} allow us to exclude the presence of close stellar companions around the primary.
An additional co-moving object, likely a physical companion, is identified in Gaia Dr2 at wider separation (12.8"). It results a very low-mass star from G band photometry. 
To our knowledge, it was not previously noticed.

\begin{figure}
\centering
\includegraphics[width=9truecm]{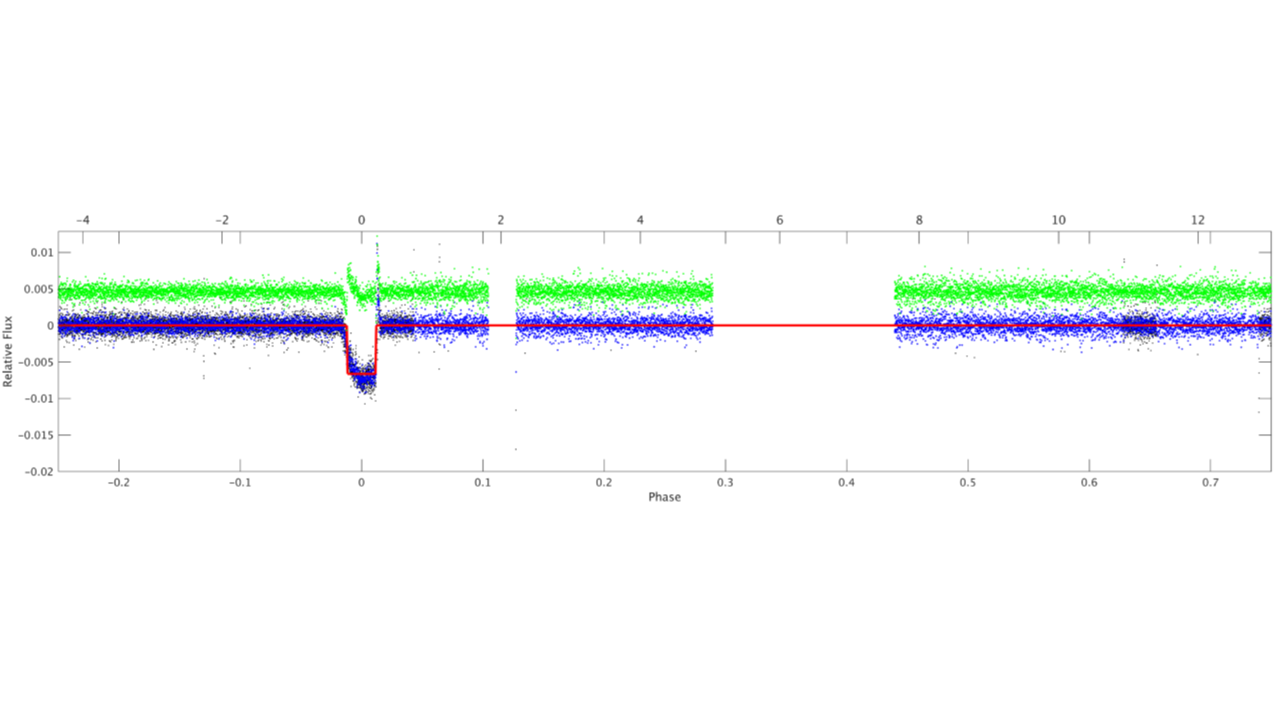}
\caption{ Results from the SPOC analysis of TESS data for TWA-24. Black dots are the pre-whitened TESS photometry phased at a period of 17.403 d; blue dots are running median values. The red line is a trapezoidal transit model. The green points are residuals.}
\label{fig:tess_twa-24}
\end{figure}

SPOC analysis of TESS data detected two possible transit candidates. However, in the following we only consider the one with a period P=17.4 d and a depth of 5362 ppm (see Fig.~\ref{fig:tess_twa-24}), because an exam of the light curve of the other one (P=10.672) shows that this has not the shape expected for a transiting object unless it is very marginal. In Gaia DR2 catalogue there are not other stars bright enough to be responsible for this transit but the two components of the binary. While the separation of the binary is not negligible, the in-transit vs. out-transit offset is very small and in a direction completely different from the secondary. We tentatively assume that the transit is on the primary. Given the stellar radius of the star (R=1.36~R$_\odot$ using COND isochrones by \citealt{Baraffe2015}), the radius of the transiting object is 0.94~R$_J$, which, given the very young age of the system, is due to a planetary object with a mass of $\sim$20-25 M$_\oplus$ (using models by \citealt{Linder2019}), that is, a Neptune-like one. Further analysis and data would be highly welcomed.

\medskip
\noindent{\bf HIP 59603 = HD 106218 }
The kinematic analysis yields a probability of 99.9\% for LCC, adopting \cite{chen2011} RVs. There is a marginal RV difference between Gaia DR2 and \citet{chen2011}. The newly detected companion is at very close separation (87.5 mas).

\medskip
\noindent{\bf HIP 60577 = HD 108035}
This is a member of LCC, as also confirmed by our kinematic analysis that included Gaia DR2 data (membership probability 99.2\%). TESS periodogram gives both short and long periods; if we interpret the long period (P=2.151 d) as due to rotation of the secondary, this is consistent with membership to LCC. 
It is flagged as binary in WDS because of the presence of a companion detected in observation performed between 1929 and 1944. The companion was instead not detected in Hipparcos, while it was detected by Gaia, with  $\Delta $G =3.59 and astrometric parameters compatible with a co-moving object. The secondary is also detected in SPHERE images and resolved for the first time into a tight equal mass pair (separation=108 mas, PA=203.4~degree, $\Delta K=0$) making the system triple. Since Gaia did not resolve the secondary, the position should refer to the photocentre of the system in the Gaia band pass. The system shows prominent X-ray emission (source 1RXS J122500.3-501121), likely originating from the late-type components.

\medskip
\noindent{\bf HIP 61087}
This is an F6 star and a confirmed member of LCC (probability 98.8\%).
The companion detected with SPHERE is a new discovery; it was not detected in previous observations by \citet{Janson2013}.
The RV of the star measured by \citet{chen2011} differ by 4.6$\pm$0.8 km/s with respect
to Gaia and by 5.0$\pm$1.8 km/s with respect to \citet{nordstrom2004}.
Two RV epochs separated by 3307 days are consistent within errors in this latter study.
The RV variability may be linked to the newly detected companion, which lies at a projected separation of just 0.054"=5.9 au.
The stellar rotation period of 1.539 d obtained from TESS data agrees with membership in the LCC association.
The star shows significant IR excess \citep{chen2014}.

\medskip
\noindent{\bf HIP 61796 = FH Mus}
This is a B8 star. Our kinematic analysis yields a membership probability of 98.5\% for LCC group. The companion detected with SPHERE is much closer than the object detected by \citet{kouwenhoven2005}. The star is flagged as RV constant in \citet{chini2012} (4 epochs). The star is classified as ellipsoidal variable in SIMBAD, but this classification was rejected by \citet{morris1985}.

\medskip
\noindent{\bf HIP 62171 = HD 110697 }
This is a confirmed LCC member (probability 99.5\% adopting \citealt{chen2011} RV). No significant RV difference between Gaia DR2 and \citet{chen2011}. New companion detected with SPHERE at 130 mas = 16 au. There is also a background object at 3" previously detected by \citet{Janson2013}.
The TESS light curve gives a period of 2.326 d; this agrees reasonably well with the expected rotational period of the secondary if the system is a member of LCC.

\medskip
\noindent{\bf HIP 62428 = HD 111102 }
This is a high probability LCC member  (99.6\% from our analysis, adopting \citealt{chen2011} RV). A very close companion (35 mas = 4 au separation) was identified for the first time with SPHERE. The spectral energy distribution was fitted by two belt components by \citet{chen2014}, one at 0.8 au and the second one at 100au. The first of these belts, if real, is then circumstellar (around one of the components), while the second one is likely circumbinary. This is an interesting configuration that suggests that this binary (with a mass ratio of $q\sim 0.6$) formed by disk fragmentation. 

\medskip
\noindent{\bf HIP 63041 =       HD 112109 }
This is a member of LCC according to \citet{rizzuto2011}. No data are available from TESS yet; however, given the F0 spectral type it is unlikely that TESS data may reveal rotation.
The target is at very low galactic latitude and projected towards a very rich field. A close companion candidate with a contrast of $\Delta J=10.13$~mag at a separation $\sim 0.285$~arcsec and PA$\sim$300 degree is a background object from the proper motion test.
A wider companion at 8.76 arcsec and $\Delta G=5.57$~mag, with very similar parallax and proper motion, is listed in Gaia DR2, but with no previous citation in the literature. Given its magnitude and colour, this outer companion is a late M star with a mass of $\sim 0.13~M_\odot$ (using isochrones by \citealt{Baraffe2015}).

We detect a new much closer low-mass companion ($M=0.17^{+0.01}_{-0.02}~M_\odot$ using isochrones by \citealt{Baraffe2015}) at 0.053 arcsec (that is 5.3 au). This secondary is only observed in the second of our two visits, because of the small rotation angle during the first one, preventing detection at this very small separation.
The star has a large PMA from comparison of Hipparcos and Gaia DR2 \citep{Kervella2019}. The PMA for Gaia DR2 and Hipparcos epochs are similar with each other, suggesting a period that is not far from 24 yr or a submultiple of it. It is unlikely that the wide companion seen by Gaia may be responsible for these features.  The close companion may well explain the PMA;  given the projected separation, the period is $>1300$~d, most likely $\sim 10$~years. This is fully compatible with Gaia DR2 and Hipparcos PMA.

\medskip
\noindent{\bf HIP 64322 = HD 114319 }
A bright companion at 2.3" is identified both in our SPHERE observations and in Gaia. 
The target was also observed by \cite{Janson2013} with the Near-Infrared Coronagraphic Imager mounted at the Gemini South Telescope \citep[NICI@GeminiS][]{nici}. It is not listed as a binary but they report in the notes that the star was observed in very poor conditions and may have a bright companion at a separation of 2.3 arcsec and position angle 170 deg, then corresponding to the one identified by SPHERE and Gaia.

The star is identified as a ScoCen member by \citet{dezeeuw1999,rizzuto2011,Pecaut2012}. BANYAN analysis yields instead a low membership probability, indicating a field object. However, the star is identified as a spectroscopic binary with period of a few years (Grandjean et al., private communication), likely causing alteration of both the RV and the astrometric parameters (the star has a large PMA from \citet{Kervella2019}; note however that the variation in RVs and the PMA are not due to the wide companion observed by us and by Gaia; hence the system is triple). Therefore, we consider membership as possible, pending the evaluation of the impact of the companion on the astrometric parameters and centre of mass velocity. 
Analysis of the TESS light curve reveals several periods, most of them being aliases of one at P=3.175 d. If this is the period of the secondary, it would confirm an age compatible with membership to LCC. The slightly off-sequence position on CMD is also compatible with the Sco-Cen age.

Conservatively, we adopt the age as a field object from isochrone fitting, with lower limit at the LCC age. 

\medskip
\noindent{\bf HIP 65219 =       HD 116038 }
These are ScoCen members, as determined in various works. Our kinematic analysis (without RV) yields 43.0\% and 43.4\% membership probability for LCC and UCL, respectively. This ambiguity is not critical for the age assignment, considering the very similar ages of the two groups. The companion identified with SPHERE at 67 mas was not previously detected \citep{kouwenhoven2005} and is likely responsible for the astrometric excess noise seen in Gaia (RUWE=4.00) and the large Gaia DR2 - Hipparcos $\Delta \mu$.
In addition to the two SHINE observations, HIP~65219 was also observed with SPHERE within programme 0103.C-0628 (PI Wagner). Since this dataset is now public and available in the archive, we used it to provide a further epoch for this target. This allowed us to cover about 60 degree in PA and to set some constraints on the orbit once we assumed the sum of the mass of the two components as given by photometry. We found that the semi-major axis is in the range 53 - 110 mas (from 6.8 to 14.1 au), the period in the range between 9.8 and 29 years, $T0$ is between 1999 and 2011, eccentricity between 0.25 and 0.51, $\Omega$ between 139 and 169 degrees, $\omega$ between 52 and 184 degrees, and the inclination between 0 and 49.6 degrees. There is a high degree of covariance between the different parameters. In Section 4.5.2 we discuss coupling of this information about the orbit with the Hipparcos-Gaia PMA found by \citet{Kervella2019} that suggests semi-major axis of 86~mas and a period of 20.3~yr, roughly in the middle of the ranges mentioned above, while the MC method described in Section 4.5.3 yields a shorter period of 14.5$^{+7.4}_{-3.8}$~yr.

\medskip
\noindent{\bf HIP 66908 = HD 119221}
Kinematic analysis (without RV) yields a membership probability of 71.6\% and 26.3\% for UCL and LCC, respectively. There are no previous detections of the companion seen in SPHERE images \citep{kouwenhoven2005}.

\medskip
\noindent{\bf HIP 67036 = V827 Cen = HD 119419}
Chemically peculiar star \citep[][and references therein]{rusomarov2018}, classified as a member of LCC in the literature. Our kinematic analysis yields 61.8\% and 38.0\% membership probability for LCC and UCL, respectively. The binarity was not previously known \citep{kouwenhoven2005}.
In addition to the SHINE observation, there is a second observation with SPHERE available in the ESO archive for this target (programme 095.C-0389, PI Apai). We downloaded this dataset and reduced it as done for the SHINE data.
In the SPHERE datasets, the secondary is itself partially resolved into a close binary with a small difference in luminosity between the two  components. We resolved it using a best fit procedure that uses the PSF obtained from the primary observation. We found that the projected separation between the two components is 29 mas (=3.7 au). While the two epochs are similar (only 76 days), there might be some orbital motion that is, however, detected at only slightly more than 1$\sigma$.
The TESS light curve gives a period of 2.632 d; this agrees reasonably well with the expected rotational period of one of the components of the secondary if the system is a member of LCC.

\medskip
\noindent{\bf HIP 70350 = HD 125912}
This is an F7 star in UCL. The membership is confirmed by our analysis (probability 99.1\%). The star is in Gaia DR2 but without astrometric solution. An extremely large Renormalised Unit Weight Error (RUWE) is derived (59.21). These facts are likely linked to the companion seen in SPHERE observations, which was not previously known. There is no significant RV variability from available measurements in the literature \citep{nordstrom2004,chen2011}. A photometric period of 0.81 days is reported by \citet{oelkers2018} and the system is detected in X-ray with ROSAT and XMM; however, the periodogram of TESS data gives no power at this period, the stronger peak being at 4.17 days. Since the system is nearly equal mass, it is not clear that the TESS photometric period may be attributed to any of the two observed components. However, this is longer than expected.

\medskip
\noindent{\bf HIP 70697 = HD 126561}
This is a confirmed member of UCL (membership probability of 99.7\% in our analysis, without including the kinematic RV by \citet{madsen2002}, which is not a true measurement). The star is listed in WDS (WDS 14276-4613) with a link to the B9 star HIP 70703 at 73", which, however, is not physically associated from Gaia DR2 astrometry. The companion seen in SPHERE images is a new discovery. It was not previously detected in the observations \citet{kouwenhoven2005}.

\medskip
\noindent{\bf HIP 70833 = HD 126838}
This target was originally selected as a ScoCen target.
The companion at 2.93" is physical, as resulting from Gaia DR2 parallax and proper motion of the individual components and the coupling with the additional imaging data from SPHERE. The Gaia DR2 parallax of 4.32$\pm$0.13 mas indicates a larger distance than the typical ScoCen targets. This value is very discrepant with respect to the Hipparcos parallax (8.27$\pm$1.26 mas). The Gaia parallax of the companion is $4.90\pm 0.32$~mas, similar to that of the primary. \citet{chen2011} spectroscopically observed both components and determined the primary and secondary to be of spectral types F3V and K3IV, respectively. This study also reveals a large and highly significant RV difference (-35.2$\pm$0.4 km/s and 6.6$\pm$0.3 km/s, for A and B, respectively). It then results that at least one of the component is itself a spectroscopic binary.

Adopting Gaia DR2 parallax, it results that the secondary is close to the main sequence \citep[][reference sequence]{pecaut2013}, while the primary is over-luminous by about 1.15 mag. The spectroscopic component may, at least partially, contribute to this excess luminosity. If such a contribution is negligible, an isochronal age of 1.6$\pm$0.1 Gyr is obtained (using PARAM + Gaia $T_{\rm eff}$). This represents the upper limit to the stellar age. On the other hand, from \citet{chen2011}, the secondary show a small amount of lithium (EW=13 m\AA). This would indicate an age of 200-400 Myr. 

HIP70833 spectral type locates it in the region where there is strong change of the rotational velocity with spectral type, related to the onset of the outer convective region. Main sequence F3 stars are expected to be slow rotators with $v~\sin{i}\sim 20$~km/s \citep{Noci1985}. This value is confirmed by a query to the $v~\sin{i}$ catalogue by \citet{Zorec2012}, which gives an average value of 23~km/s with an rms of 11~km/s for stars in the temperature range of HIP70833. On the other hand, stars of this spectral type in clusters - with ages as large as that of the Hyades - rotate faster (up to $v~\sin{i}=80-100$~km/s: \citealt{Bernacca1974}).

We downloaded from the archive eighteen HARPS spectra\footnote{Prog. ID 098.C-0739(A), 1101.C-0557(A); PI A.-M. Lagrange} of HIP70833 covering 398 days between 2018 and 2019, and analysed them using our own code (described in \citealt{Chauvin2017}) that is suited for rapidly rotating stars such as HIP70833. This code allows RV and projected rotational velocity to be determined. We obtained $v~\sin{i}=84.7\pm 1.3$~km/s for this star. The rotational velocity is very high for a field F3 main sequence star, but it is rather normal if the star is younger or as old as the Hyades.

For what concern the RV, the mean velocity ($-37.4\pm 0.8$~km/s) is similar to that measured by \citet{chen2011}, but there is a strong linear trend with a slope of 7.2 km/s/year. The scatter around this trend is only 0.12~km/s. That is, the RV measured by Chen is within the range of those measured from HARPS spectra. The epoch of the \citet{chen2011} data is not clear from their paper, but it should be either 2009 or 2011. This indicates that HIP70833 is a spectroscopic binary with a period of some years. Given the large difference with the secondary velocity, if B is a single star then A should have a very massive companion. Alternatively, B too is a spectroscopic binary.
The spectroscopic companion is also the probable responsible for the astrometric acceleration detected by Hipparcos and the astrometric excess noise from Gaia. 
While the spectroscopic companion might contribute significantly to the integrated flux, inspection of the HARPS spectra does not reveal signatures of additional objects.

From the above considerations, we adopt an age of 300 Myr from Li EW of the secondary, with limits
200-1600 Myr (upper limit from isochrone age of the primary, assuming negligible contribution from the spectroscopic companion).

\begin{figure}
\centering
\includegraphics[width=9truecm]{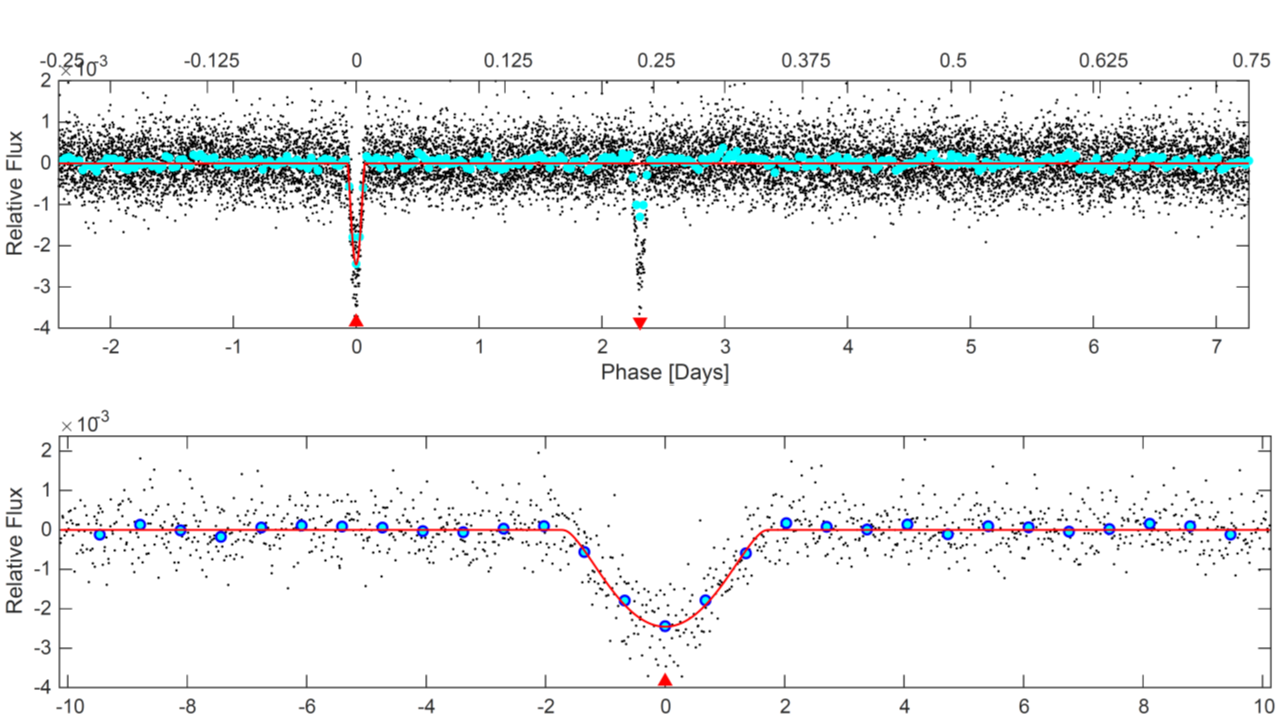}
\caption{Results from the SPOC analysis of TESS data for HIP70833. Black dots are the pre-whitened TESS photometry re-phased at a period of 9.689 d (black dots). Blue dots are running median values. The red line is a transit model. The red triangles mark the primary and secondary transits. Upper panel: full data set. Lower panel: Blow up of the primary transit.}
\label{fig:tess_hip70833}
\end{figure}

HIP 70833 is also identified as TOI-1946. TESS data clearly indicates the presence of large amplitude transits (more likely eclipses) with an amplitude of 3170 ppm (see Fig.~\ref{fig:tess_hip70833}); the SPOC analysis (that ignores that the star is a binary) yields a planetary radius of $30.4\pm 0.9$~R$_\oplus$, in the stellar regime. The in transit-off transit offset is 1.65 arcsec and PA=338 degree, clearly reminiscent of the separation and PA of the secondary (2.7 arcsec and PA=343 degree). Gaia DR2 data shows that there are not other sources within 30 arcsec bright enough to be responsible for this feature. Combined with the limited range of the primary RV on short periods, this evidence suggests that the secondary be an eclipsing binary, though both the depth and shape of the light curve suggests that in this case the eclipses are partial. Since also a secondary eclipse is clearly visible in the TESS light curve (though only once) with an amplitude not much different from that of the primary eclipse, the secondary should be itself a nearly equal mass binary; this contrasts a bit with its luminosity close to the main sequence mentioned above, but uncertainties on the colour and distance make the location on the colour-magnitude diagram a bit uncertain. If for instance the secondary is made of two objects with masses of  M$_{\rm BA}$=0.75 and M$_{\rm BB}$=0.68~$M_\odot$ (matching available data), then the amplitude of the RV curve may be as large as $\sim$50~km/s. This may well explain the RV offset between the primary and the secondary found by \citet{chen2011}. According to the SPOC analysis, the period is 10.846~d, but we cannot exclude a period of 12.238 d. In both cases, the secondary eclipse occurs far from phase 0.5, indicating that the orbit of this eclipsing binary is highly eccentric, despite the rather short period.\footnote{An alternative possibility is that the secondary is a single star and the primary is itself a triple system, made of a compact, nearly equal mass binary and of the F3 star that is seen in the optical. However, in this case in order to reproduce the offset in RV between the primary and the secondary and the observed run of the RV of the primary we should assume that the mass of the compact binary is very large (several solar masses), which requires them to be degenerate objects. We deem this possibility much less probable and in contrast with the limited astrometric signal.}

\medskip
\noindent{\bf HIP 71321 = HD 127879}
This is a confirmed member of UCL (membership probability 99.8\% excluding the RV). The new companion identified with SPHERE was not detected by \citet{kouwenhoven2005}.
The star has significant $\Delta \mu$ between Gaia and Hipparcos, and \citet{Kervella2019} found a highly significant PMA at the Gaia DR2 epoch (S/N=17.2) and a moderately significant PMA at the Hipparcos epoch (S/N=4.2). A future analysis as well as future release of the Gaia data may provide significant constraints on the orbital parameters.

\medskip
\noindent{\bf HIP 73913 = HD 133574}
This is a confirmed member of UCL (membership probability 96.6\% excluding the RV). The star was resolved as a close visual binary with separation of 88.9 mas (new detection). 
There are significant $\Delta \mu$ signatures (2.5 $\sigma$ Gaia DR2 vs. Tycho2; $>3$ sigma Gaia DR1 vs. Gaia DR2)
There is a significant RV trend from HARPS observations (Lagrange, private comm.), 
The star is included in WDS (entry WDS 15063-3524) for an object at 3.6", first observed by \citet{Janson2013} and classified as background.

\medskip
\noindent{\bf HIP 75367 = CD-40 9577}
The star is a confirmed member of UCL (98.9\%) The companion at 859 mas is a new discovery. One additional source is identified at about 5 arcsec both in SPHERE images and Gaia DR2; it is a background object. The RVs from Gaia DR2 and \cite{chen2011} are consistent within errors.

\medskip
\noindent{\bf HIP 77388 = HD 140958}
The star is a confirmed member of UCL \citep[99.7\%, from the analysis rejecting the kinematic RV by][]{madsen2002}. The companion detected with SPHERE was first identified by \citet{Janson2013} and included in WDS as WDS 15479-3816. The source is also in Gaia DR2, although without a full astrometric solution ($\Delta $ G =4.61). These data allow us to confirm the physical association
of the pair.

\medskip
\noindent{\bf HIP 77813 = HD 142113}
This is an F8 star and a probable member of US, with probability 86.8\% (field 13.1\%).
Independent of kinematics, indicators from \citet{chen2011}, the X-ray emission from ROSAT, and the rotation period from \citet{kiraga2012} confirm the young age. It was observed by \citet{Janson2013} and \citet{tokovinin2020} and reported as single. The newly detected companion is at very close separation (47.4 mas) and likely responsible for the observed large $\Delta \mu$ signature.

\begin{figure}[htbp]
    \centering
    \includegraphics[width=5.8cm,angle=90]{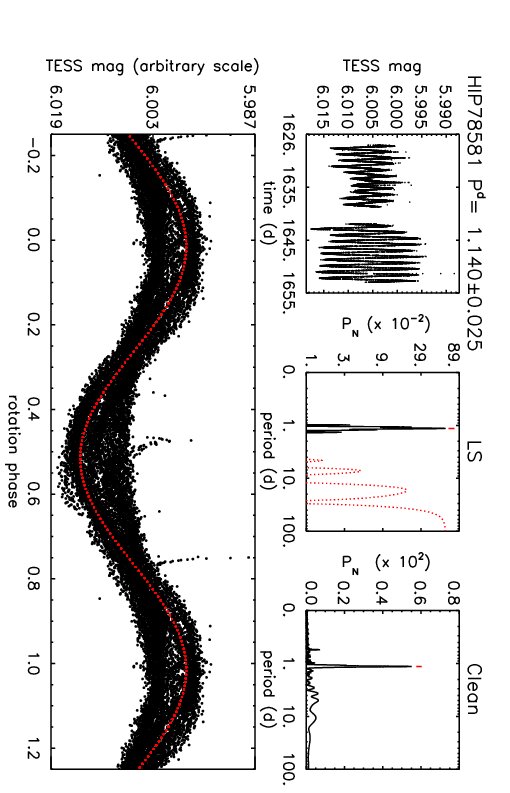}
    \caption{Photometric time sequence and periodogram for HIP 78581.}
    \label{fig:HIP78581}
\end{figure}

\medskip
\noindent{\bf HIP 78581 = HD 143637}
This is an early G star and a member of UCL (in the kinematic analysis we adopt the RV from \citealt{chen2011}, as the RV listed in SIMBAD is the astrometric one from \citealt{madsen2002}). The period indicated by TESS (P=1.137 d, a similar one was reported by \citealt{kiraga2012}) agrees with expectation for membership to UCL.
Figure\,\ref{fig:HIP78581} shows our analysis of the rotation of the star.
The star was classified as an astrometric binary by \citet{makarov2005} from Hipparcos and Tycho data and has a large PMA at the Gaia epoch \citep{Kervella2019}.
We found that this is a triple star. The outer companion (at 2.8 arcsec) was independently identified by \citet{tokovinin2020}, while the inner companion (at about 50 mas) is a new detection. Given the very long period of the outer companion, the PMA is likely due to the inner one discovered by us.

\medskip
\noindent{\bf HIP 79124 = HD 144925}
This binary star was originally discovered by \citet{kouwenhoven2005}. An additional component was later discovered by \citet{hinkley2015}, making the system triple.
The system was studied in detail by \citet{asensiotorres2019}; spectral types of M4 and M6 were assigned to the two companions. It was also observed by \citet{ruane2019}. The system is a member of US region (97.9\% membership probability).

While the period of the outer component is too long, we used {\it Orbit} to constrain the orbit of the inner binary (Aab). Assuming masses given by the photometric analysis and considering also the constraint on the PMA provided by \citet{Kervella2019}, we obtain a family of orbits yielding a good fit. The period is fixed to be longer than 50~yr, the inclination is larger than 57~degree and $\Omega$\ is between 61 and 67 degrees. The eccentricity of the orbit is more probably moderate ($<0.5$), but it is not really constrained by existing data because long period, high-eccentricity orbits fit data as well low-eccentricity orbits with  shorter period. However, such long period orbits are disfavoured because the observations should have been acquired very close to periastron, which is less probable than other circumstances, and they might be unstable, given the presence of the outer companion.

\medskip
\noindent{\bf HIP 79156 = HD 144981} 
This binary system was originally discovered by \citet{kouwenhoven2005}.
It is a member of US (99.8\% membership probability, adopting the RV from \citealt{dahm2012}, as the RV listed in SIMBAD is the astrometric one from \citealt{madsen2002}). Very recently, \citet{tokovinin2020} suggested the possible existence of a third component very close to the primary. However, this further component is not detected in our data - that are much deeper than those considered in their paper - and being below the expected limit of the SOAR data its existence is dubious.

Inspection of the Gaia DR2 catalogue reveals a wide (separation of 54 arcsec) low-mass ($0.08^{+0.1}_{-0.4}~M_\odot$) common proper motion companion (see Table~\ref{tab:known_bin}). Given its large projected separation, it is not clear if this wide companion is actually bound.
Indications of a debris disk are given by \citet{Luhman2012}.
Existing astrometric data (\citealt{kouwenhoven2005, Lafreniere2014, tokovinin2020}, and this paper) suggest a nearly circular orbit seen at low inclination with a period of about 1180 yr (see Table~\ref{tab:orbits}).

\medskip
\noindent{\bf \\ HIP 82688 = HD 152555}
This is a bona fide member of the AB Dor moving group (membership probability 95.6\%). The age indicators are fully compatible with this assignment. The lack of a long-term RV signal \citep{Grandjean2020,butler2017} is consistent with the moderately wide separation of the companion (3.795" = 172 au). The companion, which is a low-mass star of 0.19$M_\odot$, was previously identified by \citet{metchev2009,Biller2013,Brandt2014,Galicher2016} and it is also included in Gaia DR2. The differences in proper motion between the two components measured by Gaia are consistent with the orbital motion observed in high contrast imaging.

\medskip
\noindent{\bf TYC 7364-0911-1 = CD-31 13486}
This star was flagged as a possible Sco-Cen member by \citet{vianaalmeida2009}. Our kinematic analysis yields very different membership probabilities for UCL subgroup depending on the adopted parameters, as there are indication of both RV (4 km/s between \citealt{Torres2006} and Gaia DR2) and proper motion differences (2 and 5 mas/yr between Gaia DR1 and DR2). These differences are likely due to the newly detected companion at 97 mas (7.5-9.1 au), with an estimated mass 0.20 $M_\odot$. UCL membership probability is larger than 83\% for Gaia DR1 astrometric parameters and smaller than 14.7\% with Gaia DR2. The Li EW by \cite{Torres2006} is compatible with an age younger than 50 Myr.

The TESS periodogram yields a peak at 3.227 d, very close to that measured by \citet{kiraga2012} on ASAS data (3.192 d). 
This period is compatible with the age derived from lithium.
A full assessment of the kinematic properties and CMD position requires astrometric parameters that properly include the impact of the companion. We adopt the UCL age with upper limit at 50 Myr. We also note that the field of this object is rather crowded: there are 14 sources in Gaia within 10 arcsec, none being likely bound although just few of them have parallaxes and proper motions.

\begin{figure}
\centering
\includegraphics[width=9truecm]{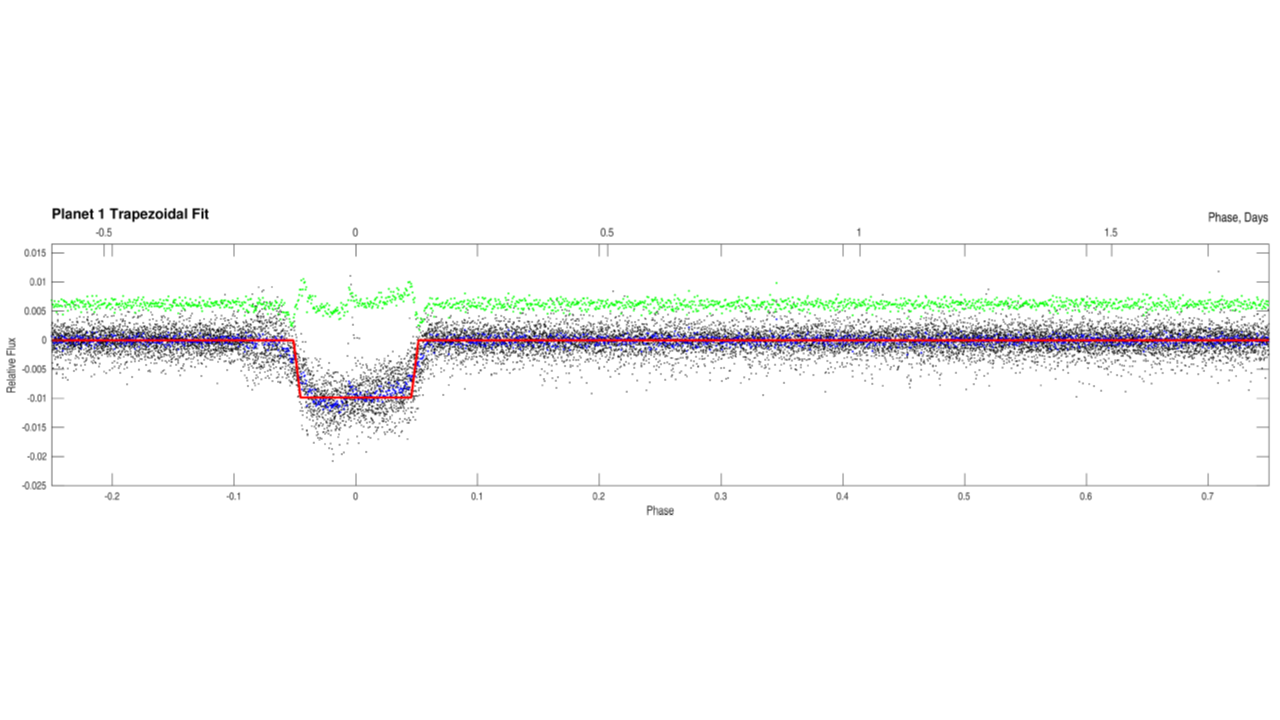}
\caption{Results from the SPOC analysis of TESS data for TYC~8332-2024-1. Black dots are the pre-whitened TESS photometry phased at a period of 2.417 d; blue dots are running median values. The red line is a trapezoidal transit model. The green points are residuals. The large offset between in transit-off transit position suggests that this feature is due to a background star and not to TYC~8332-2024-1}
\label{fig:tess_tyc-8332}
\end{figure}

\medskip
\noindent{\bf TYC 8332-2024-1}
A very close companion (separation 86 mas) was discovered with SPHERE. The star is projected on a quite crowded field (13 objects within 10 arcsec in Gaia, none confirmed to be bound).
The kinematic analysis yielded a 97.3\% membership probability in UCL, as also found by \citet{gagne2018}. Independent of this assignment, the indirect age indicators are fully consistent with a very young age, with the lithium EW yielding an upper limit at the age of $\beta$ Pic MG. The TESS period of 4.35 d is slightly longer than expected for this age, but within the uncertainties.
A different period (1.209 d) is reported by \citet{kiraga2012} but it is unlikely as it would imply pole-on orientation. We then adopt the UCL age for this system.

While there is a transit candidate (9 transits) with a rather high S/N=20.8 (see Fig.~\ref{fig:tess_tyc-8332}), the in transit-off transit offset is very large 12.3 arcsec and PA=297 degree. This offset agrees fairly well with the position of a background star in Gaia DR2 at sep=15.73 arcsec and PA=306.2 degree that has $\Delta$G=1.93 mag. We conclude that this is almost surely the object on which this transit occurs; in this case the transiting object is likely a star.

\begin{figure}
\centering
\includegraphics[width=9truecm]{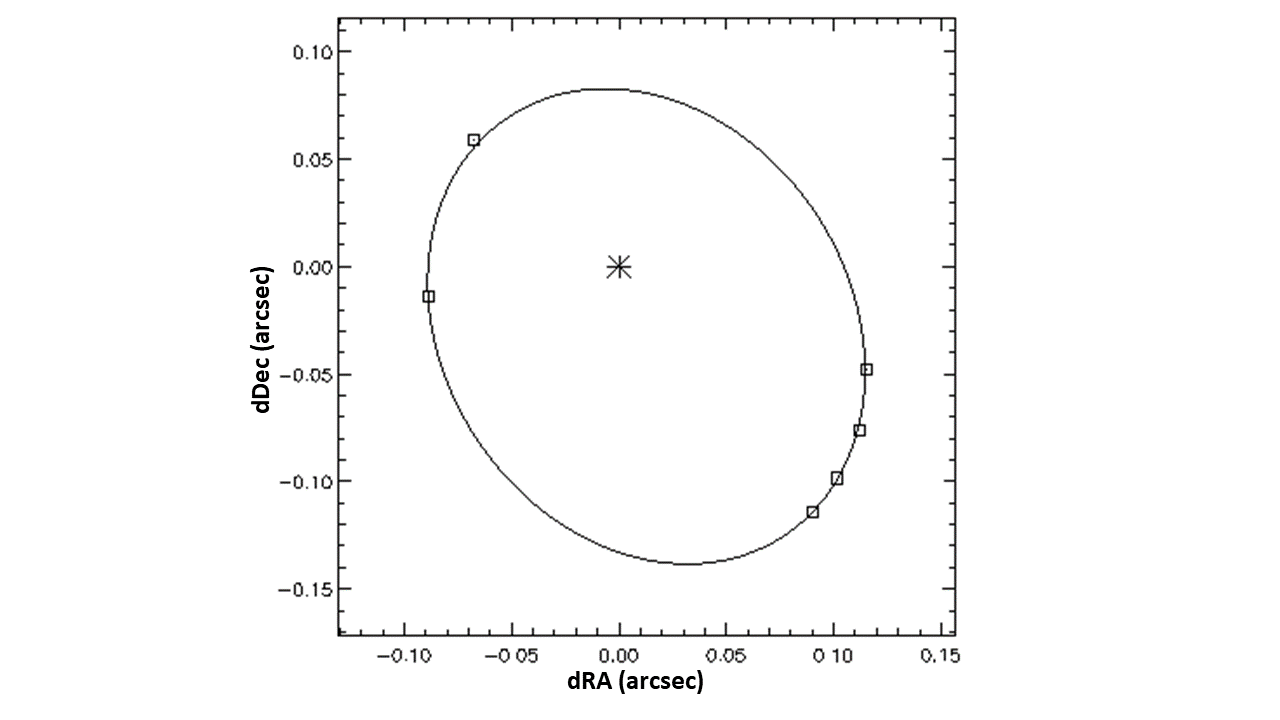}
\caption{Orbital fit for TYC~6820-0223-1. Dotted lines connect observed with predicted positions for the different observations; however, they are all shorter than the size of the symbols, so they are not evident in this figure.}
\label{fig:orbit_TYC_6820-223-1}
\end{figure}

\medskip
\noindent{\bf TYC 6820-0223-1 = CD-27 11535}
This close binary (separation 129 mas = 9-11 au) was originally discovered by \citet{Elliott2014}.
It was also detected by \citet{Tokovinin2018,Tokovinin2019}.
The components have nearly identical magnitudes so that confusion is possible between the two components in the NIR observations, while this ambiguity is less important for the speckle observations in the optical. In particular, the position of the two components is likely exchanged in  \citet{Elliott2014}.

It was classified as a member of $\beta$ Pic MG by \citet{Elliott2014} and other authors, while \citet{song2012} classified it as a Sco-Cen object.
There is a moderate range in available RV measurements, from $-1.1\pm1.8$ km/s \citep{song2012}, to $-12.3\pm3.2$ km/s (Gaia DR2), with \citet{Torres2006} and \citet{Elliott2014} determinations being  of intermediate value ($-6.4\pm1.0$ km/s and $-6.9\pm1.4$ km/s, respectively). From the magnitude difference both components should contribute to the spectrum, although the star was never classified as an SB2. The kinematic classification with BANYAN $\Sigma$ is somewhat dependent on the adopted parameters (formally significant differences in both parallax and proper motion between Gaia DR1 and DR2, beside the RV variability, but membership probability in $\beta$ Pic MG never exceeds 41\% (with Gaia DR1 astrometric parameters) and 4\% for UCL.

We then derived the age as a field object. The very large lithium EW \citep[490 m\AA,][]{Torres2006} implies an age at most as old as $\beta$ Pic MG and likely younger. The other indirect indicators such as rotation and X-ray emission are consistent with such a young age. The rotation period has been determined by \citet{kiraga2012} while there are no data  yet available from TESS. The position on the CMD is also compatible with $\beta$ Pic MG or younger ages. We thus adopted the $\beta$ Pic MG age with limits 10-30 Myr.

We obtained a nice orbital solution (reduced $\chi^2=0.74$) by combining the six available epochs (two from the SACY dataset, our own using SPHERE, and three from speckle interferometry), which covers more than half of the orbit. We fitted the orbit using the code {\it Orbit} by \citet{Tokovinin2016b} to find the best astrometric orbit. The orbital fit is shown in Fig.~\ref{fig:orbit_TYC_6820-223-1}; the parameters of this fit are listed in Table~\ref{tab:orbits}. Using these parameters, the sum of the mass of the two components is $M_A+M_B=2.14\pm 0.27$~$M_\odot$, in reasonable agreement with what we obtained from the fit of isochrones (between 1.75 and 1.82~$M_\odot$, depending on the assumed age).

\begin{figure}[htbp]
    \centering
    \includegraphics[width=5.8cm,angle=90]{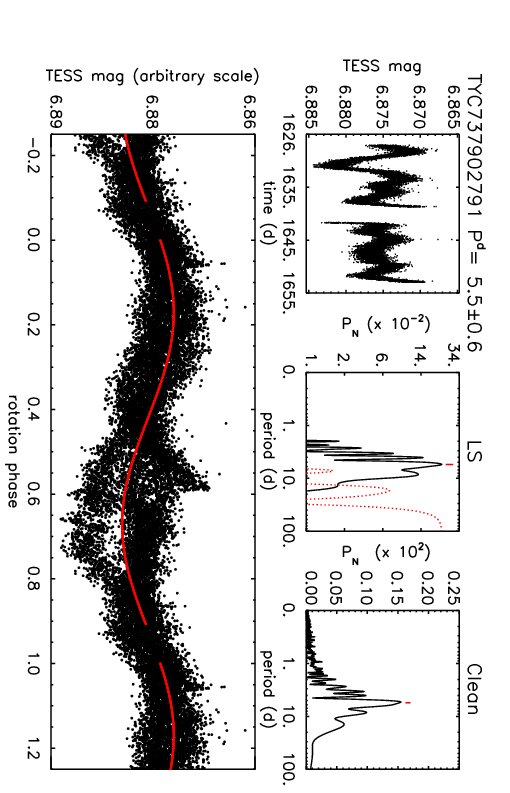}
    \caption{Photometric time sequence and periodogram for TYC 7379-0279-1.}
    \label{fig:TYC737902791}
\end{figure}

\medskip
\noindent{\bf TYC 7379-0279-1 = HD 317617}
This star was flagged as a member of AB Dor MG by \citet{torres2008} and resolved as a binary in Gaia DR2. The BANYAN kinematic analysis yields an ambiguous membership probability of 47.6\%. The age indicators are in any case compatible with AB Dor age. We measured for the first time the rotation period from the TESS photometric time series. 
(Fig.\,\ref{fig:TYC737902791}).

We found two periods (4.001~d and P=5.884) from the TESS periodogram. If we assume these are the periods for the primary and secondary, this yields ages of 88 and 231~Myr, respectively.

\medskip
\noindent{\bf HIP 87386 = HD 161935}
This is a field early-type star, with a previously undetected companion at 180 mas. This is likely responsible for $\Delta \mu$ signature. The companion is likely a K star (mass of $0.76~M_\odot$ from photometry), then contaminating significantly the photometry. Adopting the $T_{\rm eff}$ corresponding to the A9 spectral type, the isochrone age results 470$\pm$370 Myr. The lack of detection of X-ray emission from the secondary is consistent with a moderately old age.
The TESS periodogram yields a main peak at P=0.0388 (likely due to primary pulsation) and an additional long period at P=4.879. If this is interpreted as the rotation period of the secondary, this would yield a young age of $\sim 60$~Myr, which becomes 235~Myr if it is instead considered to be the first harmonic of this period. 
Some constraints on the orbit are obtained using the MC method (see Sect. 4.5.3 and Table~\ref{tab:orbits}).

\medskip
\noindent{\bf TYC 6872-1011-1}
This is a confirmed member of $\beta$ Pic MG (97\% membership probability)
Our data indicate that this is a close system with two components having a contrast of about 2-3 mag. TESS periodogram shows two very significant short period peaks consistent with both the components being fast rotators. There is one source in Gaia DR2 at 5 arcsec that results to be a background object.

Accurate PMA data are not available for this target, which is not listed in the Hipparcos catalogue. However, the three SHINE observations covers a significant fraction of the orbit and combined with the assumption about the masses of the components allow the orbit of this system to be strongly constrained. In Section 4.5.3 we used these data to provide the following orbital parameters from the MC analysis: a semi-major axis of $a=6.61^{+0.98}_{-0.90}$~au (corresponding to a period of $19\pm 4$~yr), an intermediate eccentricity of $e=0.52^{+0.06}_{-0.09}$, and a high inclination ($i=105.8\pm 1.0$~degree).

\medskip
\noindent{\bf HIP 93580 = HD 177178}
This is an early-type star, classified as a possible member of the AB Dor MG \citep{zuckerman2011}. However, the BANYAN kinematic analysis classifies it as field object (99.9\%). 
The companion seen in SPHERE images was previously identified by \citet{Rameau2013,Galicher2016}.

There is an X-ray source, 1RXS J190334.1+014838, with a nominal position at 40 arcsec from the star, quite large but still compatible with the pointing accuracy of ROSAT. Considering the spectral type of the primary, \citet{schroder2007} argued for an origin from an unknown (at that time) stellar companion. The X-ray luminosity of 2.55e29 erg/s is fully compatible for an early M object (as expected for HD 177178B) at the ages of the Pleiades, but still compatible with the scatter of Hyades at older ages \citep{magaudda2020}.
We derived the age from isochrone fitting of the primary. There is an inconsistency between the published spectral types in the literature (between A3 to A5) and the optical colours (expected to be basically unaffected by the presence of the companion), which indicates spectral types of A6-A7 when compared to the \citet{pecaut2013} sequence.  
The stars was observed with Spitzer, with no detection of IR excess \citep{zuckerman2011}.

Adding our two epochs to the previous observations \citep{Rameau2013,Galicher2016}, we have a total of four astrometric points and 12 RVs from HARPS. 
While a full orbital solution is too uncertain, quite good results can be obtained fixing the total mass of the system at the value given by photometry (2.22~$M_\odot$). The period is constrained to be about 28 yrs in order to get a reasonable agreement of the PMA with that observed by \citet{Kervella2019} for epochs 1991.25 and 2015.5. The best solution is for a low eccentric orbit ($e=0.27$) with semi-major axis of $a=0.215$~arcsec, $T0=2000.36$, $\Omega=118$~degree, $\omega=156$~degree, and $i=59.6$~degree.  

\medskip
\noindent{\bf HIP 95149 = HD 181321 = GJ 755 = HR 7330}
This nearby G2V star was shown to be an SB1 by \citet{nordstrom2004}, \citet{gunther2007}, and \citet{Grandjean2020}. These authors proposed two possible orbital solutions, with ambiguity due to a long gap in the RV time series (period 1600 days and $M~\sin{i}$ of about 0.1~$M_\odot$, or period about 3200 days with $M~\sin{i}$=0.18~$M_\odot$). We present here the first direct detection of the companion, at a projected separation between 213 and 251 mas in three epochs in 2018-2019 \citep[the companion was not detected in previous imaging efforts,][]{Biller2007}. The direct detection allows us to disentangle between the two proposed orbital solutions (see below).
The star has prominent astrometric signatures (astrometric acceleration by Hipparcos, $\Delta \mu$)
The availability of dynamical mass makes the determination of the age of system highly relevant. The star was flagged as a member of the Castor MG by \citet{ribas2003}, while BANYAN flags it as a field object  \citep[the Castor MG is not included, after considerations about its actual non-existence by][]{zuckerman2013}. 
\citet{Fuhrmann2017} noted the similar kinematic of the star to HR 2882 and 53 Aqr and proposed 
all of them as members of Octans-Near group \citep{zuckerman2013}.
An age of 320 Myr was adopted in \citet{bonavita2016}. The TESS photometry gives a period of 3.847 d. This is close to the period of Pleiades members of similar colours. To check the possibility that this is not the true period but rather its first harmonic (which would imply a significantly older gyro age), we considered the projected rotational velocity \citep[13 km/s, ][]{nordstrom2004}. For the measured period, this corresponds to an inclination close to edge-on, while a period two times longer is unphysical ($ \sin i \sim 2$).
The Li EW we measured on HARPS spectra (135 m\AA) is also compatible with Pleiades age or slightly older. X-ray and chromospheric emission are also compatible. We then conclude that the association with Octans-Near is plausible and that in any case the age indicators provide an age close to that of the Pleiades and AB Dor MG.

The orbit of HIP~95149 can be determined quite accurately combining our three position measurements with SHINE with the RVs obtained with HARPS that we downloaded from the database by \citet{Trifonov2020}. We fitted the orbit using the code {\it Orbit} by \citet{Tokovinin2016b} to find the best astrometric orbit. Results are shown in Table~\ref{tab:orbits}. This solution clearly corresponds to the long period solution of \citet{Grandjean2020}, but has a lower eccentricity. The sum of the masses of the two components is $M_A+M_B=0.93\pm 0.02$~$M_\odot$. This is lower than the sum of the mass of the two components derived from photometry ($M_A+M_B=1.28\pm 0.05$~$M_\odot$). The amplitude of the RV curve for the A component (1.70$\pm$0.07~km/s) yields the binary mass ratio that is $q=0.21\pm 0.01$, to be compared with the value of $q=0.25\pm 0.04$ determined from photometry.

\medskip
\noindent{\bf TYC 6299-2608-1 = HD 185673}
This late F star is a wide companion (sep 45") to the K1 giant 54 Sgr = HR 7476 = HIP 96808 = HD 185644.
The F type star has detectable X-ray emission, low chromospheric emission ($\log{R_{HK}}$=-5.00), and 
lithium slightly below the Hyades locus.
The space velocity is well outside the kinematic space typical of young star, although clearly within thin disk.  
We then conclude that the star is an old star, mistakenly selected as young because of the X-ray emission, lithium, and low quality B-V colour in \citet{Torres2006}.
The age derived by isochrone fitting is 4.0$\pm$1.0 Gyr. 
The age of red giant is less constrained. The mass of the red giant when allowing only this age range is 1.39 $M_\odot$. 
The red giant companion is flagged as a probable SB by \citet{demedeiros2014}.
If confirmed, this would make the system quadruple.

\medskip
\noindent{\bf HIP 97255 = HD 186704}
This G0 star was classified by \citet{zuckerman2013} as a member of the Octans-Near association. This group is not included in the current version of BANYAN $\Sigma$.
Beside the companion first detected with our observations at 0.32", there is an additional wide companion at 10" (the M type flare star V1406 Aql), making the system triple. The newly detected companion is likely the responsible for the RV variations identified by \citet{nidever2002, nordstrom2004,tremko2010,soubiran2018,Grandjean2020} and for the astrometric signature ($\Delta \mu$). To our knowledge, there are no published RV orbital solutions, although \citet{Tokovinin2014} mention a period of 3990d from a priv. comm. by D. Latham. The companion was not detected in previous imaging efforts by \citet{Biller2007}. 
\citet{bonavita2016} adopted an age of 125 Myr. There are no data from TESS yet. 
The rotation period \citep[3.51 d, ][]{kiraga2012} and the Li EW \citep[120 m\AA][]{zuckerman2013} are compatible with an age close to the Pleiades. The CMD position above Zero Age Main Sequence for V1406 Aql is also fully compatible with this age assignment.
We used the RV series from the SOPHIE and ELODIE spectrographs \citep{sophie, elodie}, our measures of the position, and the PMA measured by \citet{Kervella2019} to constrain the orbit, adopting the stellar masses given by photometry. We find that the solution is strongly constrained by these datasets (see Table~\ref{tab:orbits}). In this fit, the amplitude of the RV curve of the primary is determined independently of the assumption of the masses, simply via a Keplerian fitting the RV curve. Comfortably, the secondary mass derived from this parameter (0.261$\pm$0.005~$M_\odot$) agrees well with the mass determined from the photometry ($0.289^{+0.017}_{-0.051}~M_\odot$).

\medskip
\noindent{\bf TYC 5164-567-1  =  BD-03 4778} 
The membership probability on AB Dor MG is 36.1\%;  independent of the kinematic assignment, the age indicators are fully compatible with the age of the AB Dor MG, with the lithium larger than the median values for the Pleiades and AB Dor MG but within the distribution. There are no TESS data are available yet. We adopt the AB Dor age. The star is a close binary according to \citet{Elliott2014} and Gaia DR2, which gives two separate astrometric solutions for the two components that clearly indicate that the two components are physically linked with each others.

The epoch of our observation is not far from that of Gaia DR2; the relative position of the two components agree fairly well. On the other hand, the positions derived from \citet{Elliott2014} combined with our and Gaia DR2 ones yield a projected proper motion that exceed the expected escape velocity and disagrees with the differences in the proper motion of the two components as measured by Gaia DR2. These discrepancies might be solved if one of the two components is itself a binary, which was, however, unresolved at the epoch of the SHINE observation.

\medskip
\noindent{\bf TYC 8400-0567-1 = CD-50 12872}
This is a star with a close companion discovered by our observations with SPHERE.
It also has a large $\Delta \mu$ and possible RV variability
between Gaia DR2 and  \citet{Elliott2014}.
No data are available from TESS at present.
The Li EW measured by \citet{Torres2006} is below the mean locus of Pleiades
and AB Dor although within the observed distribution of members.
We adopt  180 Myr with limits 100-300 Myr.

\begin{figure}
\centering
\includegraphics[width=9truecm]{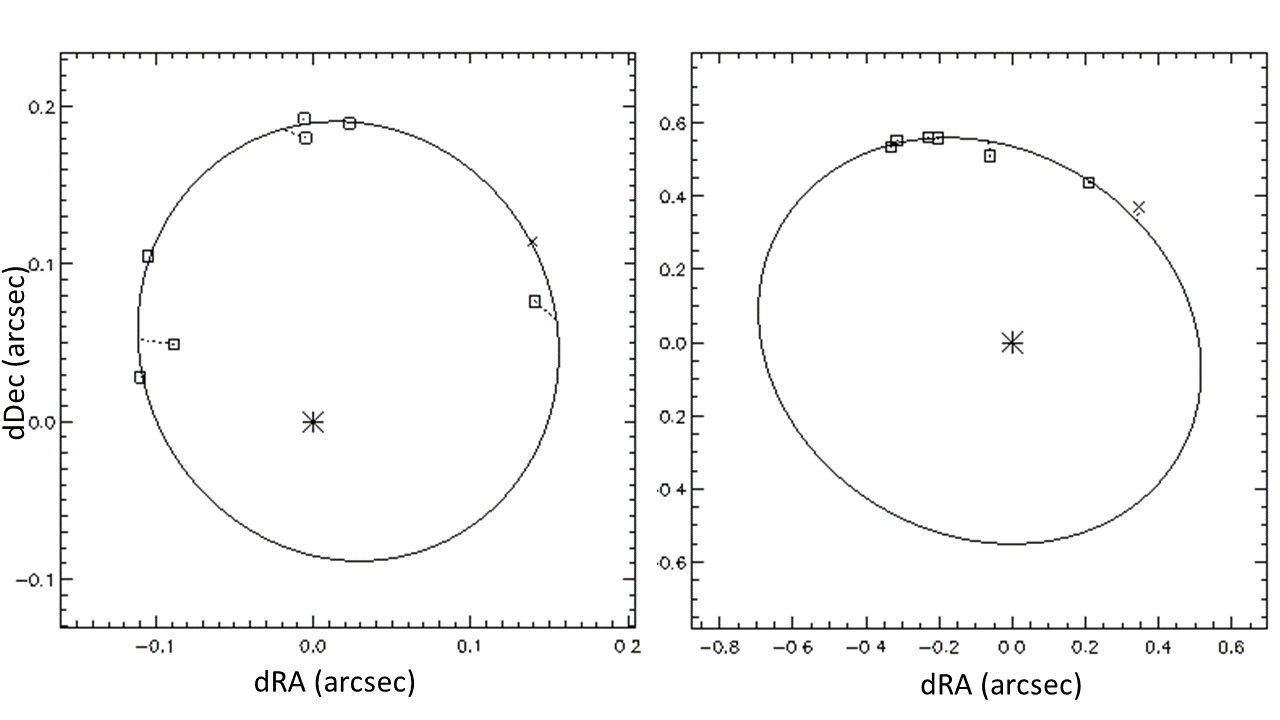}
\caption{Best orbital fit for the inner (left panel) and outer (right panel) components of HIP~107948. Positions plotted in the right panel are with respect the barycentre of the inner binary, with masses derived from photometry (see Table~\ref{tab:sys_char}). Open symbols are the data of highest accuracy; crosses are for lower quality data. Dotted lines connect observed with predicted positions for the different observations.}
\label{fig:orbit_HIP107948}
\end{figure}

\medskip
\noindent{\bf HIP 107948}
This  triple system with all components within 0.6 arcsec was discovered by \citet{Elliott2015}. Parallax and proper motion are not included in Gaia DR2, so the Hipparcos values were used; they have large error bars, presumably because of the complications related to this object being a close triple system. No data are available from TESS at present.

A good  orbital solution can be obtained for the two inner components using the ten epochs available combining our data with the literature \citep{Elliott2015, Galicher2016, Tokovinin2016, Tokovinin2018} (see Fig.~\ref{fig:orbit_HIP107948}, left panel). We used the code {\it Orbit} by \citet{Tokovinin2016b} to obtain the following parameters: $P=9.55\pm 0.09$~yr, $T0=2000.23\pm 0.19$, $e=0.410\pm 0.027$, $a=152.0\pm 4.4$~mas, $\Omega=48.4\pm 9.9$~degree, $\omega=106\pm 10.0$~degree, $i=28.0\pm 4.0$~degree. Combined with the parallax, this orbit corresponds to a sum of the masses for the two components of $M_A+M_B=1.09^{+0.57}_{-0.37}$~$M_\odot$, in good agreement with that derived from the photometry (0.96~$M_\odot$). By far the main contribution to the error in the masses is due to the uncertainties in the parallax. This orbit is in reasonable agreement with that proposed by \citet{Tokovinin2020a}, who, however, assumed $\omega=i=0$.

Since only a fraction of the orbit is covered for the outer companion, its orbit is not as well defined (see Fig.~\ref{fig:orbit_HIP107948}, right panel). Even assuming the total mass from photometry, we found solutions with similar $\chi^2$ over a quite wide range of semi-major axis (between 0.57 to 1.4 arcsec) and periods (between 70 and 260 yr). We should however notice that the system is likely unstable if the semi-major axis of the outer companion is less than 1 arcsec $\sim$ 30 au (period less than 150 yr), as suggested by a comparison with the equations by \citet{Holman1999}. This limits significantly the range of possible solutions. Orbital eccentricity is in the range 0.38 - 0.55. The orbital plane is not far from that of the inner binary, because $35<\Omega<36$ degree and $38<i<49$~degree, in agreement with what found for the majority of triple systems with projected separation below 50~au \citep{Tokovinin2017a}. 

We note that these two orbits (that were determined independently of each other) are both prograde, they are not too far from being coplanar (the mutual inclination is $16\pm 13$~degrees), have very similar eccentricity ($e \sim 0.4$) and have a quite similar longitude of the periastron. On the whole, this supports mutual interactions and a common formation within a disk.

\medskip
\noindent{\bf HIP 109285 = $\mu$ PsA = HD 210049 = HR 8431}
This is an early-type field object. It has a newly detected companion at 60 mas, with $\Delta J=2.33$ and then moderately massive (expected be a late G star). The star is characterised by a large RUWE in Gaia and rather large $\Delta \mu$. Hipparcos and Gaia parallaxes also differ formally at 2.5 $\sigma$.
This could be due to non-optimal performance of Gaia for very bright stars (V=4.495), but also on the presence of the secondary.
The isochrone age of the primary 
is 240$\pm$130 Myr. 1RXS J220824.1-325933 (nominal separation 20") is the probable X-ray counterpart of the late-type secondary \citep[the presence of a late-type companion was hypothesised by][]{schroder2007}.
The X-ray luminosity   corresponds to an age of $\sim$ 300 Myr if coming entirely from the secondary.
TESS photometry yields a period of P=5.557 d that, if interpreted as the rotational period of the secondary, would imply an age close to that of the Pleiades. Both these values are within the range
of the isochrone age of the primary.

\medskip
\noindent{\textbf{HIP 109427 = $\theta$ Peg = GJ 9771 = HD 210418 =  HR 8450} }
This field object was flagged as an SB2 by \citet{gray1987} without further details. However, RV monitoring by \citet{lagrange2009} and \citet{becker2015} did not report indications of companions, while mentioning RV variability of few hundred m/s likely of stellar origin. The null results of the interferometric observations by \citet{marion2014} cast further doubt of the existence of a bright close companion. Because of its faintness ($\Delta J =7.07$), the companion detected with our observations cannot be responsible for an SB2 appearance, though it is likely responsible for the $\Delta \mu$ signature already mentioned by \citet{makarov2005} and the Hipparcos astrometric acceleration. The newly imaged companion was not detected in previous AO observations by \citet{DeRosa2014} or \citet{stone2018}. Following the submission of the first version of our paper, \citet{Steiger2021} published an independent discovery of this companion. Most of the spectral type determinations for the primary in the literature are either A1 or A2. The optical colours are mostly intermediate between A2 and A3. The star results somewhat evolved outside Zero Age Main Sequence. The X-ray non-detection is compatible with the moderately old age and low luminosity of the companion. There is no IR excess from the Spitzer  or Herschel observations \citep{su2006,thureau2014}.

\citet{Steiger2021} published a preliminary orbit for the system. We combined our astrometric measure with theirs and the PMA measures by \citet{Kervella2019} to improve this orbit determination. At variance with \citet{Steiger2021}, we fixed the mass of the two components at the values given by photometry. The orbital parameters we derived (see Table~\ref{tab:orbits}) are quite similar to those obtained by \citet{Steiger2021}; however, since we could use more data, the error bars are reduced.

\medskip
\noindent{\bf HIP 113201}
Orbit and dynamical masses coupling imaging, HARPS RV, and astrometry and comprehensive analysis of the stellar properties will be presented in Biller et al. (in preparation). The star is considered here only for statistical purposes.

\end{appendix}

\end{document}